\documentclass[fleqn,usenatbib]{mnras}

\usepackage{newtxtext,newtxmath}

\usepackage[T1]{fontenc}
\usepackage{ae,aecompl}

\usepackage{graphicx}	% Including figure files
\usepackage{amsmath}	% Advanced maths commands
\usepackage{amssymb}	% Extra maths symbols

\usepackage[version=4]{mhchem} % Molecular typesetting
\usepackage{soul}              % Highlight new text in revised version

\newcommand{\rosetta}{{\sl Rosetta}}
\renewcommand\hl[1]{#1}       % Uncomment to turn off highlighting
\title[\ce{H2O} and \ce{O2} Absorption in the Coma of Comet~67P/C-G]{\ce{H2O} and \ce{O2} Absorption in the Coma of Comet~67P/Churyumov-Gerasimenko Measured by the Alice Far-Ultraviolet Spectrograph on \textit{Rosetta}}

\author[B. A. Keeney et~al.]{Brian A. Keeney,$^{1}$\thanks{E-mail: bkeeney@gmail.com (BAK)} S. Alan Stern,$^{1}$ Michael F. A'Hearn,$^{2}$ Jean-Loup Bertaux,$^{3}$
\newauthor
Lori M. Feaga,$^{2}$ Paul D. Feldman,$^{4}$ Richard A. Medina,$^{1}$ Joel Wm. Parker,$^{1}$ 
\newauthor
Jon P. Pineau,$^{5}$ Eric Schindhelm,$^{1}$ Andrew J. Steffl,$^{1}$ M. Versteeg,$^{6}$ and
\newauthor
Harold A. Weaver$^{7}$
\\
$^{1}$Southwest Research Institute, Department of Space Studies, Suite 300, 1050 Walnut Street, Boulder, CO 80302, USA \\
$^{2}$University of Maryland, Department of Astronomy, College Park, MD 20742, USA \\
$^{3}$LATMOS, CNRS/UVSQ/IPSL, 11 Boulevard d’Alembert, 78280 Guyancourt, France \\
$^{4}$Johns Hopkins University, Department of Physics and Astronomy, 3400 N. Charles Street, Baltimore, MD 21218, USA \\
$^{5}$Stellar Solutions, Inc., 250 Cambridge Ave., Suite 204, Palo Alto, CA 94306, USA \\
$^{6}$Southwest Research Institute, 6220 Culebra Road, San Antonio, TX 78238, USA \\
$^{7}$Johns Hopkins University Applied Physics Laboratory, 11100 Johns Hopkins Road, Laurel, MD 20723, USA
}

\date{Accepted XXX. Received YYY; in original form ZZZ}

\pubyear{2017}

\begin{document}
\label{firstpage}
\pagerange{\pageref{firstpage}--\pageref{lastpage}}
\maketitle

% Abstract of the paper
\begin{abstract}
We have detected \ce{H2O} and \ce{O2} absorption against the far-UV continuum of stars located on lines of sight near the nucleus of Comet~67P/Churyumov-Gerasimenko using the Alice imaging spectrograph on \rosetta. These stellar appulses occurred at impact parameters of $\rho=4$-20~km, and heliocentric distances ranging from $R_h=-1.8$ to 2.3~AU (negative values indicate pre-perihelion observations). The measured \ce{H2O} column densities agree well with nearly contemporaneous values measured by VIRTIS-H. The clear detection of \ce{O2} independently confirms the initial detection by the ROSINA mass spectrometer; however, the relative abundance of $\ce{O2}/\ce{H2O}$ derived from the stellar spectra \hl{(11\%-68\%, with a median value of 25\%)} is considerably larger than published values found by ROSINA. The cause of this difference is unclear, but potentially related to ROSINA measuring number density at the spacecraft \hl{position} while Alice measures column density along a line of sight that passes near the nucleus.
\end{abstract}

\begin{keywords}
comets: individual (67P) -- ultraviolet: planetary systems
\end{keywords}

\section{Introduction}
\label{intro}

One of the most significant results from the \rosetta\ mission to Comet 67P/Churyumov-Gerasimenko (67P/C-G) has been the persistent detection of \ce{O2} in the coma \citep{bieler15,fougere16} by the Double Focusing Mass Spectrometer (DFMS) of the Rosetta Orbiter Spectrometer for Ion and Neutral Analysis \citep[ROSINA;][]{balsiger07}. The initial detection by \citet{bieler15} found that the relative number density of \ce{O2} with respect to \ce{H2O} ranged from 1-10\%, with a mean of $n_{\ce{O2}}/n_{\ce{H2O}}=3.85\pm0.85$\% for measurements taken between September~2014 and March~2015. Further modeling by \citet{fougere16} found that the relative production rate of \ce{O2} with respect to \ce{H2O} is $\approx1$-2\% for measurements taken prior to February~2016. Both studies find that the number densities of \ce{O2} and \ce{H2O} are highly correlated, with Pearson correlation coefficients $>0.8$.

\begin{table*}

\caption{Journal of Targeted Stellar Appulse Observations}
\label{tab:targeted}

\begin{tabular}{llllcccccc}
\hline
Star & Sp. Type & Obs. Type & Date & UTC & Duration & $R_h$ & $\phi$    & $\theta$  & $\rho$ \\
     &          &           &      &     & (min)    & (AU)  & ($\degr$) & ($\degr$) & (km)   \\
\hline
HD~140008 & B5~V        & Appulse   & 2015 Dec 25 & 14:27:11 &  57 & 1.97 &  89.8       &   4.8-5.3   &  6.4-7.2  \\
          &             & Revisit~1 & 2016 Feb 29 &          &  37 & 2.47 &  92.9       &  88.3-88.9  &           \\
          &             & Revisit~2 & 2016 Mar 12 &          &  39 & 2.56 &  91.9-92.0  &  87.0-88.0  &           \\
HD~144294 & B2.5~V      & Appulse   & 2015 Dec 25 & 15:37:11 & 111 & 1.97 &  89.8       &   9.9-10.8  & 13.3-14.6 \\
          &             & Revisit   & 2016 Mar  4 &          & 127 & 2.51 &  91.8       & 120.0-123.8 &           \\
HD~42933  & B1/2~III    & Appulse   & 2016 Jan 10 & 07:19:29 & 164 & 2.09 &  89.6       &   5.1-6.5   &  7.0-8.9  \\
          &             & Revisit~1 & 2016 Feb 29 &          &  51 & 2.47 &  92.9       & 171.1-171.8 &           \\
          &             & Revisit~2 & 2016 Feb 29 &          &  77 & 2.48 &  92.6       & 172.9-173.8 &           \\
HD~89890  & B5~II       & Appulse   & 2016 Jan 18 & 13:28:59 & 169 & 2.16 &  60.4-60.5  &  12.1-12.9  & 17.1-18.2 \\
          &             & Revisit   & 2016 Mar 15 &          &  84 & 2.59 &  89.1       & 145.3-148.4 &           \\
HD~40111  & B0/1~II/III & Appulse~1 & 2016 Jan 25 & 17:32:33 & 222 & 2.21 &  60.2-60.4  &  11.4-12.5  & 14.0-15.4 \\
          &             & Appulse~2 & 2016 Feb  9 & 19:38:27 & 170 & 2.33 &  64.9-65.6  &   8.8-9.8   &  7.8-8.6  \\
          &             & Revisit~1 & 2016 Feb 23 &          & 131 & 2.43 &  89.2-90.1  & 120.4-122.3 &           \\
          &             & Revisit~2 & 2016 Feb 26 &          &  88 & 2.45 &  94.8       & 171.2-172.0 &           \\
HD~144206 & B9~III      & Appulse   & 2016 Feb  1 & 13:28:59 & 170 & 2.26 &  60.2-60.4  &   9.8-9.9   & 10.0-10.1 \\
          &             & Revisit   & 2016 Apr  1 &          &  62 & 2.71 & 112.1-112.9 & 177.8-178.7 &           \\
\hline

\multicolumn{10}{l}{\textit{Notes.} The phase angle is denoted by $\phi$, and the off-nadir angle by $\theta$. The last column lists the impact parameter, $\rho$.}

\end{tabular}
\end{table*}

Surprisingly, \ce{O2} is the fourth most abundant species in the coma of 67P/C-G \citep[behind \ce{H2O}, \ce{CO2}, and \ce{CO};][]{leroy15,fougere16}, despite the fact that it had never been detected in a cometary coma before \citep{bieler15}. Subsequent reanalysis of mass spectometer data from {\sl Giotto}'s visit to Oort-Cloud Comet~1P/Halley has found that $n_{\ce{O2}}/n_{\ce{H2O}}=3.7\pm1.7$\% is consistent with the measurements \citep{rubin15}, suggesting that \ce{O2} may be a common constituent of all comets, not just Jupiter Family Comets such as 67P/C-G. New theories are being developed to explain these \ce{O2} detections, such as trapping \ce{O2} in clathrates prior to agglomeration during comet formation \citep{mousis16}, astrochemical production of \ce{O2} in dark clouds or forming protoplanetary disks \citep{taquet16}, and formation of \ce{O2} during the evaporation of \ce{H2O} ice via dismutation of \ce{H2O2} \citep*{dulieu17}. 

In this Paper, we present \ce{H2O} and \ce{O2} column densities measured along lines of sight to background stars projected near the nucleus of 67P/C-G by the Alice far-UV spectrograph \citep{stern07}. These stellar sight lines allow the coma of 67P/C-G to be studied in far-UV absorption, where column densities can be measured directly. Alice's previous characterizations of the coma of 67P/C-G have primarily used emission lines from CO and atomic hydrogen, oxygen, carbon, and sulphur \citep[e.g.,][]{feldman15,feldman16}. While the strengths of these emission lines can only be used to derive molecular column densities under specific assumptions (i.e., pure resonance fluorescence), the ratios of strong, commonly observed lines can be diagnostic of physical conditions in the coma. \citet{feldman16} inferred that \ce{O2} was the primary driver of certain gaseous outbursts that exhibit a sudden increase in the O\,{\sc i}~$\lambda1356/\lambda1304$ ratio in the sunward coma without any corresponding increase in dust production. \citet{feldman16} estimate that $\ce{O2}/\ce{H2O}\geq50$\% during these outbursts, substantially higher than the mean value of $3.85\pm0.85$\% found by \citet{bieler15}. 

Several of \rosetta's instruments are capable of measuring the abundance of \ce{H2O} (as well as \ce{CO} and \ce{CO2}) in the coma of 67P/C-G. Most notably, ROSINA measures the number density of water, $n_{\ce{H2O}}$, at the spacecraft location using mass spectroscopy, while the Visible and Infrared Thermal Imaging Spectrometer \citep[VIRTIS;][]{coradini07} and the Microwave Instrument for the Rosetta Orbiter \citep[MIRO;][]{gulkis07} measure the column density of water, $N_{\ce{H2O}}$, along a specific line of sight using rotational and/or vibrational transitions. The UV-absorption spectra presented herein also allow Alice to directly measure $N_{\ce{H2O}}$, and facilitate comparisons with nearly contemporaneous measurements from ROSINA \citep{fougere16} and VIRTIS-H \citep[the high spectral resolution channel of VIRTIS;][]{bockelee-morvan16}.

In contrast to the situation with \ce{H2O}, only Alice and ROSINA are capable of directly measuring \ce{O2}. This makes the observations reported herein an important and unique confirmation of the initial \ce{O2} detections \citep{bieler15}. However, direct comparisons between ROSINA's \textit{in-situ} measurements and Alice's measurements along specific lines of sight are not straightforward. The remainder of this Paper is organized as follows: the Alice \hl{spectrograph} and stellar spectra are described in \autoref{sec:obs}; \ce{H2O} and \ce{O2} column densities are derived in \autoref{sec:fits}; our values are compared with ROSINA and VIRTIS-H measurements in \autoref{sec:discussion}; and our conclusions are presented in \autoref{sec:conclusions}.

\section{Stellar Appulse Observations}
\label{sec:obs}

Alice is a low-power, lightweight far-UV imaging spectrograph funded by NASA for inclusion on the ESA \rosetta\ orbiter \citep{stern07}. It covers the wavelength range 750-2050~\AA\ with a spectral resolution of 8-12~\AA, and has a \hl{slit that is $6\degr$ long, and} narrower in the center ($0\fdg05$ wide) than the edges \citep[$0\fdg1$ wide;][]{stern07}. Over the course of \rosetta's orbital escort mission, Alice probed the sunward coma of 67P/C-G in absorption 30 times using UV-bright stars located along lines of sight near the nucleus as background sources. Here we report on the 29~observations (``appulses'') that were not occulted by the nucleus; we will report the details of our single stellar occultation separately (B.~Keeney et~al., in prep).

Quantifying the nature of the cometary coma required re-observing, or ``revisiting'', these stars when they were far from the nucleus to characterize their intrinsic stellar spectra. This allowed us to isolate the coma absorption signature from the combined background effects of the stellar continuum and interstellar absorption. Further, there are two varieties of appulse observations, which we term ``targeted'' and ``archival'' appulses.

\begin{table*}

\caption{Journal of Archival Stellar Appulse Observations}
\label{tab:archival}

\begin{tabular}{llllcccccc}
\hline
Star & Sp. Type & Obs. Type & Date & UTC & Duration & $R_h$ & $\phi$    & $\theta$  & $\rho$ \\
     &          &           &      &     & (min)    & (AU)  & ($\degr$) & ($\degr$) & (km)   \\
\hline
HD~26912  & B3~IV           & Appulse & 2015 Apr 30 & 02:00:27 &  18 & $-1.75$ &  72.5-72.6  &         1.7 &  4.5 \\*
          &                 & Revisit & 2016 Mar 26 &          &  95 &   2.67  & 128.7-131.5 &   77.7-80.0 &      \\
HD~3901   & B2~V            & Appulse & 2015 May  3 & 21:32:21 &  11 & $-1.72$ &  60.3       &         1.7 &  4.0 \\*
          &                 & Revisit & 2016 Aug  5 &          &  12 &   3.52  &  ---        &        44.9 &      \\
HD~29589  & B8~IV           & Appulse & 2015 May 27 & 03:19:43 &  46 & $-1.55$ &  65.9       &         1.3 &  7.1 \\*
          &                 & Revisit & 2016 Jul 22 &          &  24 &   3.44  &  88.8       &        99.0 &      \\
HD~174585 & B3~IV           & Appulse & 2015 Jun  8 & 00:40:54 &  16 & $-1.48$ &  87.4       &         1.5 &  5.4 \\*
          &                 & Revisit & 2016 Aug  5 &          &  12 &   3.52  &  90.1       &        92.9 &      \\
HD~180554 & B4~IV           & Appulse & 2015 Jun 28 & 00:16:04 &   4 & $-1.36$ &  89.2       &         2.4 &  7.7 \\*
          &                 & Revisit & 2016 Aug  5 &          &  12 &   3.52  &  92.6       &        96.9 &      \\
HD~191692 & B9.5~III        & Appulse & 2015 Jul 12 & 22:59:29 &  14 & $-1.30$ &  88.8       &         2.5 &  6.8 \\*
          &                 & Revisit & 2016 Apr 19 &          &  17 &   2.84  &  86.4       &        35.2 &      \\
HD~195810 & B6~III          & Appulse & 2015 Jul 25 & 08:56:47 &  11 & $-1.26$ &  90.0       &         2.0 &  6.5 \\*
          &                 & Revisit & 2016 Apr 19 &          &  17 &   2.84  &  86.5       &        41.1 &      \\
HD~192685 & B3~V            & Appulse & 2015 Jul 26 & 08:54:44 &  11 & $-1.26$ &  89.9       &         2.5 &  7.4 \\*
          &                 & Revisit & 2016 Jun 27 &          &  17 &   3.29  &  93.8       &        99.9 &      \\
HD~68324  & B2~V            & Appulse & 2015 Aug  9 & 19:39:33 &  20 & $-1.24$ &  89.0       &         1.3 &  7.0 \\*
          &                 & Revisit & 2016 Jun  6 &          &  30 &   3.15  &  67.9-68.2  &   88.2-88.4 &      \\
HD~66006  & B2/3            & Appulse & 2015 Aug 10 & 04:28:49 &  21 & $-1.24$ &  89.0       &         1.0 &  5.7 \\*
          &                 & Revisit & 2016 Jun  6 &          &  31 &   3.15  &  69.1-69.5  &   86.6-87.0 &      \\
HD~64722  & B2~IV           & Appulse & 2015 Aug 10 & 18:45:04 &  25 & $-1.24$ &  89.2       &         2.5 & 14.2 \\*
          &                 & Revisit & 2016 Jun 27 &          &  21 &   3.29  &  93.8       &   47.7-48.1 &      \\
HD~39844  & B6~V            & Appulse & 2015 Aug 13 & 00:57:11 &  14 &   1.24  &  89.3       &         2.2 & 12.6 \\*
          &                 & Revisit & 2016 Jun 27 &          &  17 &   3.29  &  93.8       &        35.4 &      \\
HD~207330 & B3~III          & Appulse & 2015 Aug 27 & 03:18:10 &  12 &   1.26  &  79.7       &         1.5 & 10.4 \\*
          &                 & Revisit & 2016 Apr  4 &          & 116 &   2.67  &  83.1-83.3  & 149.5-150.8 &      \\
HD~109387 & B6~III          & Appulse & 2015 Sep  1 & 07:04:57 &  10 &   1.27  &  70.4       &         1.2 &  8.6 \\*
          &                 & Revisit & 2016 Jun  5 &          &  22 &   3.15  &  85.3-85.6  & 135.2-135.7 &      \\
HD~124771 & B3~V            & Appulse & 2015 Sep 10 & 04:52:33 &  12 &   1.29  & 119.9       &         3.6 & 20.1 \\*
          &                 & Revisit & 2016 Jun  6 &          &  18 &   3.15  &  68.8       &        50.0 &      \\
HD~21428  & B3~V            & Appulse & 2015 Nov  2 & 16:07:12 &  10 &   1.58  &  60.2       &         2.8 & 12.8 \\*
          &                 & Revisit & 2016 Aug  5 &          &  12 &   3.52  &  94.9       &        28.2 &      \\
HD~32249  & B3~IV           & Appulse & 2015 Nov  6 & 06:42:59 &   9 &   1.60  &  61.3       &         2.6 & 10.7 \\*
          &                 & Revisit & 2016 Jul 22 &          &   9 &   3.44  &  88.9       &        77.1 &      \\
HD~33328  & B2~IV           & Appulse & 2015 Nov  6 & 09:41:26 &   4 &   1.60  &  61.6       &         1.4 &  5.9 \\*
          &                 & Revisit & 2016 Jul 22 &          &   9 &   3.44  &  88.7       &        81.6 &      \\
HD~106625 & B8~III          & Appulse & 2015 Nov 13 & 08:16:25 &   6 &   1.65  &  61.1       &         1.6 &  4.7 \\*
          &                 & Revisit & 2016 Jul 22 &          &  24 &   3.44  &  89.1       &        70.7 &      \\
HD~27376  & B9~V            & Appulse & 2015 Nov 27 & 22:04:50 &   8 &   1.76  &  90.0       &         3.4 &  8.1 \\*
          &                 & Revisit & 2016 Jul 22 &          &  14 &   3.44  &  88.9       &        48.7 &      \\
HD~23466  & B3~V            & Appulse & 2015 Dec 16 & 22:38:45 &   7 &   1.90  &  89.8       &         2.9 &  5.2 \\*
          &                 & Revisit & 2016 Aug  2 &          &  12 &   3.51  &  78.5       &        60.3 &      \\
HD~144217 & B1~V            & Appulse & 2015 Dec 26 & 06:30:02 &   8 &   1.98  &  89.8       &         3.2 &  4.4 \\*
          &                 & Revisit & 2016 Aug  5 &          &  12 &   3.52  &  91.5       &       130.4 &      \\
\hline

\multicolumn{10}{l}{\textit{Notes.} The phase angle is denoted by $\phi$, and the off-nadir angle by $\theta$. The last column lists the impact parameter, $\rho$.}

\end{tabular}
\end{table*}

For the targeted appulses, we actively searched during operations planning for upcoming opportunities where a known bright star would be located within a few degrees of the nucleus. Inertial pointings were designed that facilitated long stares at these stars during the appulses, at the expense of a time-varying distance to the nucleus over the course of each observation. These targeted appulses were observed between 2015~Dec~25 and 2016~Feb~1 at heliocentric distances of $R_h=1.97$-2.26~AU, and are characterized by long exposure times (typically 12~Alice spectral images with exposure times of 10-20~minutes each were obtained per appulse), large off-nadir angles ($\theta\approx5$-$10\degr$), and large $R_h$ compared to their archival counterparts.

To complement the targeted appulses, we also searched the extensive Alice archive ($\sim40,000$ exposures include the nucleus in the field-of-view) for instances where we serendipitously observed a UV-bright star near the nucleus as part of normal operations. This search returned hundreds of candidates that were prioritized by the star's brightness and proximity to the nucleus, as well as the duration of the appulse and its proximity to the comet's perihelion passage on 2015~Aug~12, when coma activity was near its peak \citep{fougere16}. Since our typical pointing during normal operations was fixed with respect to the nucleus (i.e., not an inertial reference frame), we do not know the exact duration of the archival appulses because the star is moving with respect to the slit; however, we can estimate their durations with uncertainties of $\sim10$\% using NAIF/SPICE \citep{acton96}. The archival appulses were observed between 2015~Apr~29 and 2015~Dec~26 at $R_h=1.24$-1.98~AU, and typically have shorter durations (10-20~min), smaller off-nadir angles ($\theta<5\degr$), and smaller $R_h$ than their targeted counterparts. However, the smaller off-nadir angles for the archival appulses are somewhat counteracted by the large spacecraft-comet distance, $\Delta$, near perihelion, which led to similar impact parameters ($\rho=\Delta\sin{\theta}\approx5$-20~km) for all appulses.

\autoref{tab:targeted} and \autoref{tab:archival} list the properties of the 7~targeted and 22~archival appulses, respectively. The following information is listed by column: (1) the name of the star; (2) the stellar type and luminosity class as listed by SIMBAD \citep{wenger00}; (3) the observation type (either ``appulse'' or ``revisit''); (4) the date of observation; (5) the total exposure time, in minutes; (6) the heliocentric distance, $R_h$, in AU, where negative values indicate that the observation occurred prior to perihelion on 2015~Aug~12; (7) the phase angle, $\phi$, in degrees; (8) the off-nadir angle, $\theta$, in degrees; and (9) the impact parameter, $\rho$, in km. The impact parameter is only listed for appulse observations, not revisits, and the entries are ordered by appulse date. 

Appulse observations have small off-nadir angles by construction ($\theta=0\degr$ implies we are looking straight at the nucleus), and revisits were constrained to have $\theta>30\degr$, although most were acquired when $\theta>90\degr$. Most of the appulses and revisits were observed at $\approx90\degr$ phase, with occasional deviations up to $\pm30\degr$ from this value. Note that one of the targeted stars, HD~40111, has two distinct appulses separated by $\sim2$~weeks (see \autoref{tab:targeted}).

All exposures for a given appulse or revisit were flux-calibrated using spectrophotometric standard stars. Stellar spectra were then extracted from the spectral images and background subtracted. Spectra extracted from individual exposures were combined to improve the signal-to-noise ratio after first being normalized to have the same median flux from 1800-1900~\AA. This range was chosen because both \ce{H2O} and \ce{O2} have very small absorption cross sections in this region \citep{chung01,yoshino05}, but the stellar spectra still have sufficient signal-to-noise to allow a robust flux measurement \citep[the effective area of Alice decreases rapidly for wavelengths $>1800$~\AA;][]{stern07}. 

Next, the co-added revisit spectrum was scaled to have the same median flux from 1800-1900~\AA\ as the co-added appulse spectrum. Finally, the appulse spectrum was divided by the scaled revisit (i.e., unocculted) spectrum to create a normalized spectrum in which the intrinsic stellar flux and interstellar absorption have been removed and only the differences in foreground coma absorption between the appulse and revisit spectra remain. By normalizing the spectra in this manner we also make ourselves insensitive to the uncertainty in the amount of time the star was in the slit.

\begin{figure}
\includegraphics[width=\columnwidth]{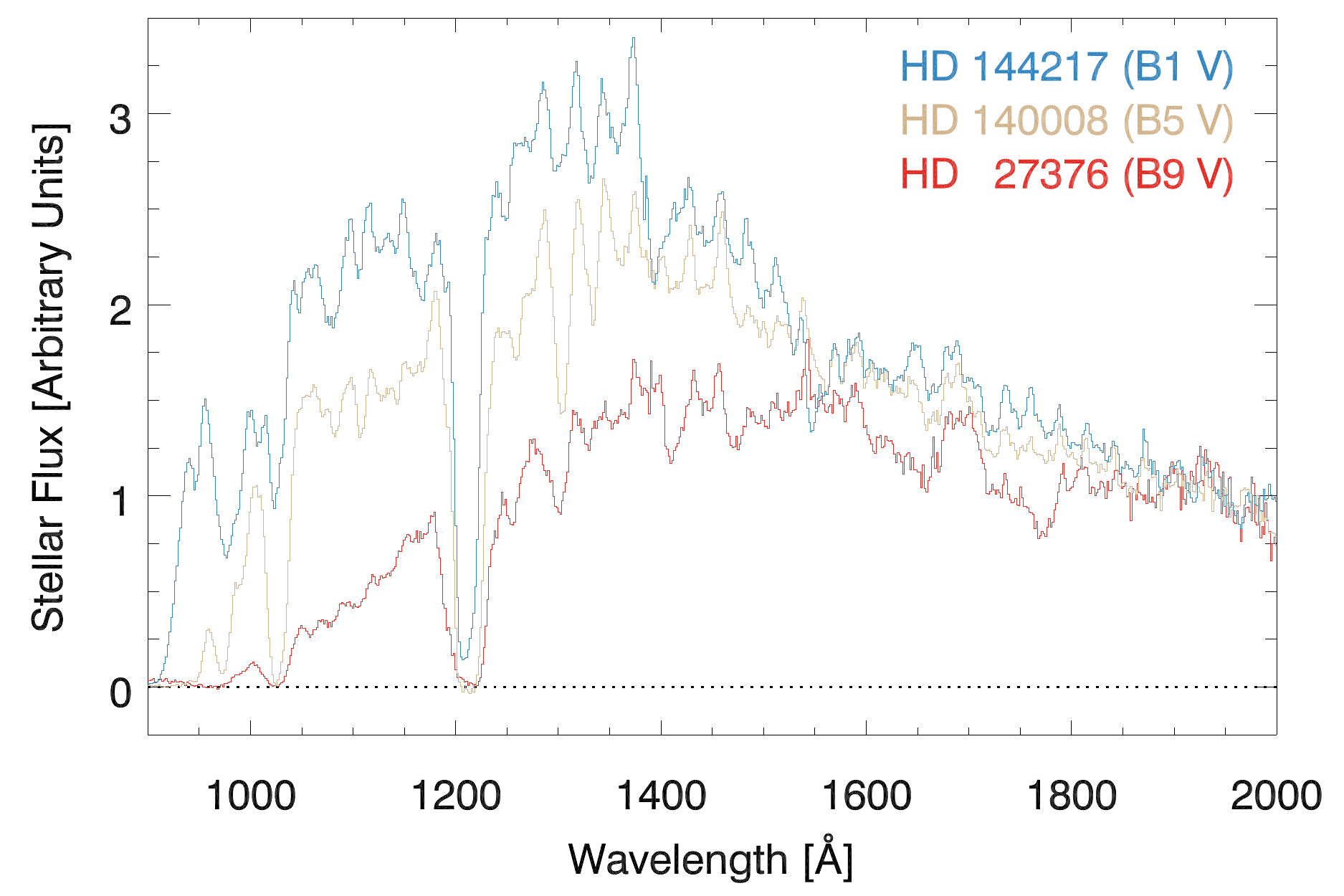}
\vspace{-2em}
\caption{\hl{Revisit spectra for three main sequence stars. The spectra are normalized to have the same flux from 1800-1900~\hbox{\AA} to emphasize the differences between early- and late-type \hbox{$B$} stars at far-UV wavelengths.}
\label{fig:stars}}
\end{figure}

\hl{\hbox{\autoref{fig:stars}} displays co-added revisit spectra for three main sequence stars that span the range of stellar types observed. All three stars have sufficient flux at \hbox{$\lambda>1400$~\AA} to create normalized spectra with reasonable signal-to-noise, but the early- and mid-type \hbox{$B$} stars have considerably more flux at shorter wavelengths than the late-type \hbox{$B$} star does. Thus, normalized spectra for late-type appulse stars are inherently noisier at bluer wavelengths than normalized spectra for earlier-type stars.}

We note that in a few cases we normalized the spectra from 1400-1450~\AA\ when normalization from 1800-1900~\AA\ was problematic. While 1400-1450~\AA\ has small \ce{H2O} absorption cross sections \citep{chung01}, it is the region where \ce{O2} absorption cross sections are largest \citep{yoshino05}. The 1400-1450~\AA\ region is therefore not ideal for spectral normalization, since using it reduces our sensitivity for \ce{O2} absorption. Fits to spectra where we had to use this normalization region are not used for detailed analyses (see \autoref{sec:fits} for details).

\section{Analysis of \ce{H2O} and \ce{O2} Absorption}
\label{sec:fits}

We have searched for optically-thin absorption from \ce{H2O} and \ce{O2} in the normalized stellar spectra as described above. For a given molecule, $i$, we model the optical depth, $\tau_i$, as a function of wavelength, $\lambda$, as:
\begin{equation}
\tau_i(\lambda) = N_i\,\sigma_i(\lambda), 
\label{eqn:tau}
\end{equation}
where $N_i$ is the column density of species $i$ and $\sigma_i(\lambda)$ is the absorption cross section of species $i$ as a function of wavelength. Combining absorption from several different species yields an expected (normalized) model flux of
\begin{equation}
F(\lambda) = \mathrm{e}^{-\sum\tau_i(\lambda)}.
\label{eqn:flux}
\end{equation}
This model spectrum can then be compared to the normalized stellar spectrum to constrain the column densities of interest.

\begin{table}

\caption{Molecular Cross Sections}
\label{tab:xsec}

\begin{tabular}{lccl}

\hline
Species & $\lambda$ (\AA) & $T$ (K) & Reference \\
\hline
\ce{H2O}  & 1400-1898 & 250 & \citet{chung01}     \\
          & 1148-1939 & 298 & \citet{mota05}      \\
          &  850-1110 & 298 & \citet{watanabe64}  \\
          & 1060-1860 & 298 & \citet{watanabe53}  \\
\ce{O2}   & 1300-1752 & 295 & \citet{yoshino05}   \\
          &   41-1771 & 298 & \citet{brion79}     \\
          & 1163-2000 & 298 & \citet{ackerman70}  \\
\ce{CO}   &  584-1038 & 298 & \citet{cairns65}    \\
\ce{CO2}  & 1061-1187 & 295 & \citet{stark07}     \\
          & 1187-1755 & 295 & \citet{yoshino96}   \\
          &   61-1450 & 298 & \citet{chan93}      \\
          &  155-1550 & 298 & \citet{hitchcock80} \\
\ce{CH4}  & 1380-1600 & 295 & \citet{mount78}     \\
          &  952-1306 & 295 & \citet{sun55}       \\
          &  773-1370 & 298 & \citet{ditchburn55} \\
\ce{C2H2} & 1050-2011 & 298 & \citet{nakayama64}  \\
          &  600-1000 & 298 & \citet{metzger64}   \\
\ce{C2H6} & 1380-1600 & 295 & \citet{mount78}     \\
          & 1200-1380 & 298 & \citet{okabe63}     \\
          & 1160-1200 & 298 & \citet{lombos67}    \\
          &  354-1127 & --- & \citet{koch71}      \\
\ce{C2H4} &  500-1200 & --- & \citet{schoen62}    \\
          & 1065-1960 & --- & \citet{zelikoff53}  \\
\ce{C4H2} & 1210-1730 & 296 & \citet{okabe81}     \\
          & 1600-2600 & 295 & \citet{fahr94}      \\
\ce{H2CO} &  600-1760 & --- & \citet{mentall71}   \\
          & 1760-1850 & --- & \citet{gentieu70}   \\
\hline

\end{tabular}
\end{table}

\begin{figure}
\includegraphics[width=\columnwidth]{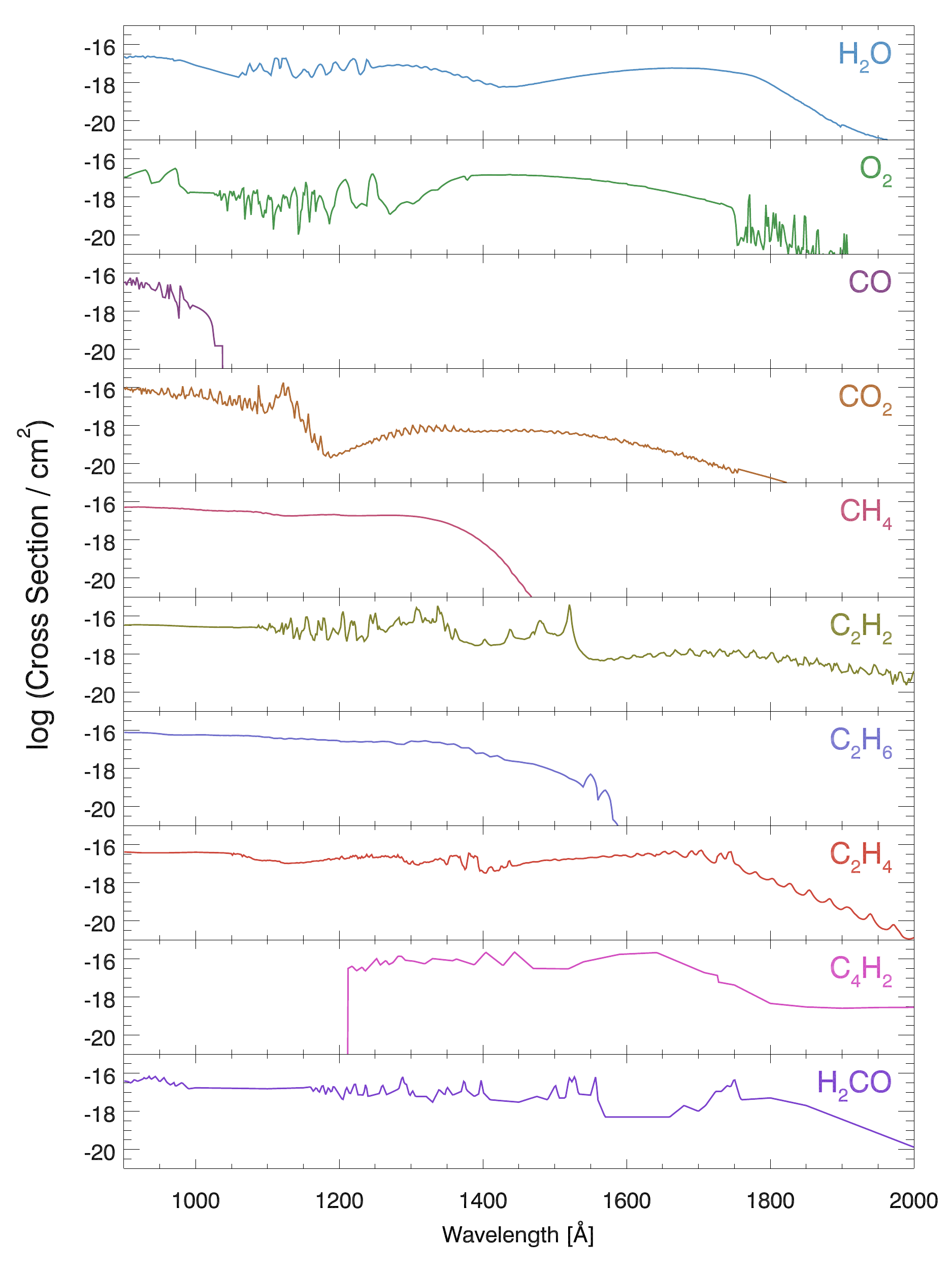}
\vspace{-2em}
\caption{\hl{Molecular absorption cross sections used in this work.}
\label{fig:xsec}}
\end{figure}

\autoref{tab:xsec} lists the ten species that we model in our analysis. While we are primarily interested in \ce{H2O} and \ce{O2}, other abundant species must be included to robustly constrain the range of permissible \ce{H2O} and \ce{O2} column densities. All species with $>0.5$\% abundance relative to \ce{H2O} in the coma of 67P/C-G in \citet{leroy15} with available far-UV absorption cross sections are tabulated. \autoref{tab:xsec} lists the following information by column: (1) species; (2) wavelength range; (3) measurement temperature; and (4) measurement reference. The adopted cross sections were downloaded from the PHoto Ionization/Dissociation RATES website\footnote{\url{http://phidrates.space.swri.edu}} \citep{huebner15}; for most species, they are composites of several different measurements covering the wavelength range 900-2000~\AA. The molecular cross sections in \autoref{tab:xsec} are displayed in \autoref{fig:xsec}.

All of the cross section measurements were performed near room temperature and laboratory measurements are not consistently available for all species in \autoref{tab:xsec} at any other temperature; \hl{however, the gas kinetic temperature in the coma of 67P/C-G varies considerably. \hbox{\citet{barucci16}} found that exposed water ice on the nucleus has $T\approx160$-220~K, while \hbox{\citet{lee15}} found that the temperature of the coma decreases as $T\propto\rho^{-1}$ until it reaches a terminal temperature of $T\approx50$-75~K.} The discrepancy between the temperature of the gas whose cross section was measured and the temperature of the absorbing coma gas introduces a systematic uncertainty in our model column densities that is not quantified by our modeling procedure. \hl{The peak \hbox{\ce{O2}} cross section decreases by $\sim0.1$~dex as the temperature decreases from 295 to 78~K \hbox{\citep{yoshino05}}; thus, by assuming room-temperature cross sections we are systematically under-estimating the \hbox{\ce{O2}} column density required to match the observed absorption. Unfortunately, no \hbox{\ce{H2O}} cross sections are available at $T<250$~K, so we are unable to estimate the magnitude of the systematic variation in \hbox{$\ce{O2}/\ce{H2O}$}.}

\begin{figure}
\includegraphics[width=\columnwidth]{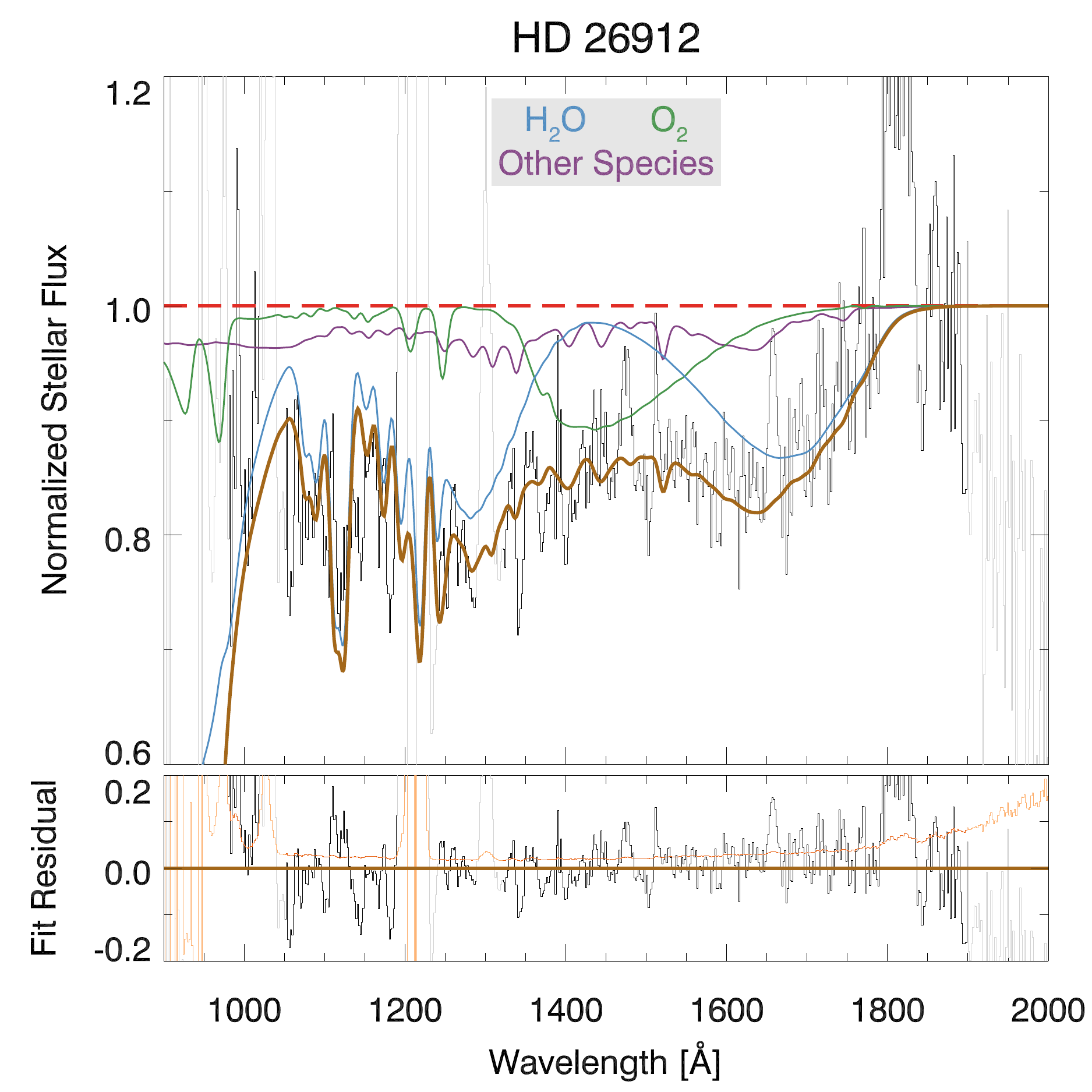}
\vspace{-2em}
\caption{Fits to the appulse absorption of HD~26912 ($\mathrm{FQ}=2$). \textit{Top:} The normalized stellar flux (black) with best-fit ensemble absorption (brown) overlaid. Individual absorption from \hl{\hbox{\ce{H2O}} (blue), \hbox{\ce{O2}} (green), and other species (purple; ensemble sum of \hbox{\ce{CO}}, \hbox{\ce{CO2}}, \hbox{\ce{CH4}}, etc. from \hbox{\autoref{tab:xsec}}) are also shown.} \textit{Bottom:} The residual of the ensemble fit\hl{, with 1$\sigma$ flux uncertainty (orange) overlaid. Masked regions are shown in lighter hues in both panels; these regions are not used to constrain the fits.} Absorption fits for all targeted and archival stellar appulses are shown in Appendix~\ref{app:fits}.
\label{fig:fit}}
\end{figure}

\begin{table*}

\caption{Stellar Appulse Column Densities}
\label{tab:fits}

\begin{tabular}{lcccccc}
\hline
     &     &    & \multicolumn{2}{c}{\it Best-Fit Values} & \multicolumn{2}{c}{\it Adopted Values}   \\
Star & S/N & FQ & $\log{N_{\ce{H2O}}}$ & $\ce{O2}/\ce{H2O}$ & $\log{N_{\ce{H2O}}}$ & $\ce{O2}/\ce{H2O}$ \\
\hline
HD~26912     &  33 & 2 & $16.40\pm0.01$ & $0.327\pm0.024$ & $16.40\pm0.04$ & $ 0.315\pm0.056$ \\
HD~3901      &  19 & 4 & $16.16\pm0.02$ &  0.000          & $16.14\pm0.09$ & $<0.179        $ \\
HD~29589     &  48 & 2 & $17.03\pm0.01$ & $0.442\pm0.015$ & $17.03\pm0.03$ & $ 0.435\pm0.046$ \\
HD~174585    &  17 & 3 & $16.49\pm0.02$ & $0.038\pm0.021$ & $16.49\pm0.06$ & $<0.155        $ \\
HD~180554    &  13 & 4 & $16.37\pm0.03$ &  0.000          & $16.36\pm0.09$ & $<0.164        $ \\
HD~191692    &  28 & 2 & $16.76\pm0.01$ & $0.123\pm0.009$ & $16.76\pm0.04$ & $ 0.123\pm0.035$ \\
HD~195810    &  27 & 2 & $16.80\pm0.01$ & $0.223\pm0.013$ & $16.80\pm0.04$ & $ 0.219\pm0.040$ \\
HD~192685    &  30 & 1 & $16.75\pm0.01$ & $0.087\pm0.014$ & $16.75\pm0.04$ & $<0.123        $ \\
HD~68324     &  45 & 2 & $16.85\pm0.01$ & $0.161\pm0.016$ & $16.85\pm0.03$ & $ 0.155\pm0.042$ \\
HD~66006     &  39 & 1 & $17.08\pm0.01$ & $0.111\pm0.009$ & $17.08\pm0.03$ & $ 0.109\pm0.030$ \\
HD~64722     &  37 & 2 & $16.76\pm0.01$ & $0.324\pm0.015$ & $16.76\pm0.03$ & $ 0.321\pm0.044$ \\
HD~39844     &  14 & 2 & $16.73\pm0.01$ & $0.188\pm0.013$ & $16.72\pm0.05$ & $ 0.190\pm0.048$ \\
HD~207330    &  39 & 2 & $16.80\pm0.01$ & $0.150\pm0.010$ & $16.80\pm0.04$ & $ 0.149\pm0.033$ \\
HD~109387    &  27 & 1 & $16.78\pm0.01$ & $0.154\pm0.013$ & $16.78\pm0.04$ & $ 0.151\pm0.041$ \\
HD~124771    &  24 & 1 & $16.56\pm0.01$ & $0.288\pm0.017$ & $16.55\pm0.04$ & $ 0.285\pm0.049$ \\
HD~21428     &  20 & 4 & $15.81\pm0.03$ &  0.000          & $15.82\pm0.12$ & $<0.272        $ \\
HD~32249     &  28 & 2 & $16.50\pm0.01$ & $0.569\pm0.025$ & $16.50\pm0.04$ & $ 0.560\pm0.066$ \\
HD~33328     &  25 & 2 & $16.47\pm0.01$ & $0.128\pm0.023$ & $16.46\pm0.04$ & $<0.167        $ \\
HD~106625    &  49 & 1 & $16.72\pm0.01$ & $0.309\pm0.009$ & $16.72\pm0.04$ & $ 0.308\pm0.031$ \\
HD~27376     &  23 & 3 & $16.18\pm0.03$ &  0.000          & $16.16\pm0.09$ & $<0.145        $ \\
HD~23466     &  15 & 4 & $16.07\pm0.04$ &  0.000          & $16.06\pm0.11$ & $<0.224        $ \\
HD~140008    &  49 & 2 & $15.95\pm0.03$ & $0.355\pm0.048$ & $15.94\pm0.06$ & $ 0.334\pm0.072$ \\
HD~144294    & 102 & 3 & $15.63\pm0.02$ & $0.590\pm0.069$ & $15.61\pm0.05$ & $ 0.563\pm0.089$ \\
HD~144217    &  53 & 3 & $16.00\pm0.03$ &  0.000          & $15.98\pm0.08$ & $<0.116        $ \\
HD~42933     & 119 & 3 & $15.60\pm0.02$ & $0.441\pm0.050$ & $15.58\pm0.05$ & $ 0.412\pm0.081$ \\
HD~89890     &  62 & 4 & $15.33\pm0.04$ &  1.000          & $15.30\pm0.11$ & $>0.733        $ \\
HD~40111 (A) &  86 & 3 & $15.59\pm0.02$ & $0.702\pm0.053$ & $15.57\pm0.06$ & $ 0.678\pm0.089$ \\
HD~144206    &  35 & 2 & $15.80\pm0.03$ & $0.521\pm0.056$ & $15.78\pm0.08$ & $ 0.495\pm0.092$ \\
HD~40111 (B) &  76 & 3 & $15.27\pm0.04$ & $0.297\pm0.060$ & $15.18\pm0.12$ & $<0.338        $ \\
\hline

\multicolumn{7}{l}{\textit{Notes.} Entries are ordered chronologically, and all column densities have units of ${\rm cm}^{-2}$.} \\
\multicolumn{7}{l}{Uncertainties are quoted at the $1\sigma$ level, and limits are quoted at the $3\sigma$ level.}

\end{tabular}
\end{table*}

We estimate the molecular column densities using nonlinear least-squares regression of \autoref{eqn:flux} with MPFIT\footnote{\url{http://purl.com/net/mpfit}} \citep{markwardt09}. The free parameters of the fit are the logarithm of the \ce{H2O} column density, in units of $\mathrm{cm}^{-2}$, and the relative column densities of \ce{O2}, \ce{CO}, \ce{CO2}, etc. with respect to water (e.g., $\ce{O2}/\ce{H2O}\equiv N_{\ce{O2}}/N_{\ce{H2O}}$). The \ce{O2}, \ce{CO}, and \ce{CO2} columns are constrained to lie in the range 0-100\% relative to \ce{H2O}, and all other species are constrained to the range 0-1\%.  We model the wavelength range 950-1900~\AA, with regions near strong coma emission lines (e.g., H\,{\sc i} Ly$\alpha$, H\,{\sc i} Ly$\beta$, and the O\,{\sc i} 1304~\AA\ multiplet, where residuals from background subtraction are often present) and regions with very low S/N masked out. The fit to the appulse of HD~26912 is presented in \autoref{fig:fit}, which shows the normalized stellar spectrum compared to ensemble and individual-species absorption in the top panel, and the ensemble fit residual in the bottom panel. Fits to all targeted and archival appulses are presented in Appendix~\ref{app:fits}.

The best-fit values of $\log{N_{\ce{H2O}}}$ and $\ce{O2}/\ce{H2O}$ for all of the stellar appulses are shown in \autoref{tab:fits}, which lists the following information by column: (1) star name; (2) median S/N in the wavelength range 1250-2000~\AA; (3) Fit Quality (FQ) flag; (4) logarithm of the best-fit \ce{H2O} column density, in $\mathrm{cm}^{-2}$; (5) best-fit value of the relative column density of \ce{O2} relative to \ce{H2O}; (6) logarithm of the adopted \ce{H2O} column density, in $\mathrm{cm}^{-2}$; and (7) adopted value of the relative column density of \ce{O2} relative to \ce{H2O}. The ``adopted'' values in Columns~6 and 7 are described in more detail in \autoref{sec:fits:adopted}. All quantities in Columns 4-7 are listed with $1\sigma$ uncertainties. 

The FQ flag in Column~3 is a subjective measure of the quality of the absorption line fit for a given star, with lower values indicating higher quality. \hl{Stars with \hbox{$\mathrm{FQ}=1$} are reasonably fit over the full wavelength range 900-2000~\hbox{\AA} (see \hbox{\autoref{fig:fit_hd192685}}). Stars with $\mathrm{FQ}=2$ have some regions of very low S/N (see \hbox{\autoref{fig:fit_hd191692}}), or mild discrepancies between the observed and model fluxes (see \hbox{\autoref{fig:fit_hd29589}}). Stars with $\mathrm{FQ}=3$ have large regions with systematic discrepancies between the observed and model fluxes.} All stars that were normalized from 1400-1450~\AA\ instead of the default 1800-1900~\AA\ region (see \autoref{sec:obs}) were assigned $\mathrm{FQ}=4$. \hl{We also assigned $\mathrm{FQ}=4$ to the appulse of HD~89890, whose fit preferred \hbox{$N_{\ce{O2}}>N_{\ce{H2O}}$} and had systematic discrepancies throughout the fitting range.} Only stars with $\mathrm{FQ} \leq 3$ are used in subsequent analyses.

There are two notable features of the best-fit column densities in \autoref{tab:fits}. The first is that the formal fitting uncertainties are very small. The second is that the $\ce{O2}/\ce{H2O}$ values are considerably higher than those in \citet{bieler15}, who found a mean value of $3.85\pm0.85$\% over an approximately 7~month period when $R_h=-3.4$ to $-2$~AU. It is possible that seasonal variations can account for some of this difference since the dates of our appulses do not overlap with the dates of the \citet{bieler15} measurements. However, \citet{bieler15} find no evidence of systematically increasing $\ce{O2}/\ce{H2O}$ in their measurements, almost all of which have $\ce{O2}/\ce{H2O}<0.1$, and several of the best-fit values in \autoref{tab:fits} have $\ce{O2}/\ce{H2O}>0.5$.

\subsection{Adopted Values of $N_{\ce{H2O}}$ and \hbox{$\ce{O2}/\ce{H2O}$}}
\label{sec:fits:adopted}

We tested our fitting procedure by forward modeling simulated data with pre-defined, ``true'', values of S/N, $N_{\ce{H2O}}$, and $\ce{O2}/\ce{H2O}$. We began with a flat-spectrum source ($F(\lambda)=1$ at all wavelengths) upon which we superimposed $\ce{H2O}$ absorption with a column density uniformly drawn from the range $15<\log{N_{\ce{H2O}}}<17.5$, \ce{O2}, \ce{CO}, and \ce{CO2} absorption with a column density relative to water uniformly drawn from the range 0-100\%, and \ce{CH4}, \ce{C2H2}, \ce{C2H6}, \ce{C2H4}, \ce{C4H2}, and \ce{H2CO} absorption with a column density relative to \ce{H2O} uniformly drawn from the range 0-1\%. These are the same ranges that were used in the fits to the appulse observations.

\begin{figure}
\includegraphics[width=\columnwidth]{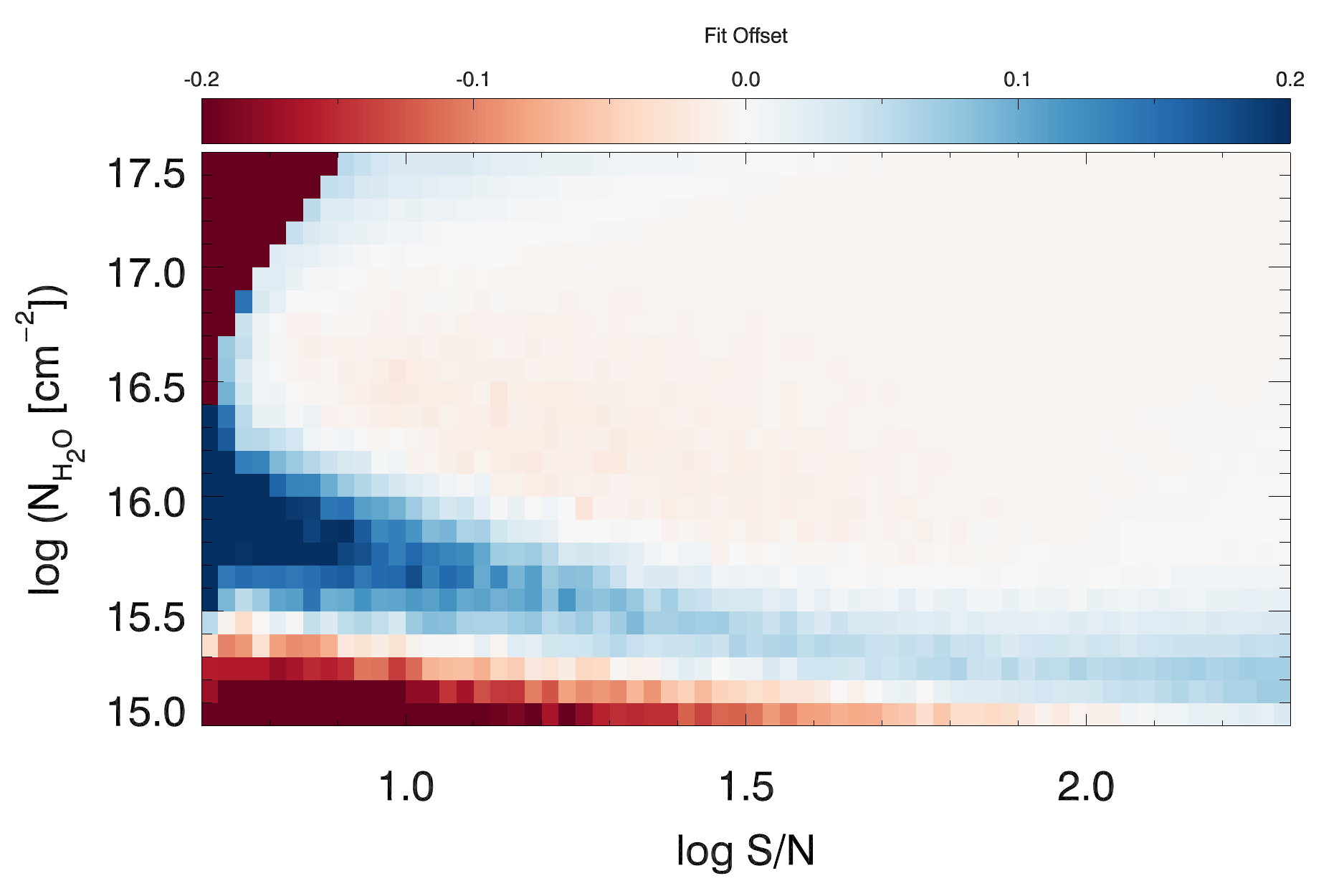}
\includegraphics[width=\columnwidth]{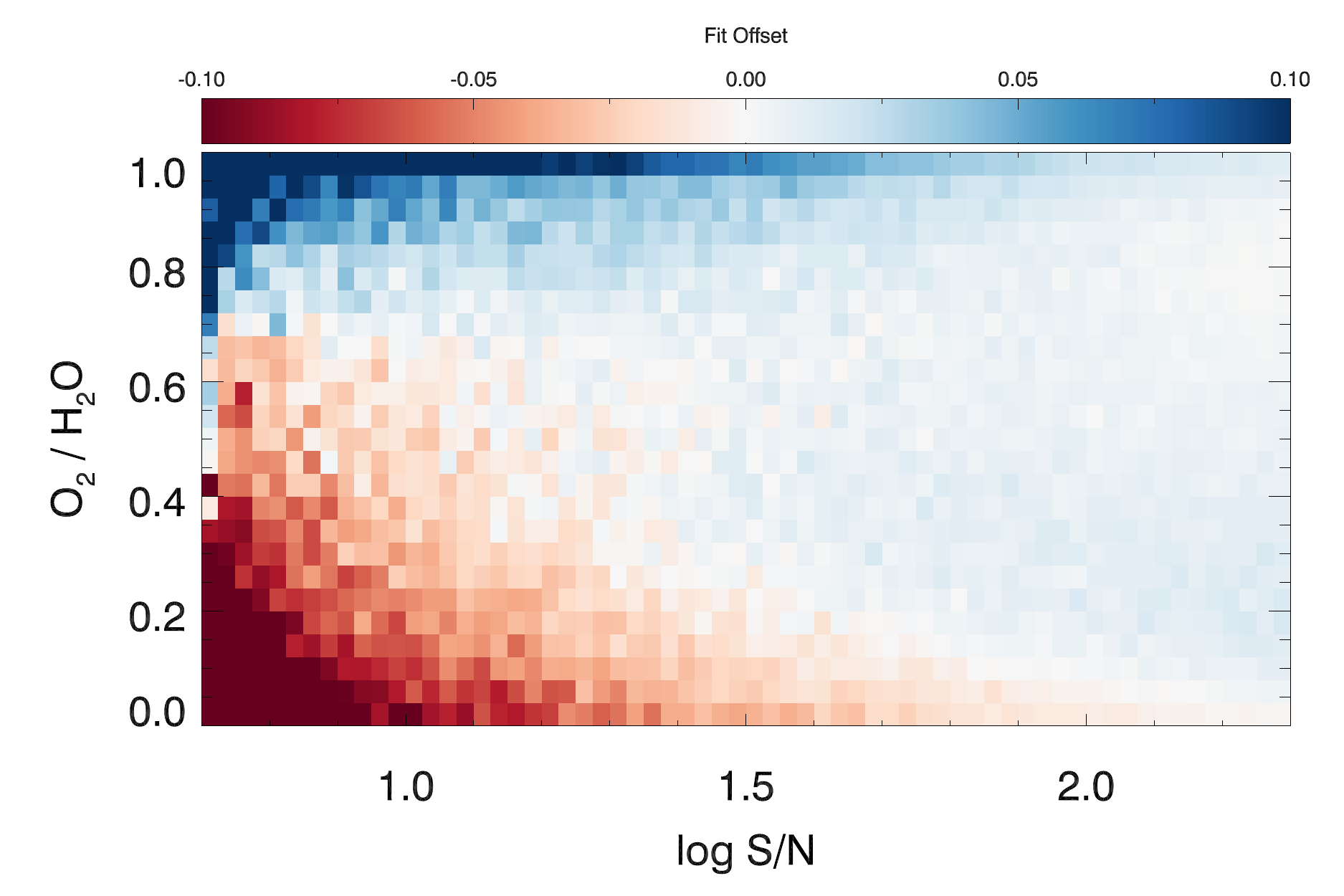}
\vspace{-2em}
\caption{Average offsets between the true and best-fit values of $\log{N_{\ce{H2O}}}$ (\textit{top}) and $\ce{O2}/\ce{H2O}$ (\textit{bottom}) as a function of $\log{\rm S/N}$. When $\mathrm{S/N}>10$, the magnitude of the $\log{N_{\ce{H2O}}}$ offset is typically $\la0.05$~dex, and the magnitude of the $\ce{O2}/\ce{H2O}$ offset is $\la0.02$.
\label{fig:offset}}
\end{figure}

\begin{figure}
\includegraphics[width=\columnwidth]{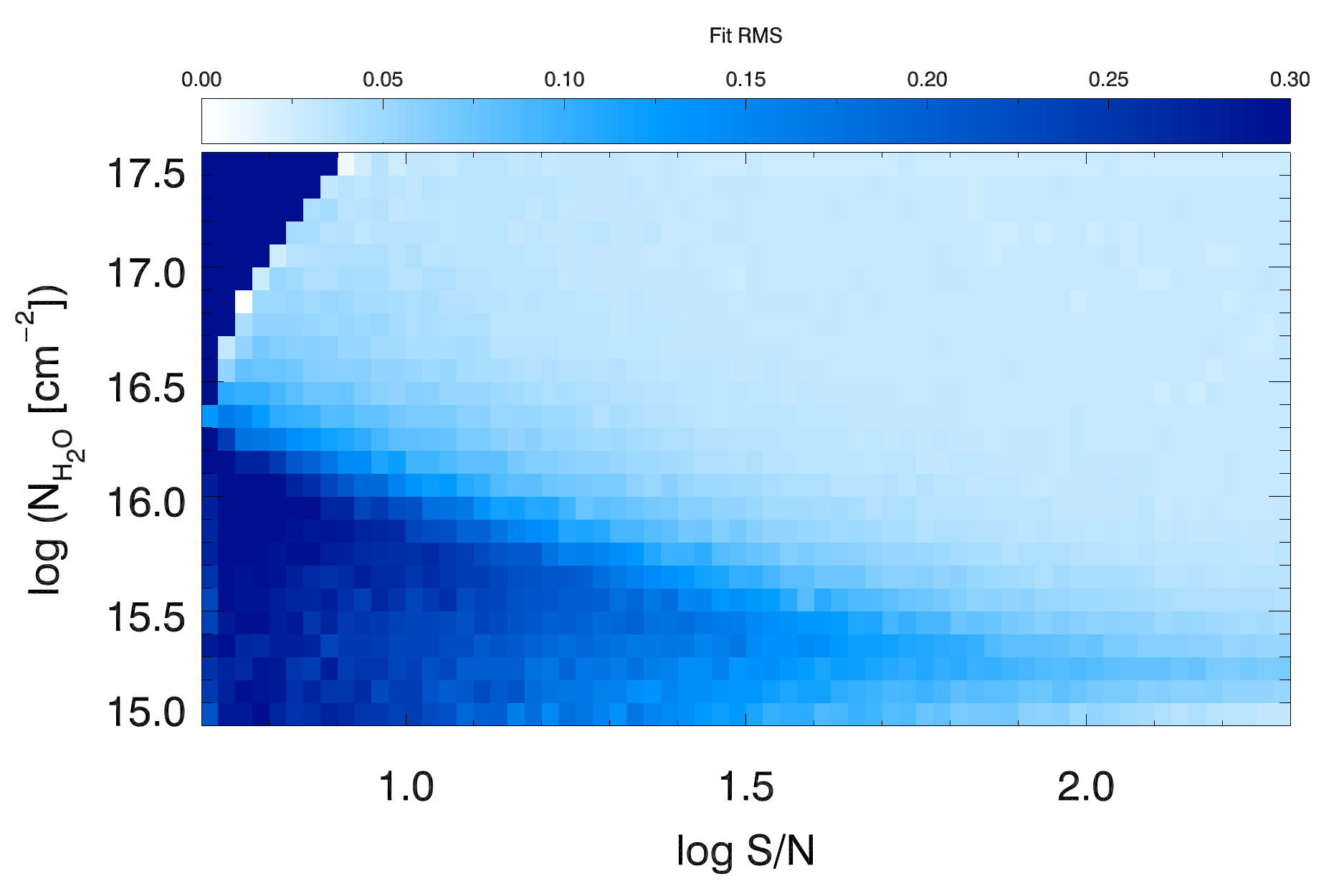}
\includegraphics[width=\columnwidth]{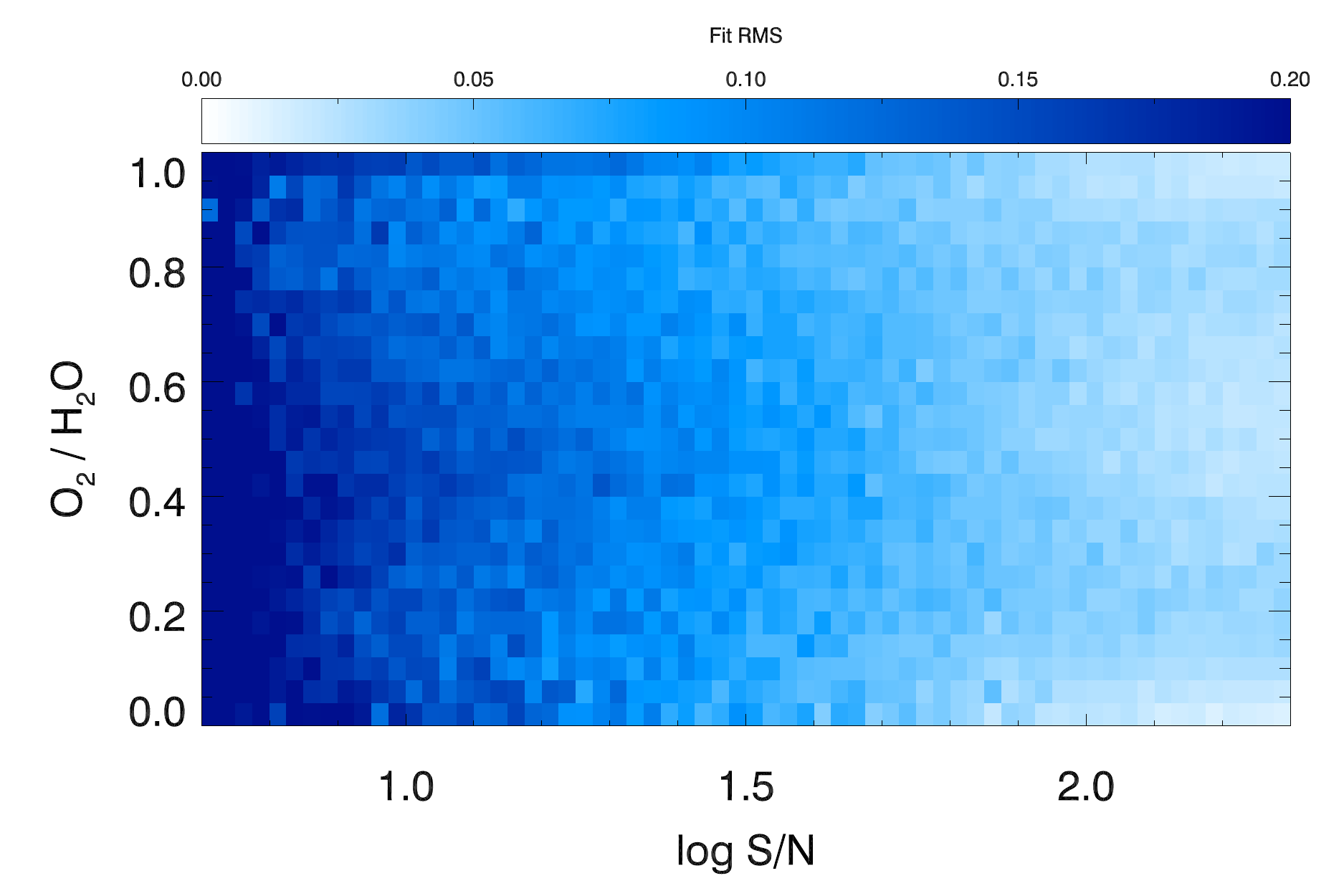}
\vspace{-2em}
\caption{RMS deviations between the true and best-fit values of $\log{N_{\ce{H2O}}}$ (\textit{top}) and $\ce{O2}/\ce{H2O}$ (\textit{bottom}) as a function of $\log{\rm S/N}$, after correcting for the systematic offsets in \autoref{fig:offset}. When $\mathrm{S/N}>10$, the RMS of $\log{N_{\ce{H2O}}}$ is typically 0.05-0.10~dex, and the RMS of $\ce{O2}/\ce{H2O}$ is $\sim0.05$.
\label{fig:rms}}
\end{figure}

Next, we added to the spectrum Poisson noise that had a median S/N in the 1250-2000~\AA\ range chosen uniformly from $0.7<\log{\rm S/N}<2.3$, bracketing the observed values. \hl{A template for the S/N as a function of wavelength was derived from the revisit (i.e., unocculted) spectra of our appulse targets by normalizing each spectrum to have the same median S/N from 1250-2000~\hbox{\AA}. Then at each wavelength we chose the median ``normalized S/N'' value from all of the spectra to form the S/N profile of a ``typical'' appulse star. This template achieves peak S/N at \hbox{$\sim1350$~\AA} and varies by a factor of \hbox{$\sim10$} over the wavelength range 950-2000~\AA.}

This noisy, simulated spectrum was then treated just like the stellar appulse observations; i.e., it was normalized to have $\langle F(\lambda) \rangle = 1$ from 1800-1900~\AA\ and then fit with the same procedure described above. The best-fit column densities and uncertainties were then saved along with the true values used to generate the \hl{simulated} spectrum, and the process was repeated 500,000 times to thoroughly sample the full range of parameter space.

The best-fit and true values of $N_{\ce{H2O}}$ and $\ce{O2}/\ce{H2O}$ are compared as a function of S/N in \autoref{fig:offset}. These images are two-dimensional histograms, where the color bars display the mean offset between the best-fit and true values in a given bin. Systematic offsets are present in both $N_{\ce{H2O}}$ and $\ce{O2}/\ce{H2O}$ when $\mathrm{S/N}<10$, but are quite modest at the higher S/N values typical of our appulse observations (see \autoref{tab:fits}). \autoref{fig:rms} is similar to \autoref{fig:offset}, except its color bars display the RMS deviations between the best-fit and true values in a given bin after correcting for the systematic offsets in \autoref{fig:offset}. These deviations quantify the spread in true values that are associated with a particular best-fit value.

The ``adopted'' values of $N_{\ce{H2O}}$ and $\ce{O2}/\ce{H2O}$ are derived from our Monte Carlo simulations by identifying the 1,000 simulated spectra with S/N and best-fit values closest to those measured for a given observation, and fitting a Gaussian to the distribution of true values. We treat the mean of this Gaussian as the adopted value, and its standard deviation as the $1\sigma$ uncertainty. Since our fits constrain the allowable range of $\ce{O2}/\ce{H2O}$, we quote limits whenever the adopted value is $<3\sigma$ from these boundaries.

\begin{figure}
\includegraphics[width=\columnwidth]{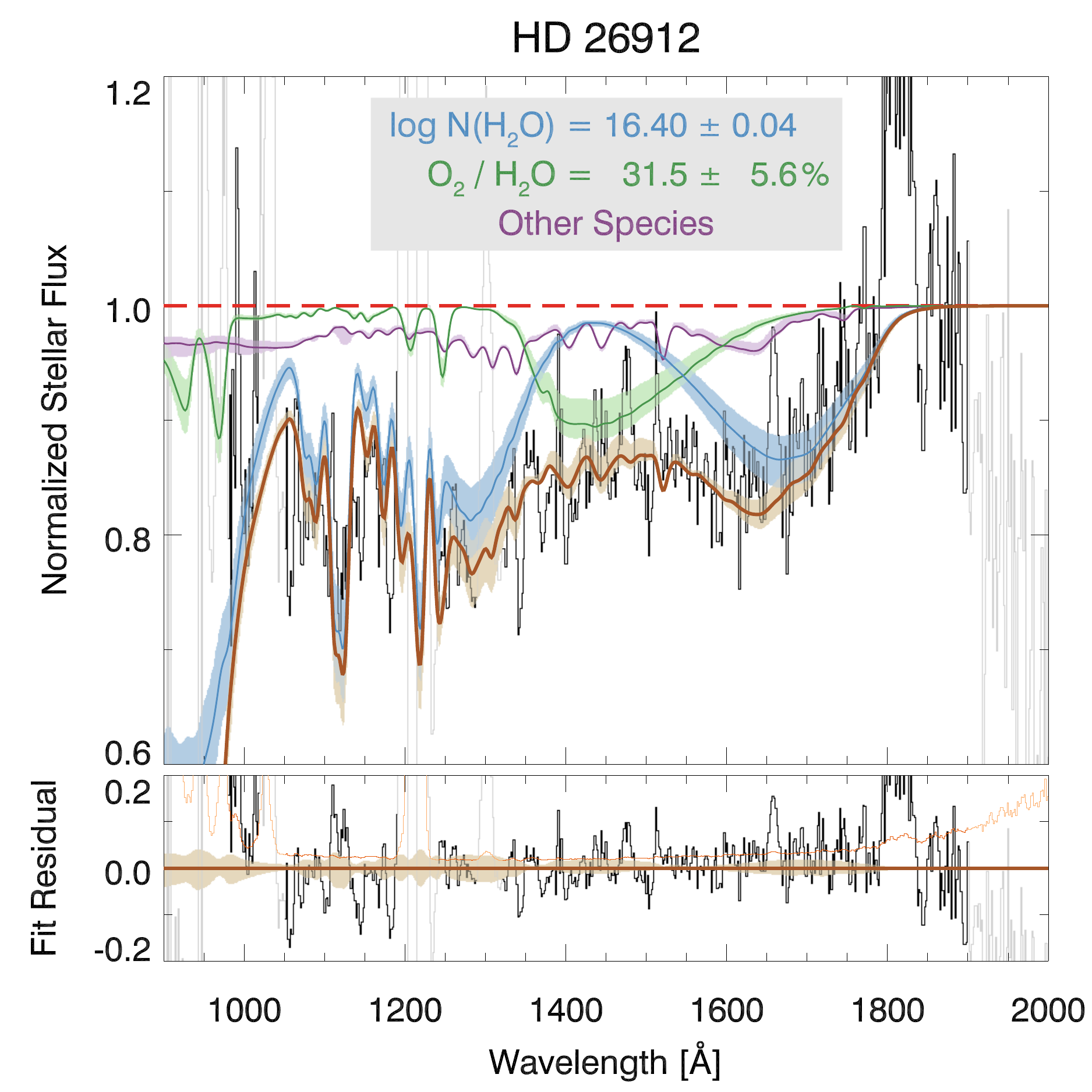}
\vspace{-2em}
\caption{Adopted column densities for the appulse of HD~26912 ($\mathrm{FQ}=2$), with 95\% ($2\sigma$) confidence bands. \textit{Top:} The normalized stellar flux, with ensemble fit (brown) and individual-species absorption overlaid using the adopted column densities of \ce{H2O} and \ce{O2} from \autoref{tab:fits}. \textit{Bottom:} The residual of the ensemble fit\hl{, with 1$\sigma$ flux uncertainty (orange) overlaid. Masked regions are shown in lighter hues in both panels; these regions are not used to constrain the fits.} Adopted column densities for all targeted and archival stellar appulses are shown in Appendix~\ref{app:conf}.
\label{fig:conf}}
\end{figure}

The last two columns of \autoref{tab:fits} list the adopted values of $\log{N_{\ce{H2O}}}$ and $\ce{O2}/\ce{H2O}$, respectively, for our stellar appulse observations. \autoref{fig:conf} shows absorption profiles of \ce{H2O} and \ce{O2} and their associated 95\% ($2\sigma$) confidence bands using the adopted values for the appulse of HD~26912. These profiles are overlaid on the normalized stellar spectrum as in \autoref{fig:fit}, along with confidence bands for the sum of all other modeled species and the total absorption from all species. These profiles clearly show that the adopted values of $\log{N_{\ce{H2O}}}$ and $\ce{O2}/\ce{H2O}$ are consistent with the data. Absorption profiles derived from the adopted values for all targeted and serendipitous appulses are presented in Appendix~\ref{app:conf}.

\section{Discussion}
\label{sec:discussion}

\subsection{\ce{H2O} Column Densities}
\label{sec:discussion:H2O}

\begin{figure}
\includegraphics[width=\columnwidth]{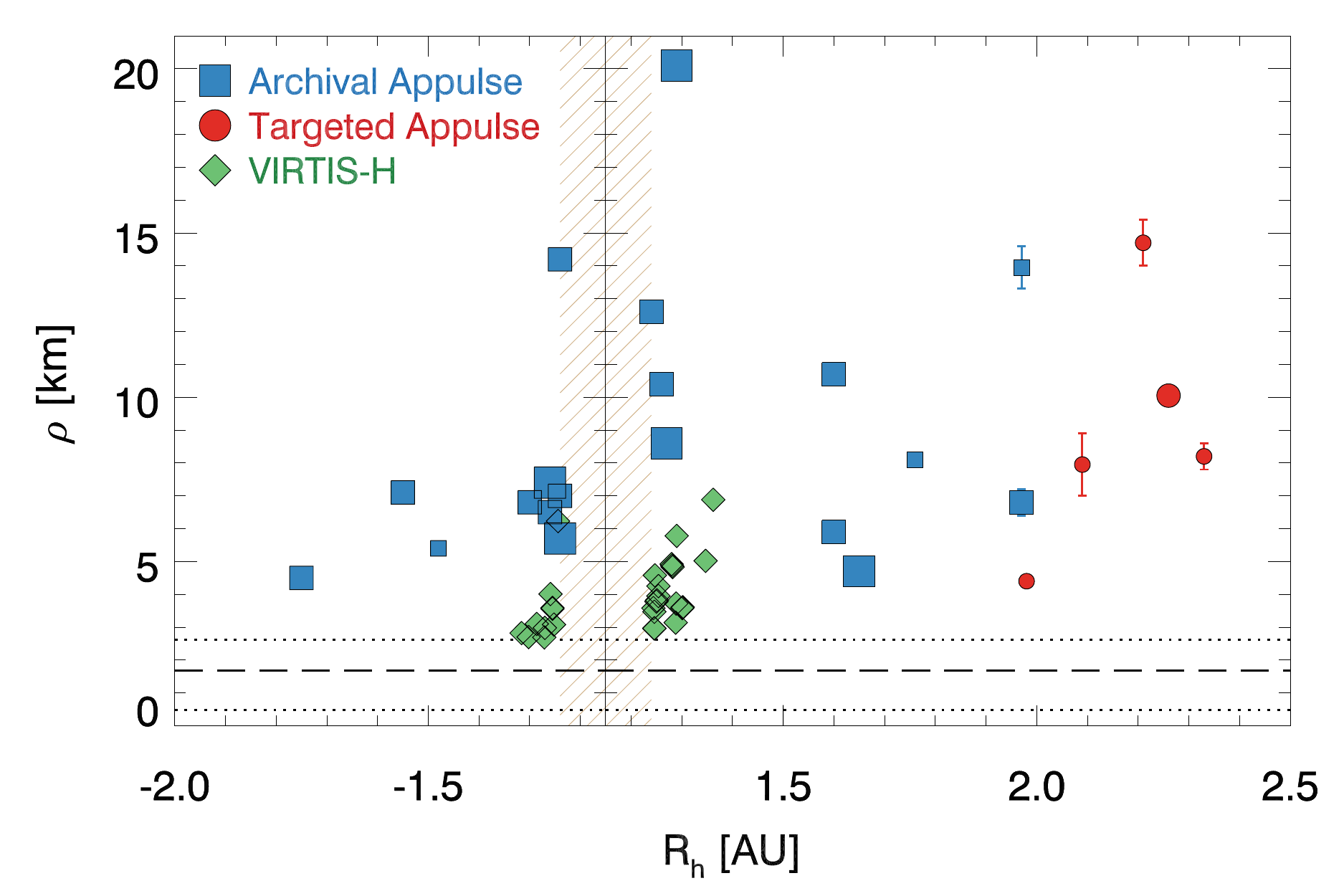}
\includegraphics[width=\columnwidth]{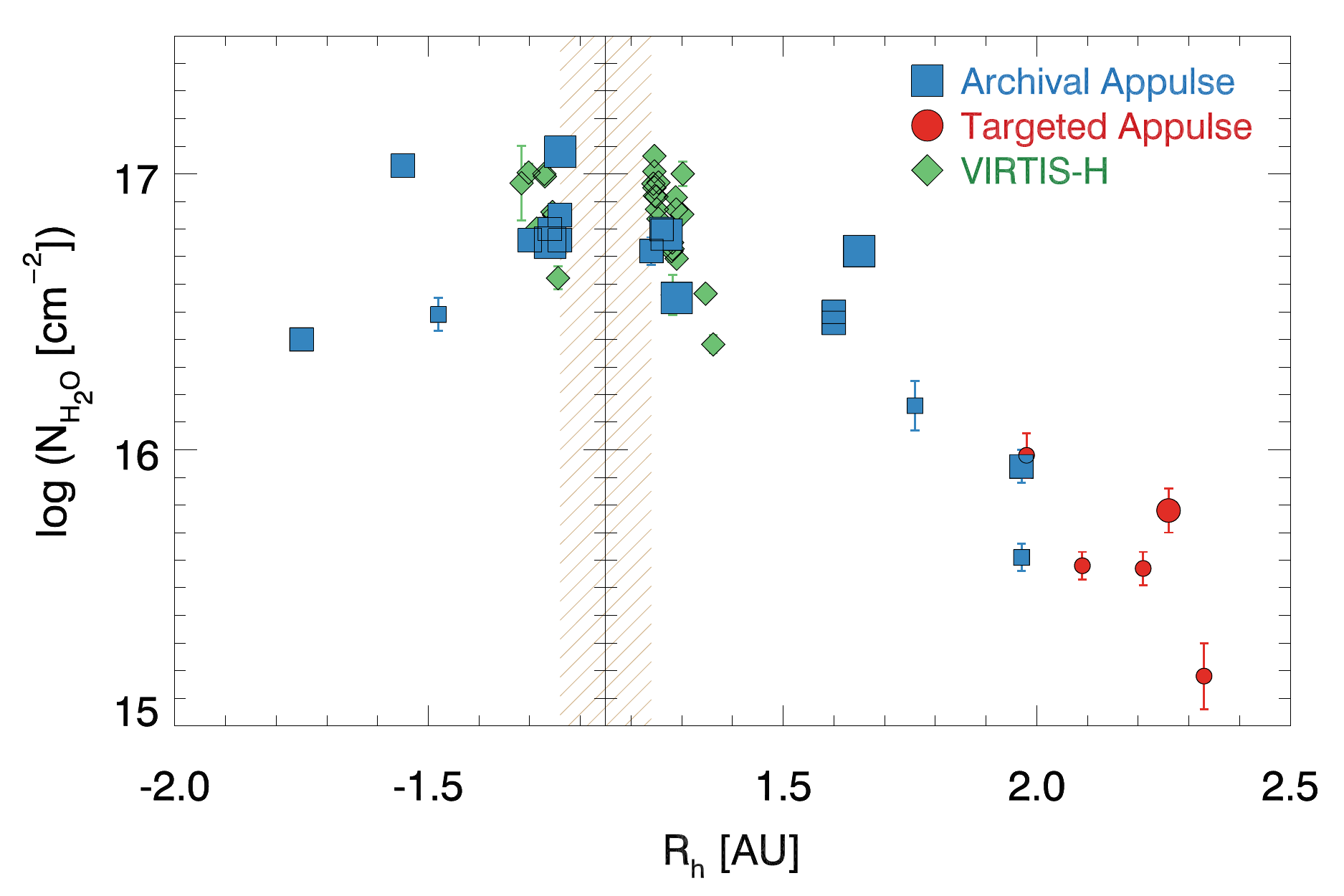}
\includegraphics[width=\columnwidth]{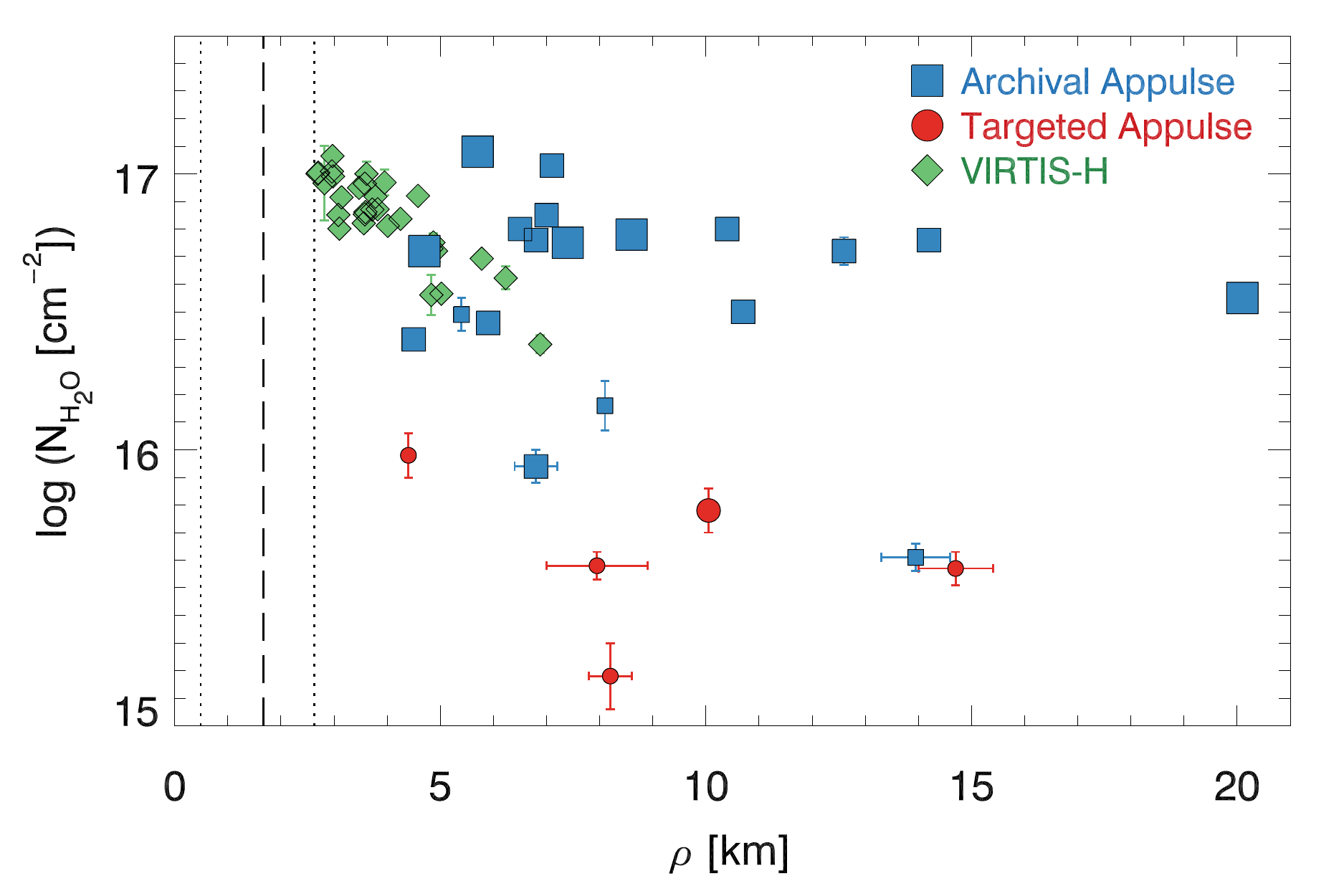}
\vspace{-2em}
\caption{\textit{Top}: Distribution of the stellar appulse observations with heliocentric radius ($R_h$) and impact parameter ($\rho$). \textit{Middle}: Distribution of the adopted values of $N_{\ce{H2O}}$ with $R_h$. \textit{Bottom}: Distribution of the adopted values of $N_{\ce{H2O}}$ with $\rho$. The VIRTIS-H measurements of \citet{bockelee-morvan16} are also plotted in all three panels. Heliocentric distances in the beige hatched region are smaller than the perihelion distance of 1.24~AU. The dashed  line in the top and bottom panels is the effective radius of the nucleus, and the dotted lines indicate its minimum and maximum radii. \hl{Symbol size encodes fit quality in all panels, with higher quality fits (lower FQ values) having larger symbols; points with no error bars have uncertainties smaller than the symbol size.}
\label{fig:NH2O}}
\vspace{-2em}
\end{figure}

The Monte Carlo simulations presented in \autoref{sec:fits:adopted} are one way to gain confidence in the validity of our absorption fits. Another is to compare our \ce{H2O} column densities with values measured by other instruments on \rosetta\ at similar times. \autoref{fig:NH2O} shows the adopted values of $\log{N_{\ce{H2O}}}$ for our stellar appulse observations as a function of $R_h$, compared to the VIRTIS-H measurements of \citet{bockelee-morvan16}. Despite the large scatter in the column densities for a given value of $R_h$, the adopted values for our appulse observations are reassuringly similar to the measured values from VIRTIS. One reason for the differences that do exist is the fact that the VIRTIS measurements were taken at systematically lower impact parameters than the appulses, as shown in the top panel of \autoref{fig:NH2O}.

While the $N_{\ce{H2O}}$ values from Alice and VIRTIS are in good agreement, there may be discrepancies with the ROSINA measurements. \citet{fougere16} present a sophisticated Direct Simulation Monte Carlo (DSMC) model of the major species (\ce{H2O}, \ce{CO2}, \ce{CO}, and \ce{O2}) in the coma of 67P/C-G, which derives molecular production rates from a non-uniform surface activity distribution. The DSMC model does a remarkable job of reproducing the \textit{in-situ} ROSINA measurements of the number density of these species for all data taken before March~2016 \citep{fougere16}. However, when the model production rates are used to predict the $N_{\ce{H2O}}$ values along the lines of sight probed by \citet{bockelee-morvan16}, it finds model column densities that are four times higher than those measured by VIRTIS \citep{fougere16}. The cause of this discrepancy is unclear, which illustrates the difficulty of directly comparing measurements from \textit{in-situ} instruments such as ROSINA to those from remote-sensing instruments such as VIRTIS and Alice.

\begin{figure}
\includegraphics[width=\columnwidth]{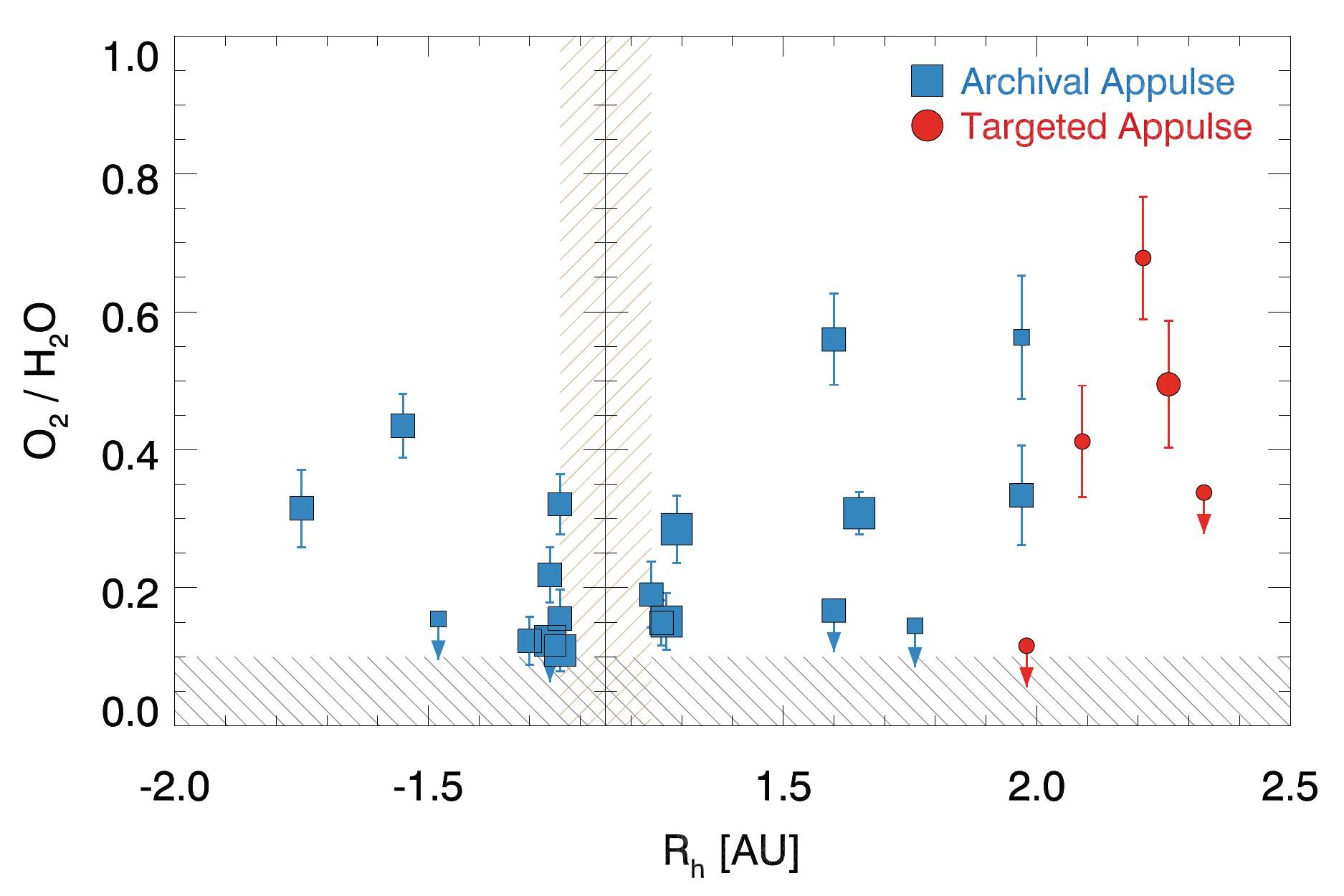}
\includegraphics[width=\columnwidth]{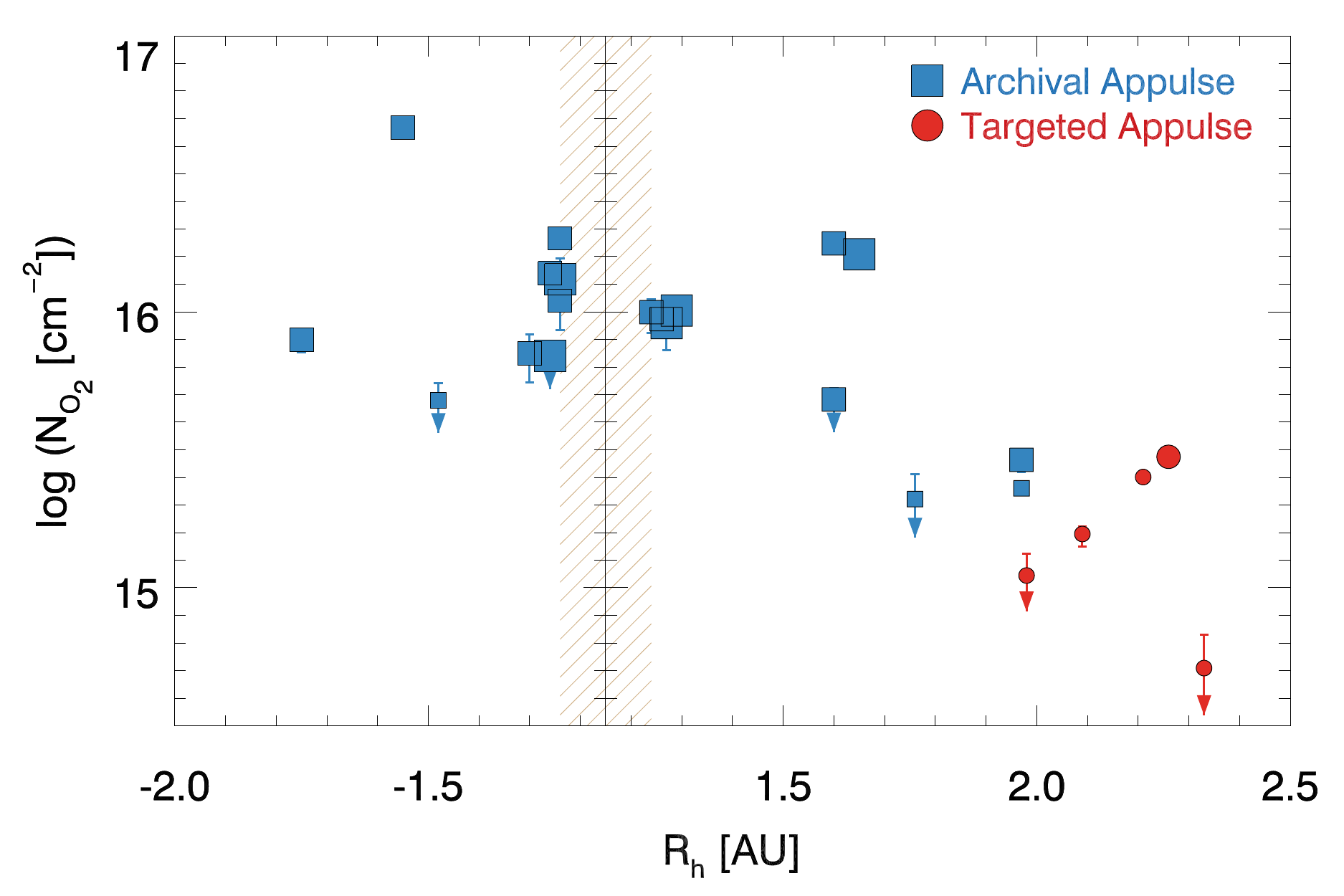}
\vspace{-2em}
\caption{$\ce{O2}/\ce{H2O}$ (\textit{top}) and $N_{\ce{O2}}$ (\textit{bottom}) as a function of $R_h$. Symbol sizes are the same as in \autoref{fig:NH2O}. Heliocentric distances in the beige hatched region are smaller than the perihelion distance of 1.24~AU. The gray hatched region in the top panel indicates typical values of $\ce{O2}/\ce{H2O}$ measured by ROSINA \citep{bieler15,fougere16}.
\label{fig:Rh_O2}}
\end{figure}

\begin{figure}
\includegraphics[width=\columnwidth]{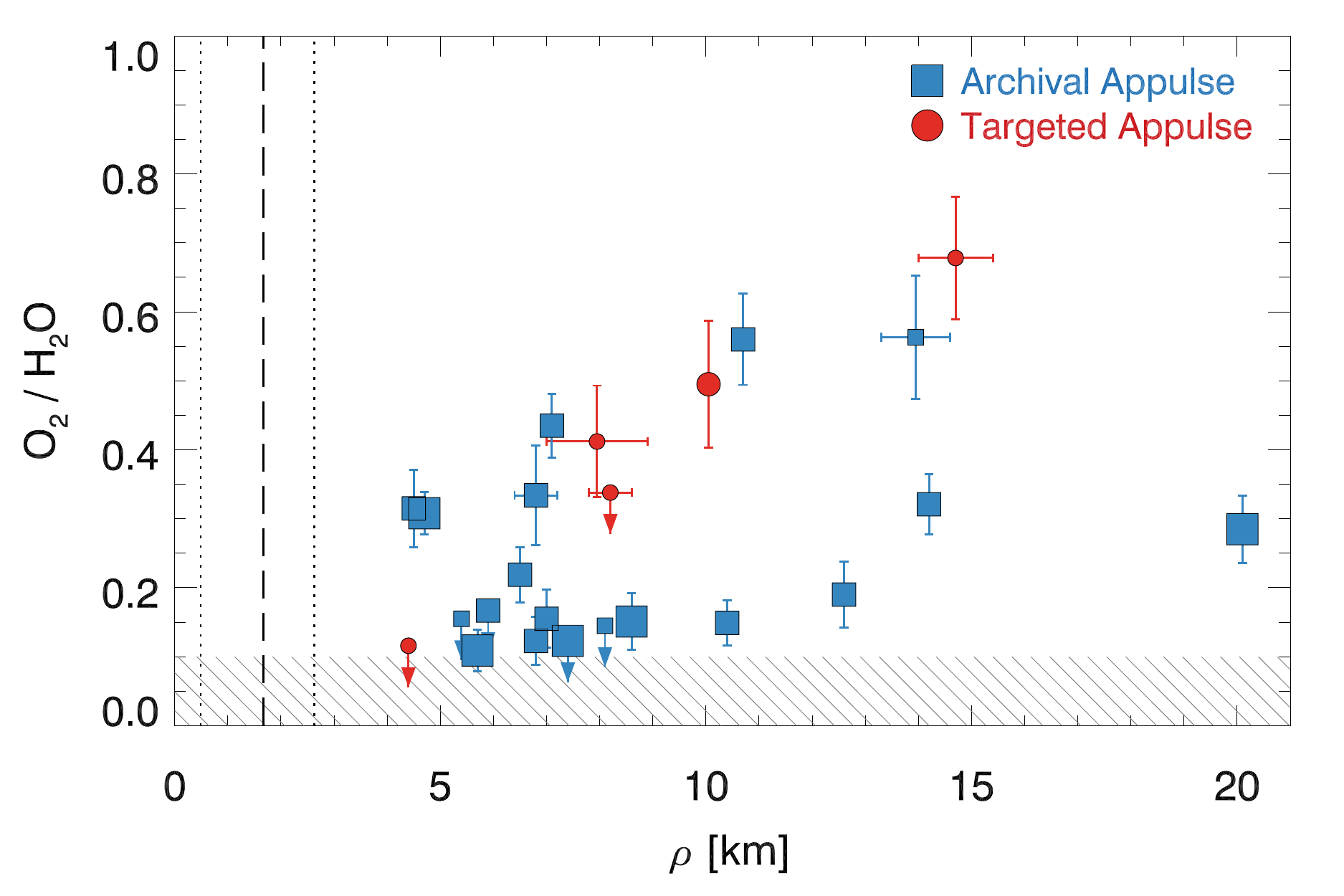}
\includegraphics[width=\columnwidth]{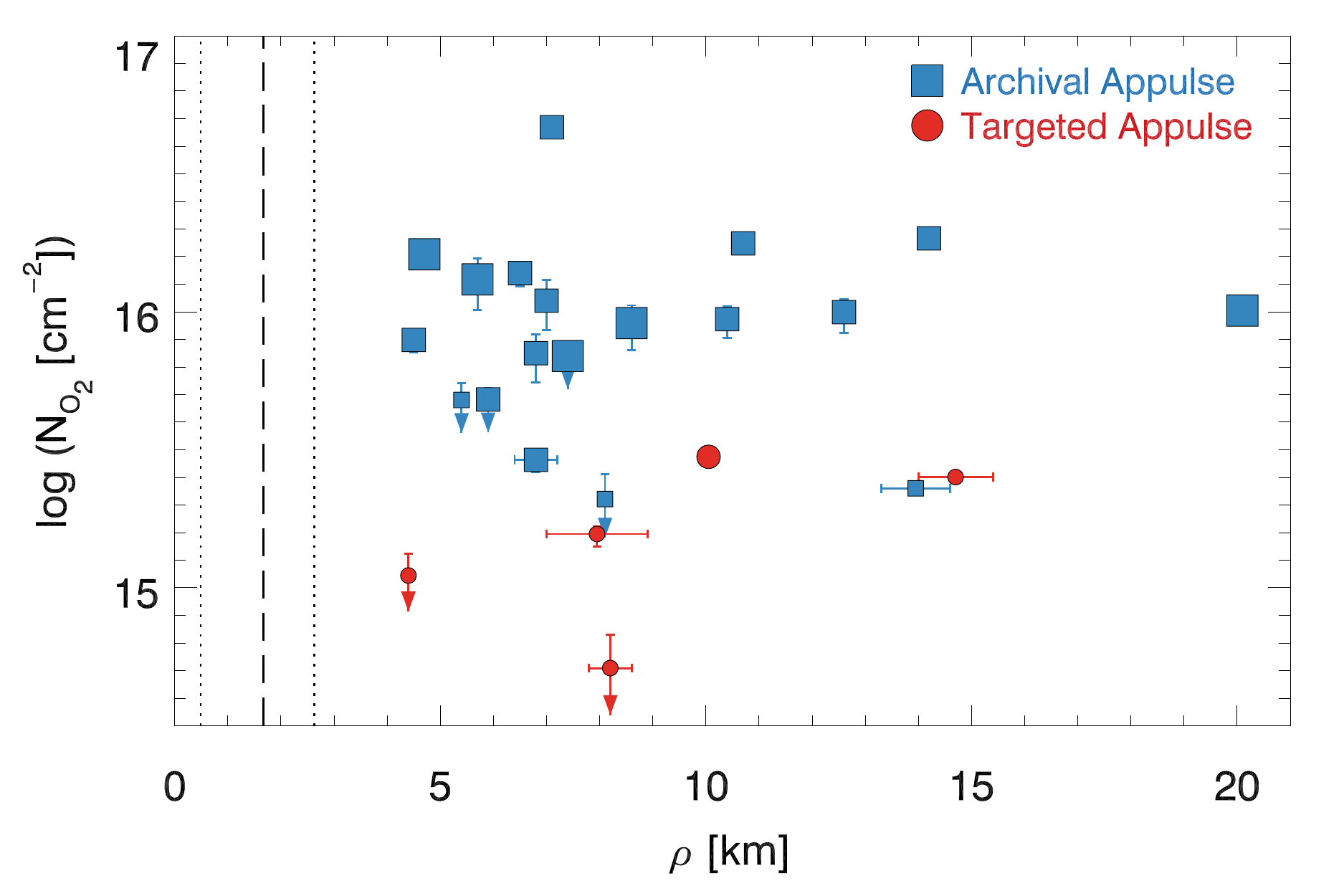}
\vspace{-2em}
\caption{$\ce{O2}/\ce{H2O}$ (\textit{top}) and $N_{\ce{O2}}$ (\textit{bottom}) as a function of impact parameter. Symbol sizes are the same as in \autoref{fig:NH2O}. The dashed vertical line is the effective radius of the nucleus, and the dotted vertical lines indicate its minimum and maximum radii. 
\label{fig:offnad_O2}}
\end{figure}

\subsection{$\ce{O2}/\ce{H2O}$}
\label{sec:discussion:fO2}

\autoref{fig:Rh_O2} shows the relative abundance of \ce{O2} with respect to \ce{H2O} (\textit{top panel}) and the column density of \ce{O2} (\textit{bottom panel}) as a function of $R_h$. \autoref{fig:offnad_O2} shows the same quantities as a function of impact parameter. The relative $\ce{O2}/\ce{H2O}$ abundance tends to increase with increasing heliocentric distance and increasing impact parameter. These correlations ($3.9\sigma$ and $2.5\sigma$ significance, respectively, according to Kendall's tau test) cause the distributions of $N_{\ce{O2}}$ as a function of $R_h$ and $\rho$ to be flatter than the corresponding distributions of $N_{\ce{H2O}}$ shown in \autoref{fig:NH2O}. 

The relatively flat distribution of $N_{\ce{O2}}$ as a function of $\rho$ is particularly interesting, as it suggests a distributed source of \ce{O2}. This would seem to argue against the variety of mechanisms that \citet{mousis16} suggest for trapping \ce{O2} in the icy \ce{H2O} matrix of 67P/C-G. Formation of \ce{O2} through the dismutation of \ce{H2O2} during the evaporation of \ce{H2O} ice, as suggested by \citet{dulieu17}, might be able to explain the shape of the $\ce{O2}/\ce{H2O}$ distribution as a function of $\rho$. Interestingly, ROSINA detects \ce{H2O2} in the coma of 67P/C-G (see Figure~4 of \citealp{leroy15}, and Figure~4 of \citealp{bieler15}), but with a relative abundance of $\ce{H2O2}/\ce{O2}<0.1$\% \citep{bieler15}, far less than the ratio of $\ce{H2O2}/\ce{O2}=2$ predicted by the dismutation reaction \citep{dulieu17}. 

\citet{feldman16} used Alice to study gaseous outbursts in the coma of 67P/C-G. These outbursts exhibit no increase in long-wavelength solar reflected light that would indicate an increase in dust production, and are characterized by a sudden increase in the brightness ratio of O\,{\sc i}~$\lambda1356/\lambda1304$ in the sunward coma. \citet{feldman16} infer that these outbursts are driven by \ce{O2} release, and estimate that $\ce{O2}/\ce{H2O} \geq 50$\% during the outbursts. 

Coincidentally, our earliest archival appulse (HD~26912; see \autoref{fig:fit} and \autoref{fig:conf}) occurred during the onset of one of the \citet{feldman16} outbursts (see their Section~2.5). We adopt $\ce{O2}/\ce{H2O}=31.5\pm5.6$\% for this appulse (see \autoref{tab:fits}), which is somewhat lower than the \citet{feldman16} estimate.  This apparent discrepancy is likely a result of timing differences; i.e., the adopted value from the appulse measures the ambient $\ce{O2}/\ce{H2O}$ in the coma just prior to outburst, whereas the \citet{feldman16} value measures the peak $\ce{O2}/\ce{H2O}$ over the $\sim30$-minute duration of the outburst.

As mentioned in \autoref{sec:fits}, the $\ce{O2}/\ce{H2O}$ values in \autoref{tab:fits} are generally higher, and have considerably larger scatter, than the values found by ROSINA-DFMS. \citet{bieler15} found $n_{\ce{O2}}/n_{\ce{H2O}}=3.85\pm0.85$\% in data taken between August~2014 and March~2015, and \citet{fougere16} found $Q_{\ce{O2}}/Q_{\ce{H2O}}\approx2$\% throughout the time frame of our appulse observations. Notably, neither \citet{bieler15} nor \citet{fougere16} list a single observation where $\ce{O2}/\ce{H2O}>15$\%, but we find a median value of 25\%.

As discussed in \autoref{sec:discussion:H2O}, comparisons between the \textit{in-situ} measurements of ROSINA and the line-of-sight measurements of Alice and VIRTIS are not straightforward, even with a sophisticated coma model \citep{fougere16}. Nonetheless, the large values of $\ce{O2}/\ce{H2O}$ derived from the Alice data are surprising. While we have included several minor species in our absorption fits (see \autoref{sec:fits}), there are many species detected in the coma of 67P/C-G by ROSINA for which we were unable to find absorption cross sections \citep[e.g., \ce{HS}, \ce{S2}, \ce{CH4O};][]{leroy15}. Some of these ``missing'' species could have cross sections large enough to cause measurable far-UV absorption, even for very small column densities, causing the current $\ce{O2}/\ce{H2O}$ values to be over-estimated. Quantifying the magnitude of these systematic uncertainties is exceedingly difficult without additional laboratory data for far-UV molecular absorption cross sections.

Further, even if our fits currently include all of the relevant species, the absorption cross sections we use were all measured at $T\approx300$~K (see \autoref{tab:xsec}). Since the absorbing coma gas is expected to be at lower temperature, variations in the absorption cross sections with temperature could lead us to infer incorrect values of the column density with our current procedure. \hl{However, the scant existing data suggest that our procedure under-estimates the amount of low-temperature \hbox{\ce{O2}} present by assuming room-temperature cross sections \hbox{\citep[see discussion in \autoref{sec:fits};][]{yoshino05}}, which would serve to increase the discrepancy between our results and those of ROSINA.}

\section{Conclusions}
\label{sec:conclusions}

Using the Alice far-UV imaging spectrograph aboard \rosetta, we have independently verified the presence of \ce{O2} in the coma of Comet~67P/C-G. \ce{O2} was detected for the first time in the coma of a comet by \rosetta's ROSINA mass spectrometer \citep{bieler15,fougere16}. In the present study, both \ce{O2} and \ce{H2O} were detected in far-UV absorption against the continuum of stars located near the nucleus of 67P/C-G, at impact parameters of 4-20~km. These stellar appulses occurred at heliocentric distances of $-1.8$ to 2.3~AU, where negative distances indicate pre-perihelion observations. The main results of our analysis are as follows:
\begin{enumerate}
\item The \ce{H2O} column densities derived from the stellar spectra are in good agreement with VIRTIS-H measurements from the same time period taken at similar impact parameters \citep{bockelee-morvan16}.
\item The median value for the relative abundance of \ce{O2} with respect to \ce{H2O} derived from the stellar spectra is $\ce{O2}/\ce{H2O}=25$\%. This value is considerably higher than those reported by ROSINA; \citet{bieler15} and \citet{fougere16} found mean values of $\ce{O2}/\ce{H2O}<5$\%.
\end{enumerate}

We see no simple explanation for the difference in $\ce{O2}/\ce{H2O}$ measured by Alice and ROSINA, unless it is related to the unmodeled species and $T=300$~K cross sections discussed at the end of \autoref{sec:discussion:fO2}. The Alice \ce{H2O} measurements are consistent with values published by other remote-sensing instruments on \rosetta; and while this does not guarantee that our \ce{O2} values are correct it does suggest that our measurements are reasonably robust. The ROSINA measurements, on the other hand, are performed \textit{in situ} at the spacecraft location, and the sophisticated coma model of \citet{fougere16} is designed to reproduce these measurements. This same model has difficulty reproducing the \ce{H2O} column densities of \citet{bockelee-morvan16}, which were measured very close to perihelion \citep{fougere16}. There is clearly much future work to be done to reconcile these differences.

\section*{Acknowledgements}
\rosetta\ is an ESA mission with contributions from its member states and NASA. We thank the members of the \rosetta\ Science Ground System and Mission Operations Center teams, in particular Richard Moissl and Michael K\"uppers, for their expert and dedicated help in planning and executing the Alice observations. The Alice team acknowledges continuing support from NASA via Jet Propulsion Laboratory contract 1336850 to the Southwest Research Institute. This research has made use of the SIMBAD database, operated at CDS, Strasbourg, France.

\appendix
\section{Best-Fit Absorption Profiles}
\label{app:fits}

\begin{figure}
\centering\includegraphics[width=0.8\columnwidth]{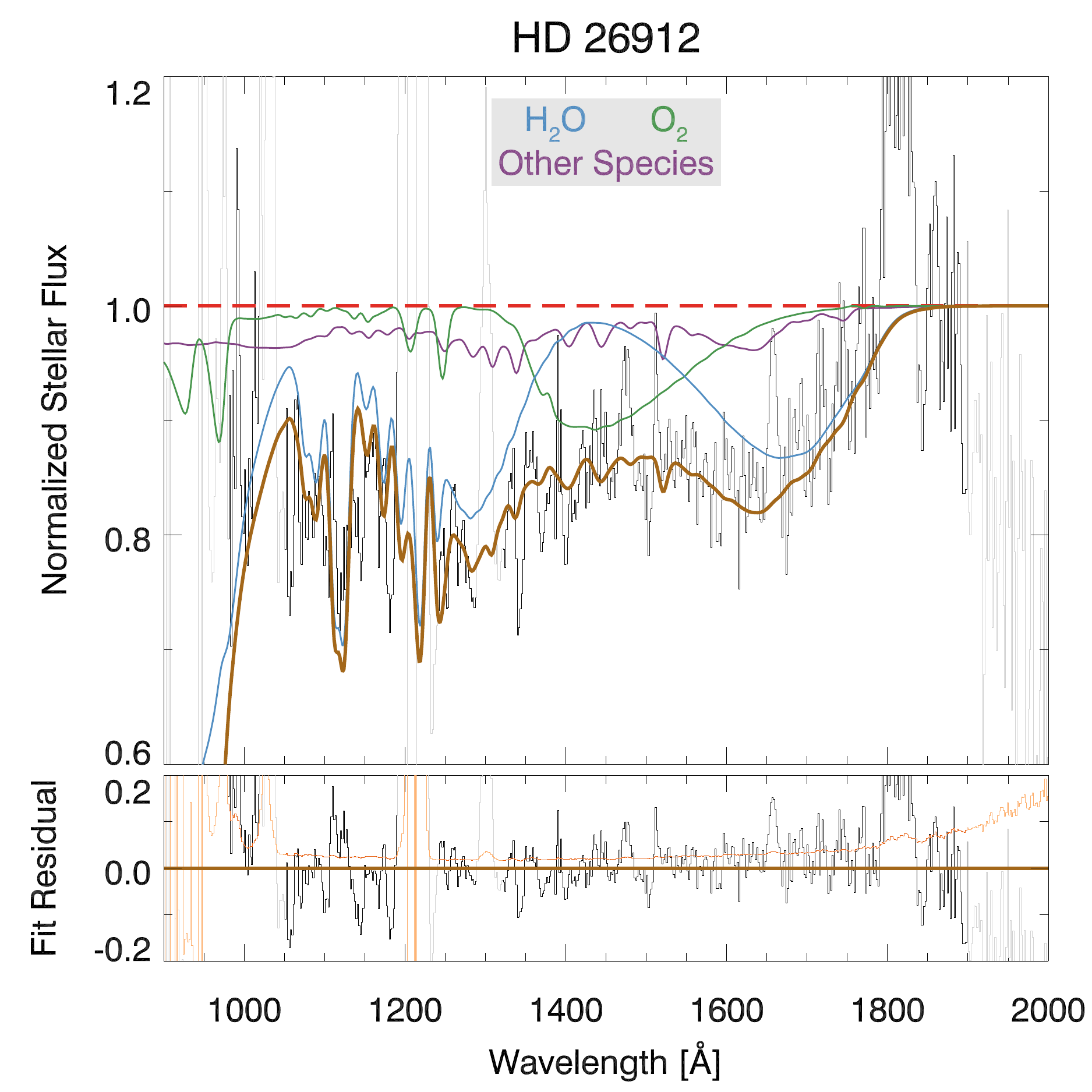}
\vspace{-1em}
\caption{Fits to the appulse absorption of HD~26912 ($\mathrm{FQ}=2$). 
\label{fig:fit_hd26912}}
\end{figure}

\begin{figure}
\centering\includegraphics[width=0.8\columnwidth]{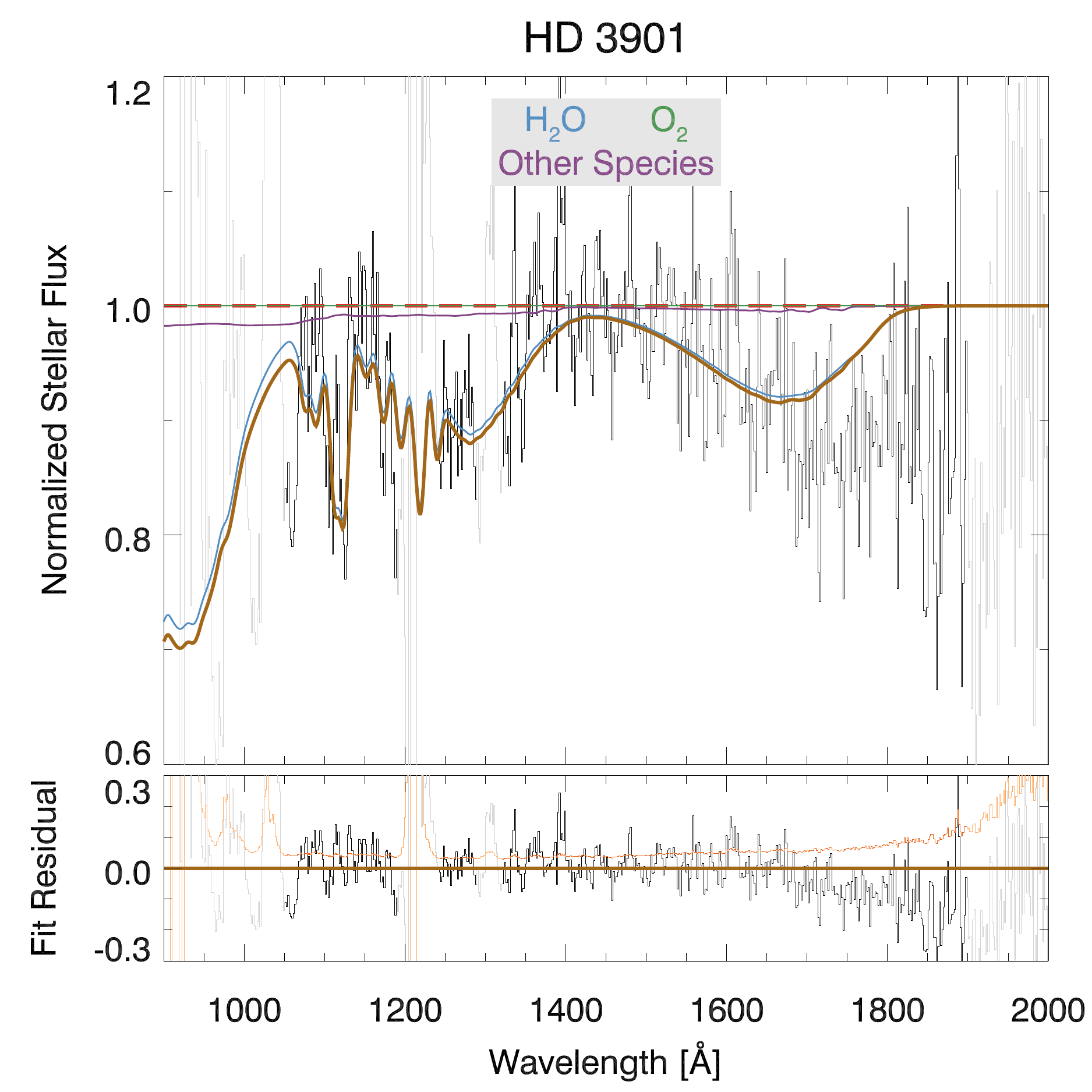}
\vspace{-1em}
\caption{Fits to the appulse absorption of HD~3901 ($\mathrm{FQ}=4$).
\label{fig:fit_hd3901}}
\end{figure}

\begin{figure}
\centering\includegraphics[width=0.8\columnwidth]{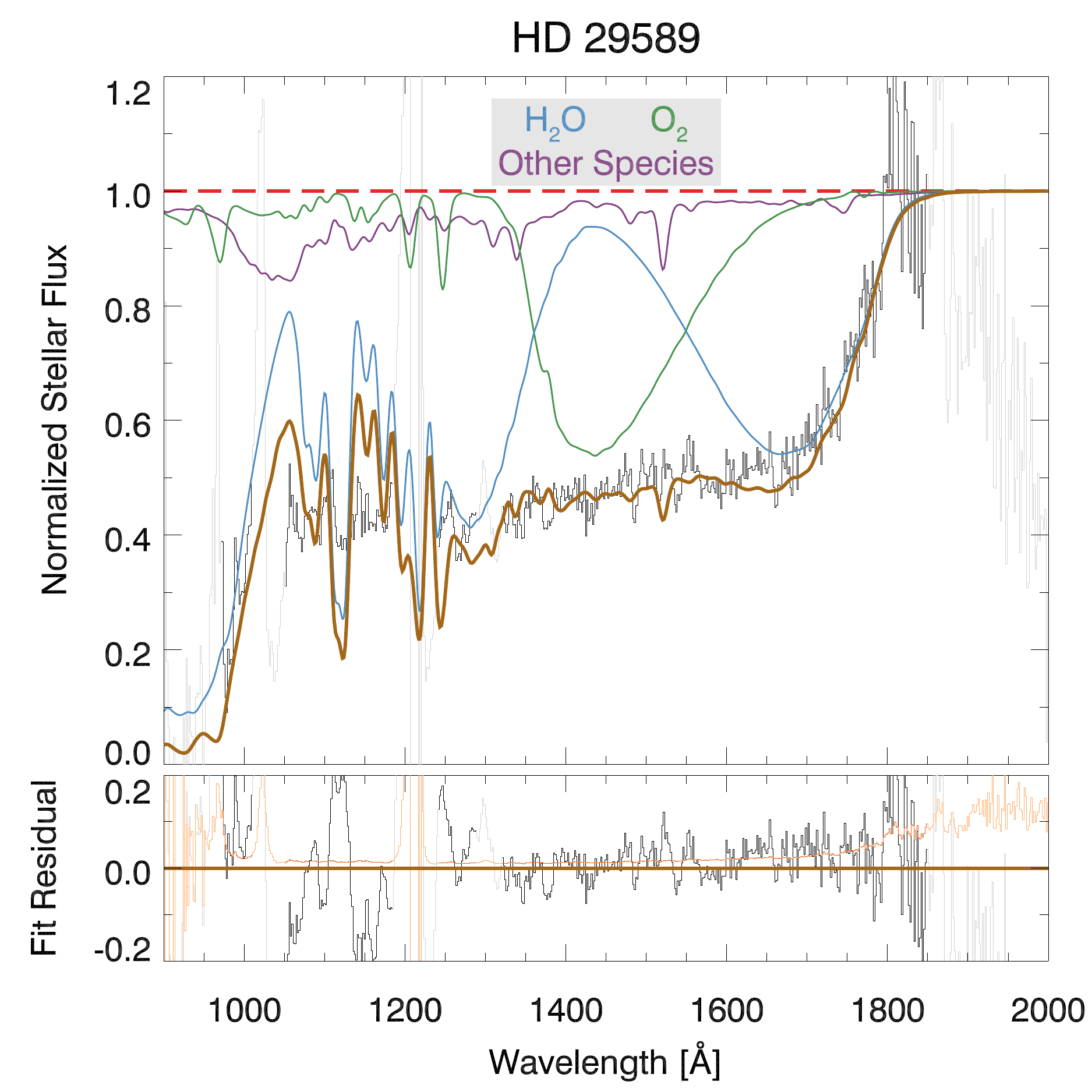}
\vspace{-1em}
\caption{Fits to the appulse absorption of HD~29589 ($\mathrm{FQ}=2$). 
\label{fig:fit_hd29589}}
\end{figure}

\begin{figure}
\centering\includegraphics[width=0.8\columnwidth]{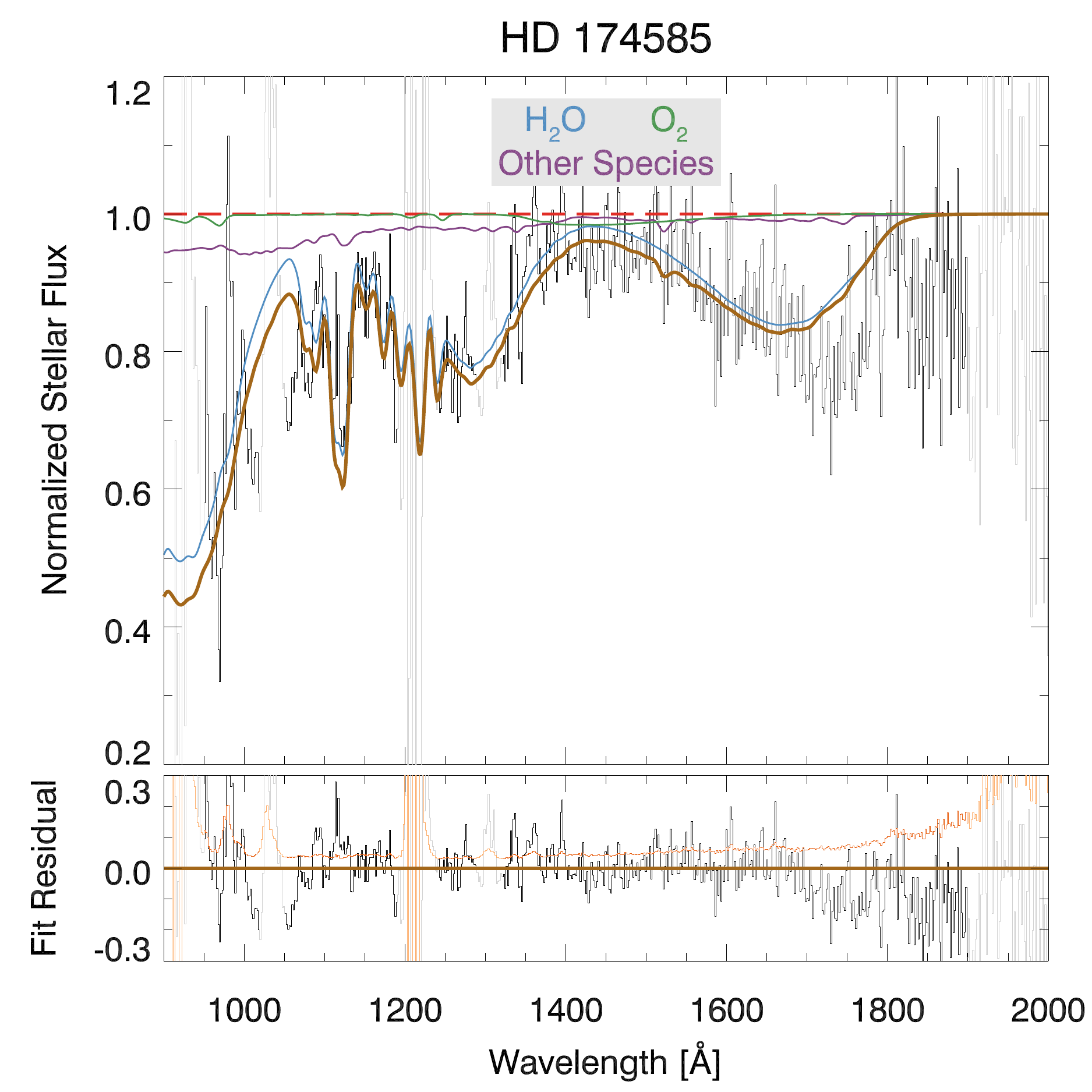}
\vspace{-1em}
\caption{Fits to the appulse absorption of HD~174585 ($\mathrm{FQ}=3$). 
\label{fig:fit_hd174585}}
\end{figure}

\begin{figure}
\centering\includegraphics[width=0.8\columnwidth]{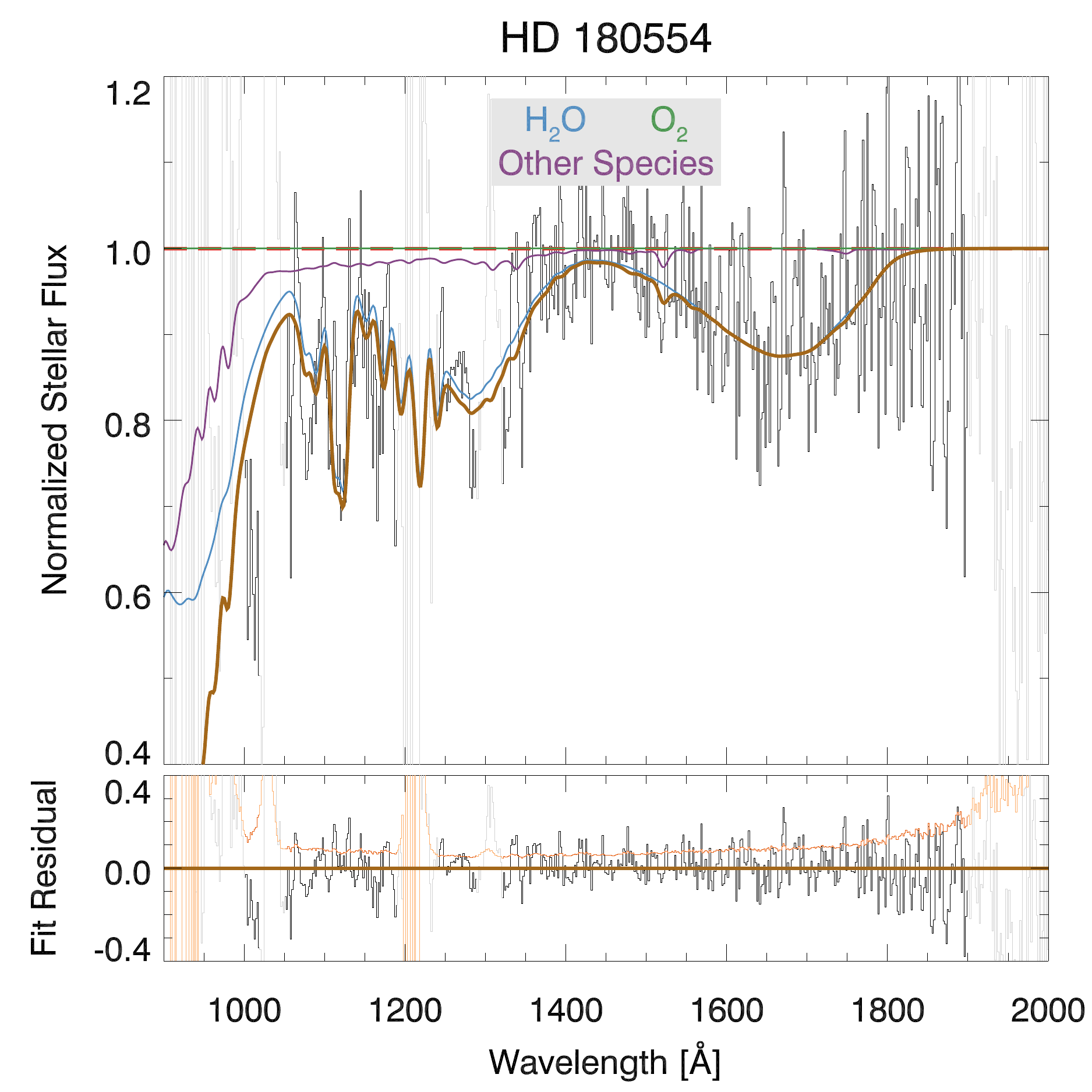}
\vspace{-1em}
\caption{Fits to the appulse absorption of HD~180554 ($\mathrm{FQ}=4$). 
\label{fig:fit_hd180554}}
\end{figure}

\begin{figure}
\centering\includegraphics[width=0.8\columnwidth]{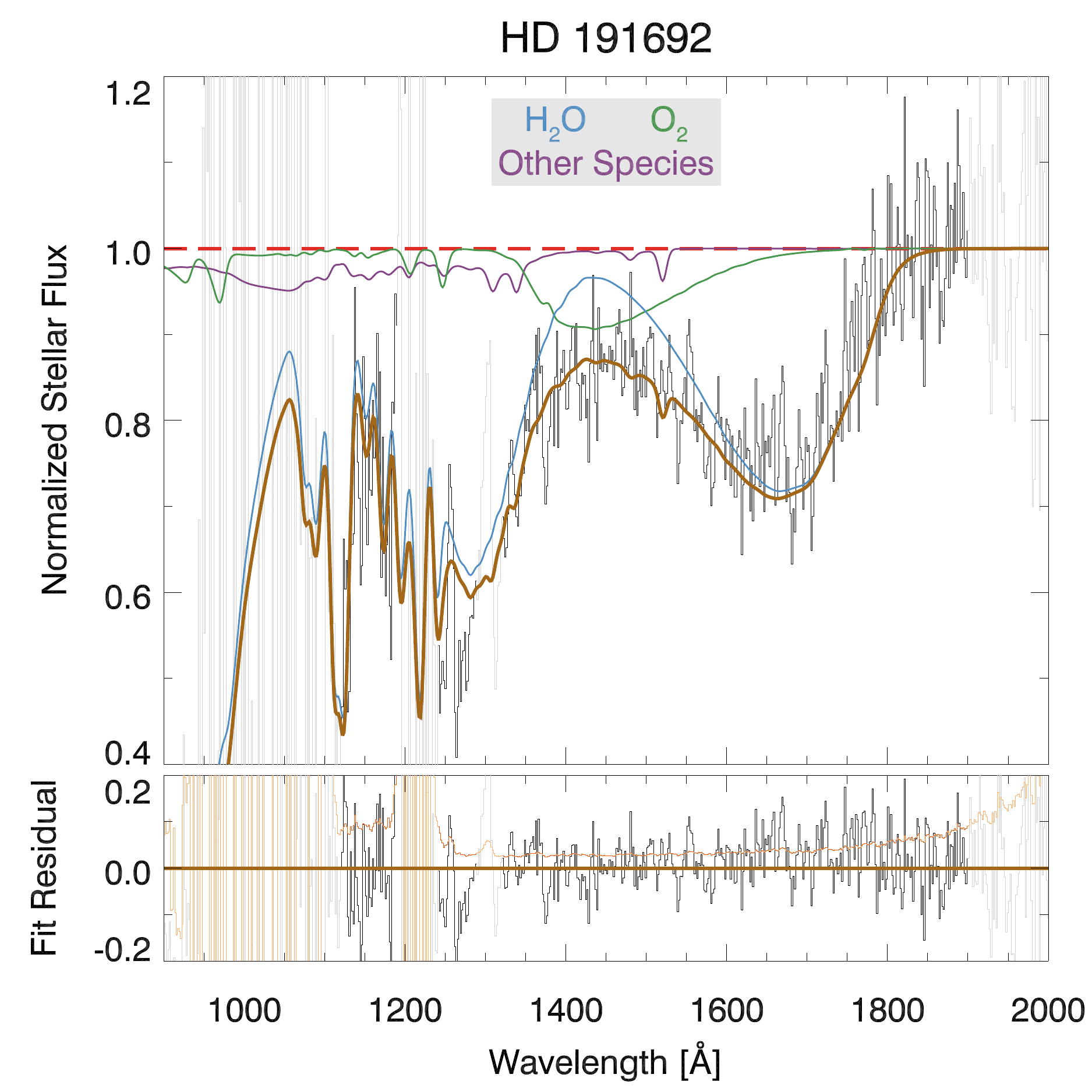}
\vspace{-1em}
\caption{Fits to the appulse absorption of HD~191692 ($\mathrm{FQ}=2$). 
\label{fig:fit_hd191692}}
\end{figure}

\begin{figure}
\centering\includegraphics[width=0.8\columnwidth]{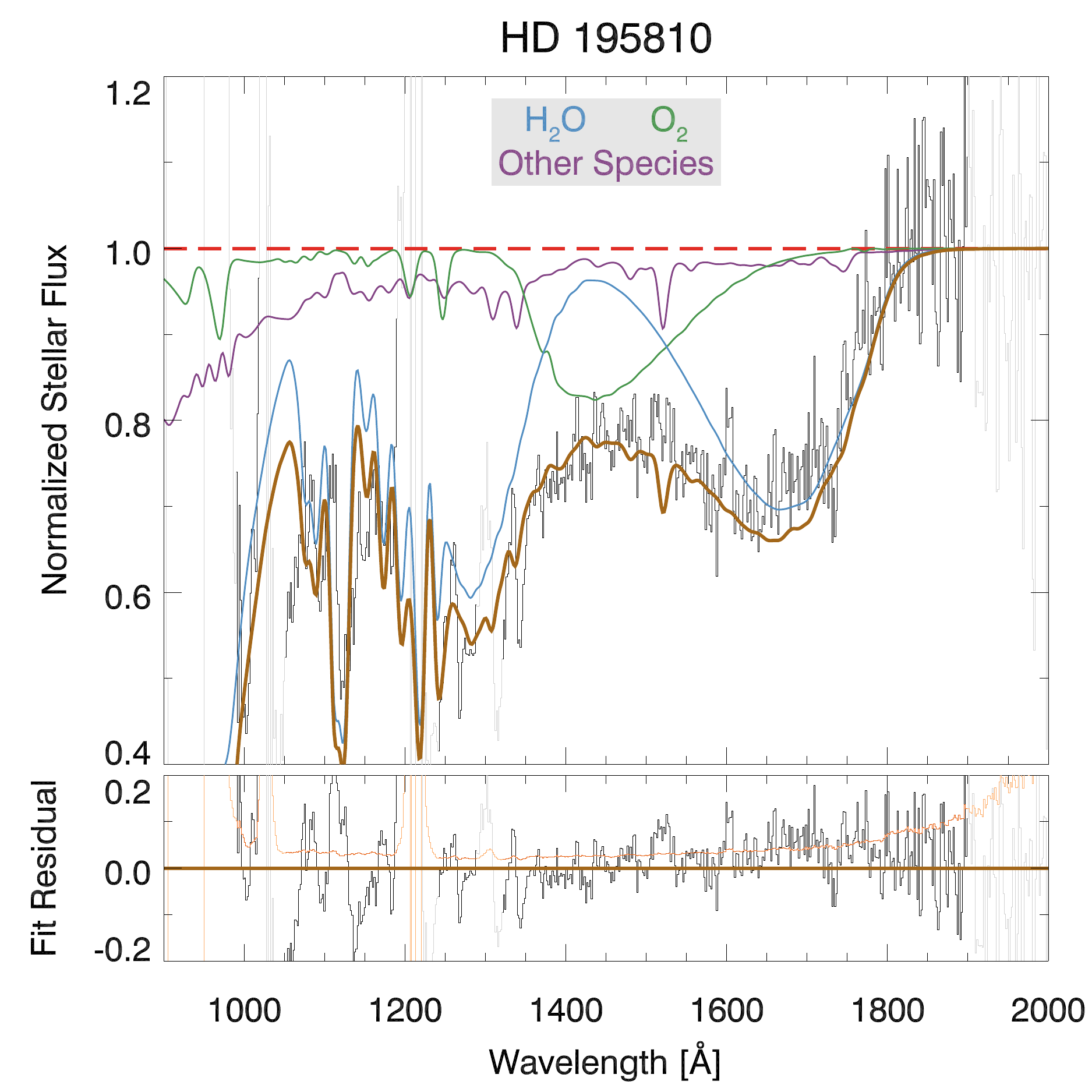}
\vspace{-1em}
\caption{Fits to the appulse absorption of HD~195810 ($\mathrm{FQ}=2$). 
\label{fig:fit_hd195810}}
\end{figure}

\begin{figure}
\centering\includegraphics[width=0.8\columnwidth]{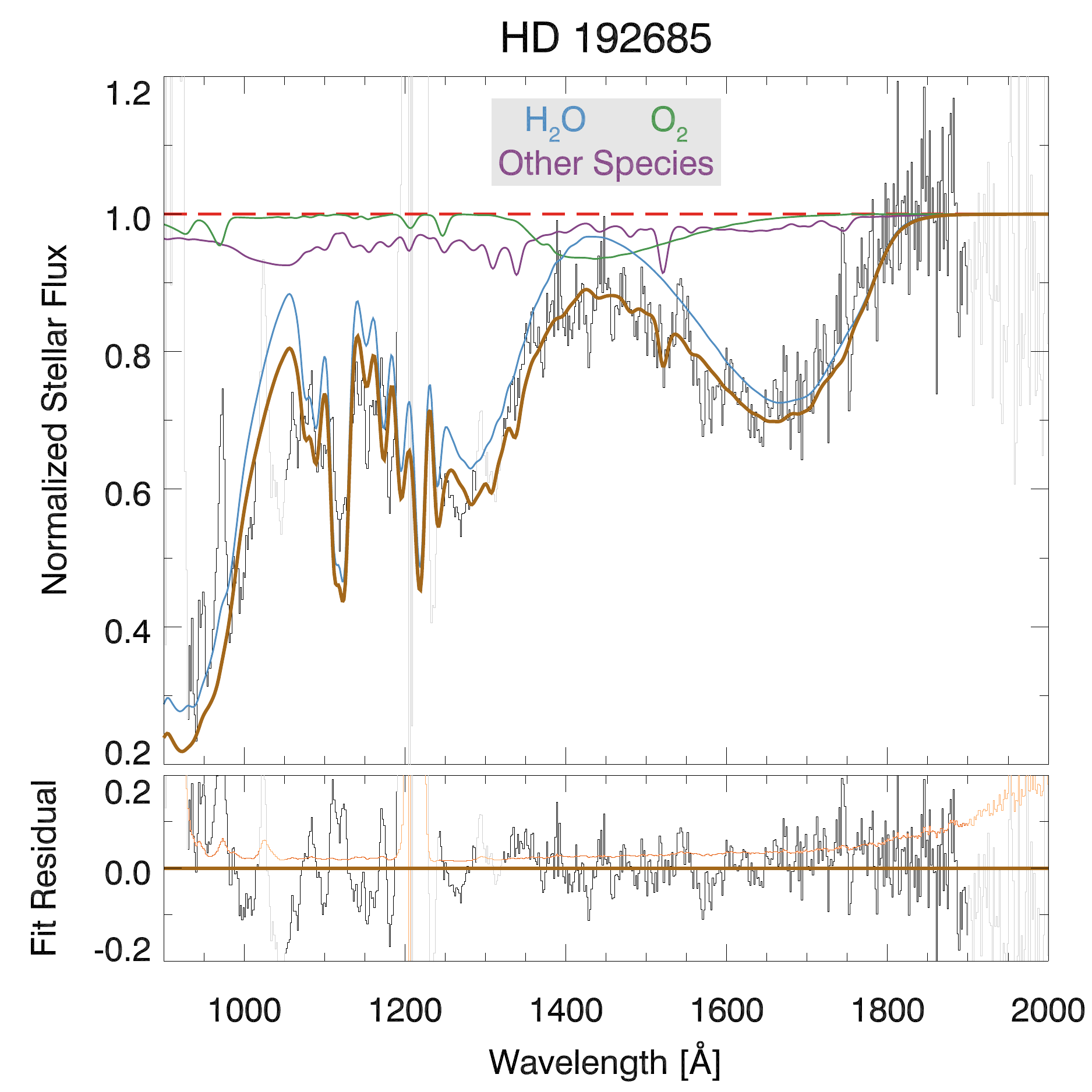}
\vspace{-1em}
\caption{Fits to the appulse absorption of HD~192685 ($\mathrm{FQ}=1$). 
\label{fig:fit_hd192685}}
\end{figure}

\begin{figure}
\centering\includegraphics[width=0.8\columnwidth]{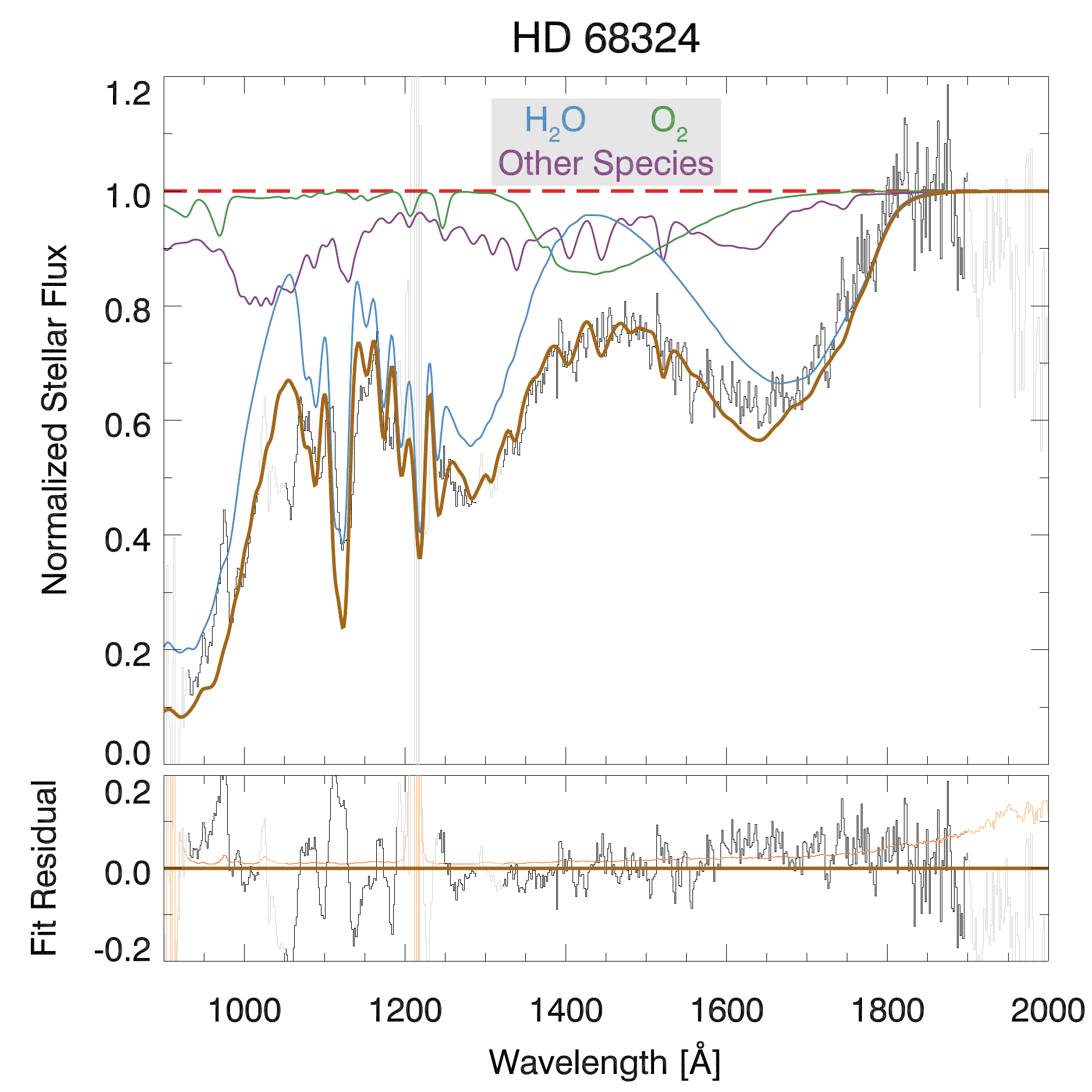}
\vspace{-1em}
\caption{Fits to the appulse absorption of HD~68324 ($\mathrm{FQ}=2$). 
\label{fig:fit_hd68324}}
\end{figure}

\begin{figure}
\centering\includegraphics[width=0.8\columnwidth]{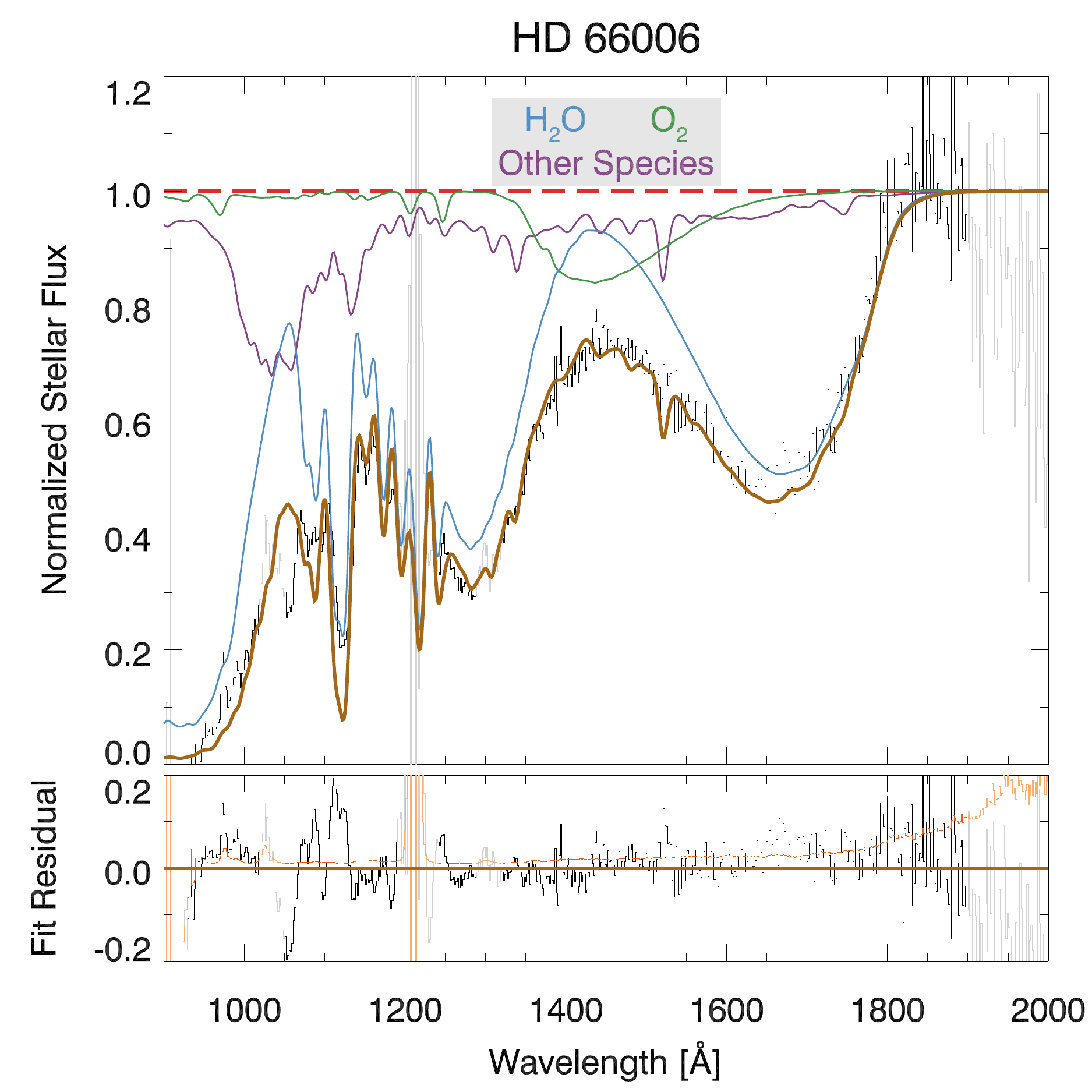}
\vspace{-1em}
\caption{Fits to the appulse absorption of HD~66006 ($\mathrm{FQ}=1$). 
\label{fig:fit_hd66006}}
\end{figure}

\begin{figure}
\centering\includegraphics[width=0.8\columnwidth]{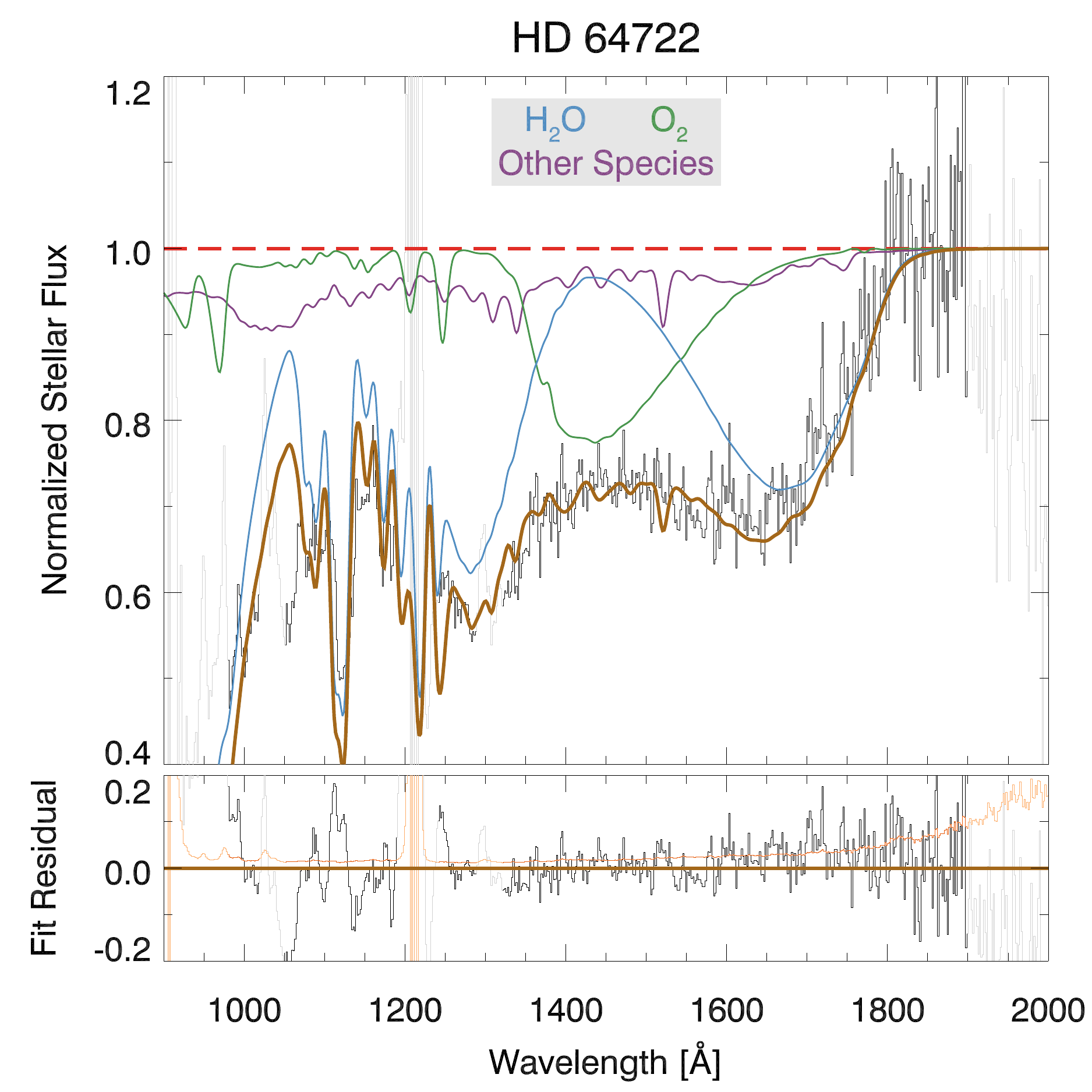}
\vspace{-1em}
\caption{Fits to the appulse absorption of HD~64722 ($\mathrm{FQ}=2$). 
\label{fig:fit_hd64722}}
\end{figure}

\begin{figure}
\centering\includegraphics[width=0.8\columnwidth]{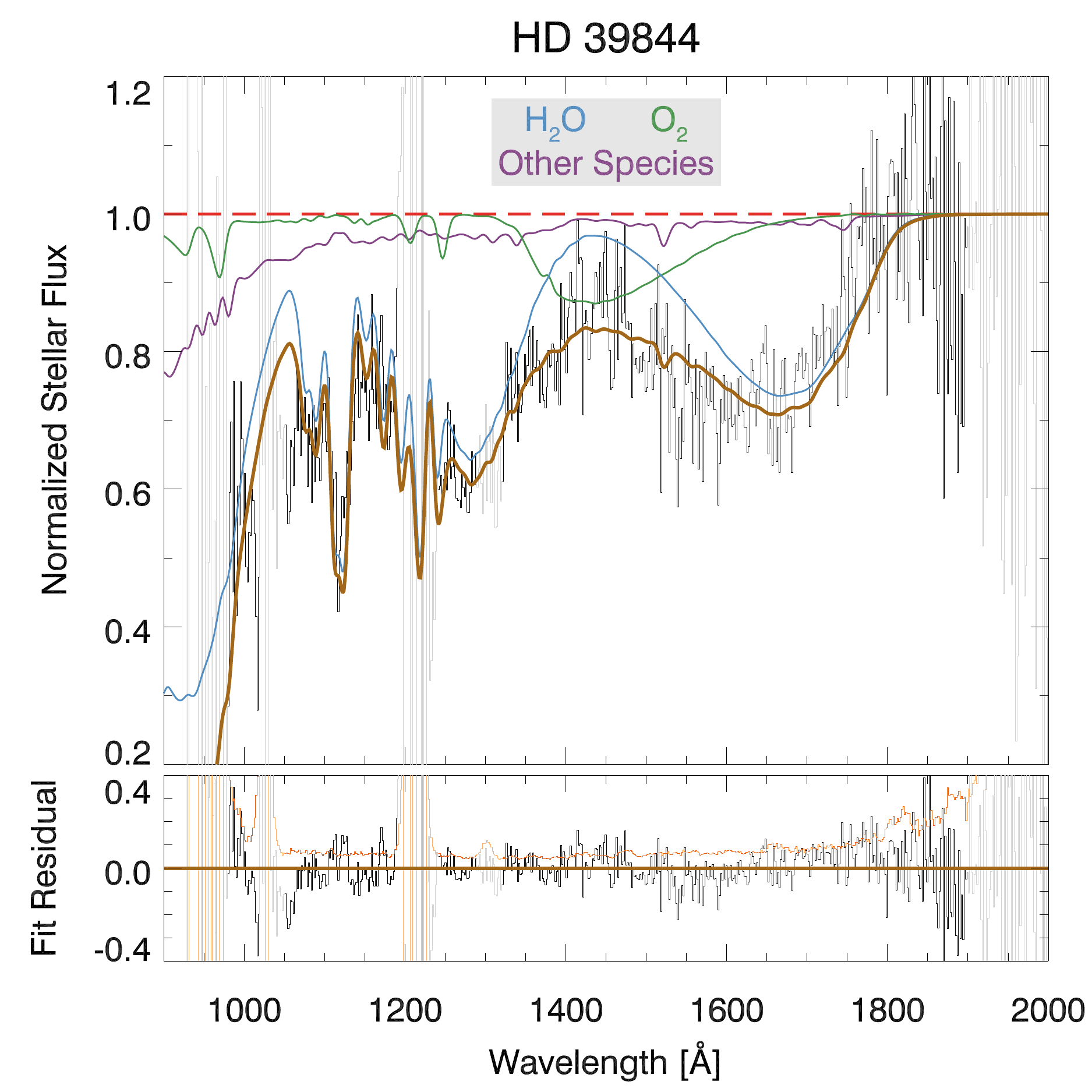}
\vspace{-1em}
\caption{Fits to the appulse absorption of HD~39844 ($\mathrm{FQ}=2$). 
\label{fig:fit_hd39844}}
\end{figure}

\begin{figure}
\centering\includegraphics[width=0.8\columnwidth]{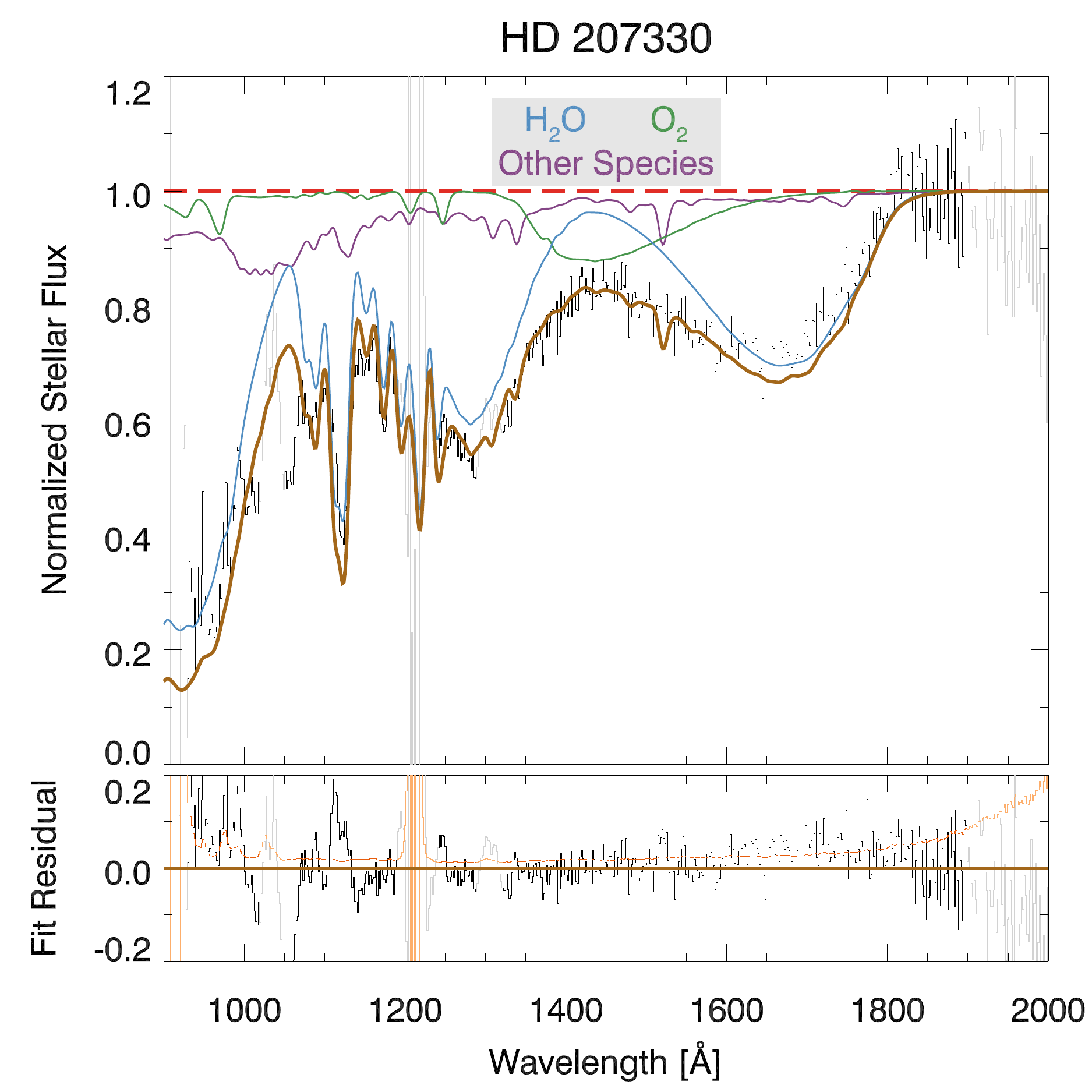}
\vspace{-1em}
\caption{Fits to the appulse absorption of HD~207330 ($\mathrm{FQ}=2$). 
\label{fig:fit_hd207330}}
\end{figure}

\begin{figure}
\centering\includegraphics[width=0.8\columnwidth]{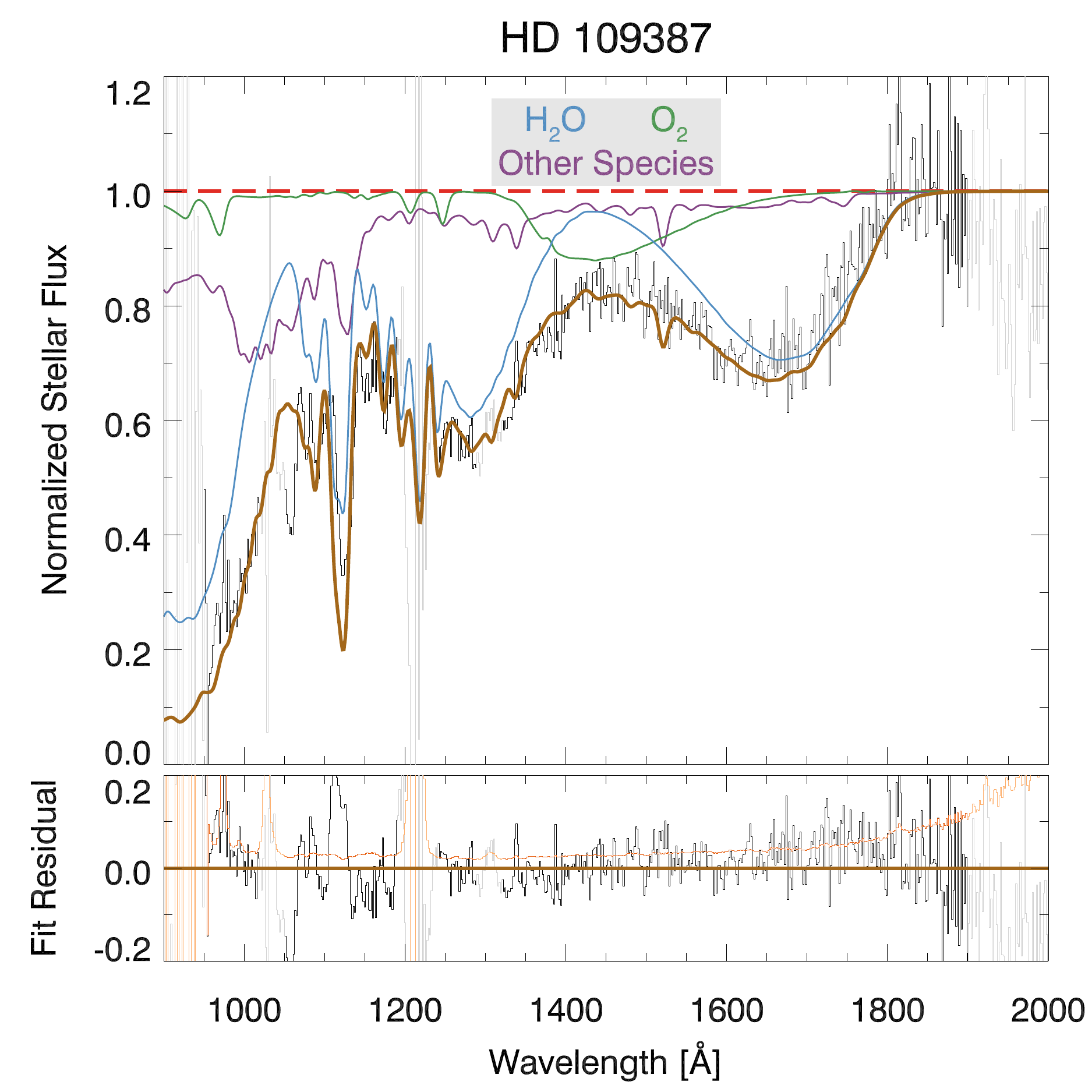}
\vspace{-1em}
\caption{Fits to the appulse absorption of HD~109387 ($\mathrm{FQ}=1$). 
\label{fig:fit_hd109387}}
\end{figure}

\begin{figure}
\centering\includegraphics[width=0.8\columnwidth]{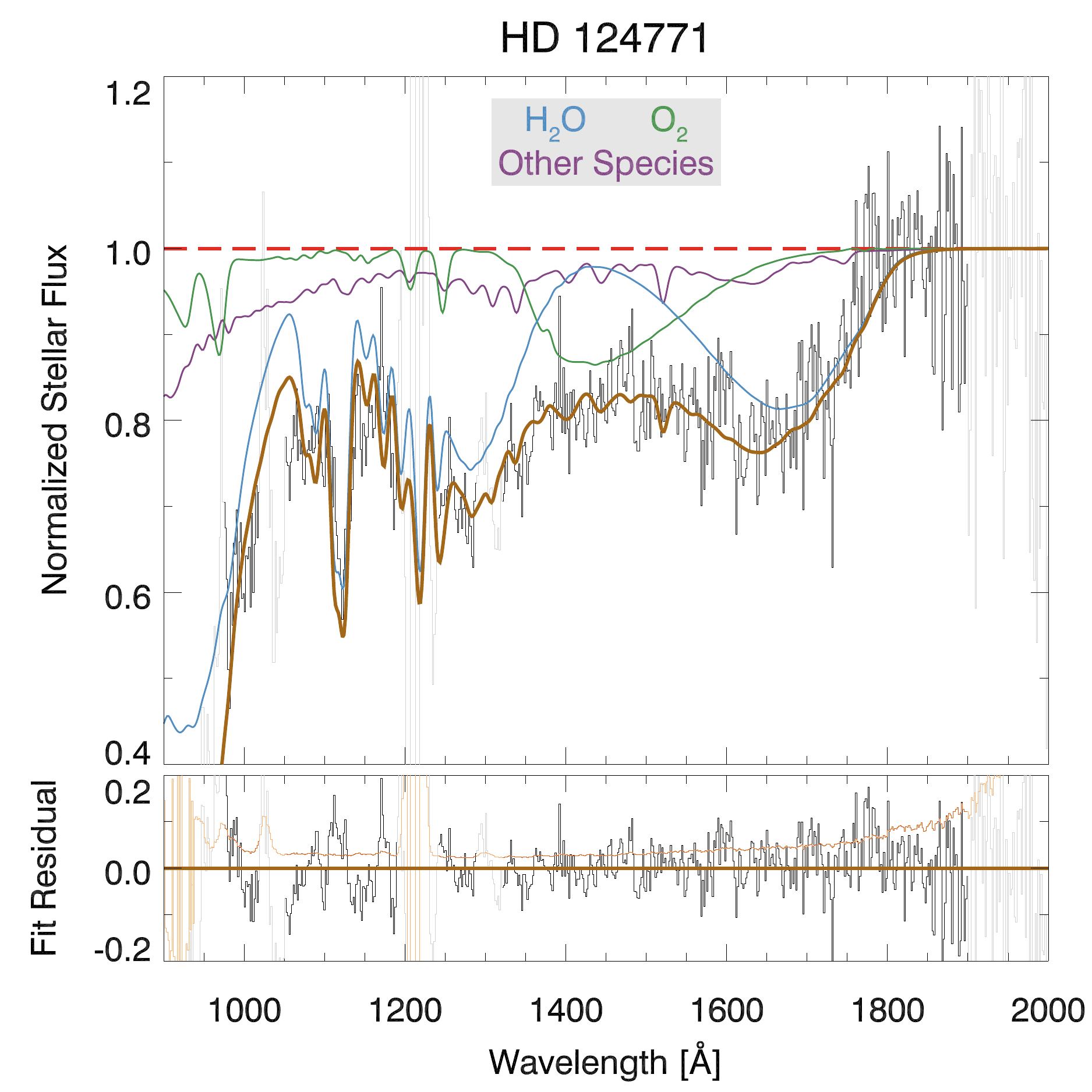}
\vspace{-1em}
\caption{Fits to the appulse absorption of HD~124771 ($\mathrm{FQ}=1$). 
\label{fig:fit_hd124771}}
\end{figure}

\begin{figure}
\centering\includegraphics[width=0.8\columnwidth]{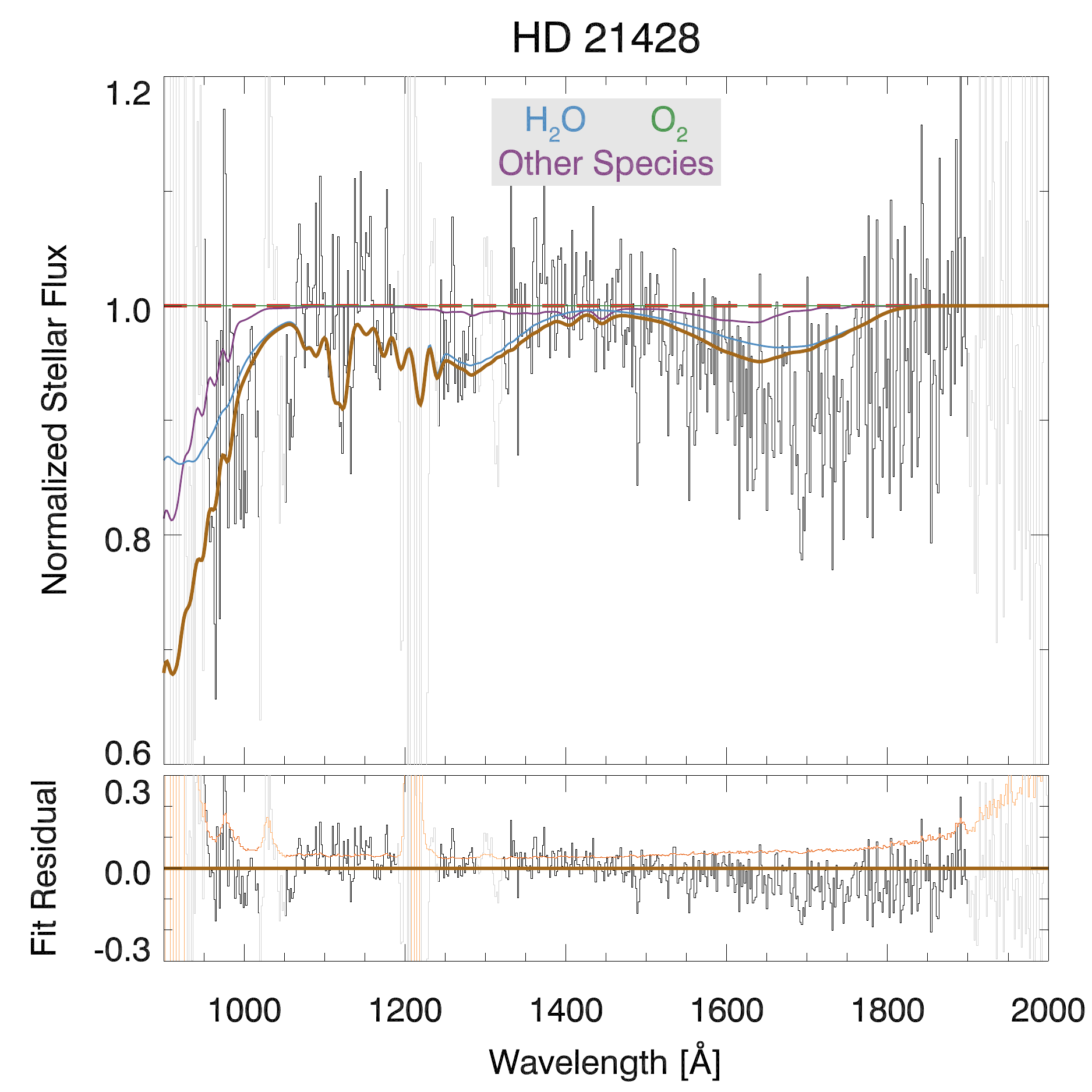}
\vspace{-1em}
\caption{Fits to the appulse absorption of HD~21428 ($\mathrm{FQ}=4$). 
\label{fig:fit_hd21428}}
\end{figure}

\begin{figure}
\centering\includegraphics[width=0.8\columnwidth]{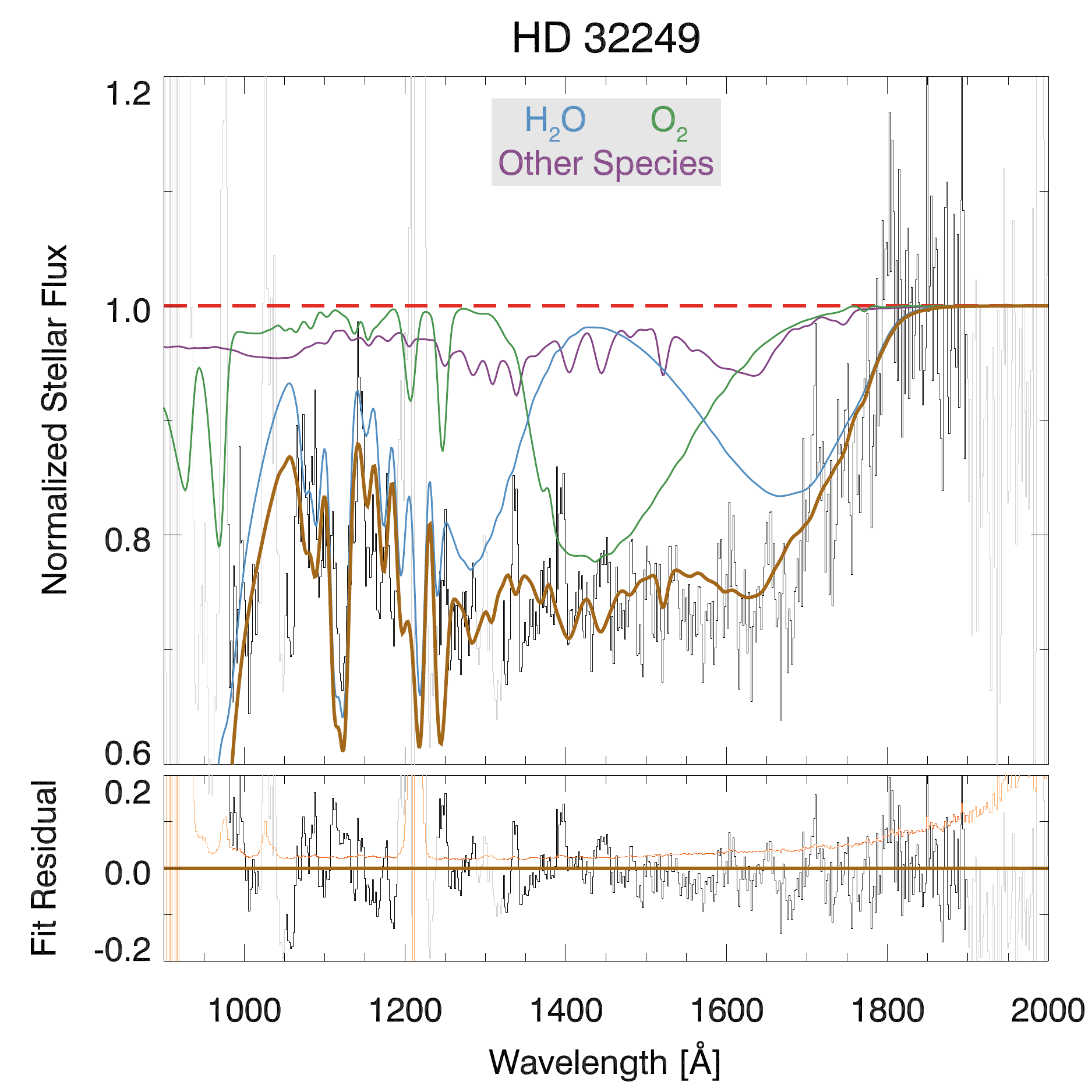}
\vspace{-1em}
\caption{Fits to the appulse absorption of HD~32249 ($\mathrm{FQ}=2$). 
\label{fig:fit_hd32249}}
\end{figure}

\begin{figure}
\centering\includegraphics[width=0.8\columnwidth]{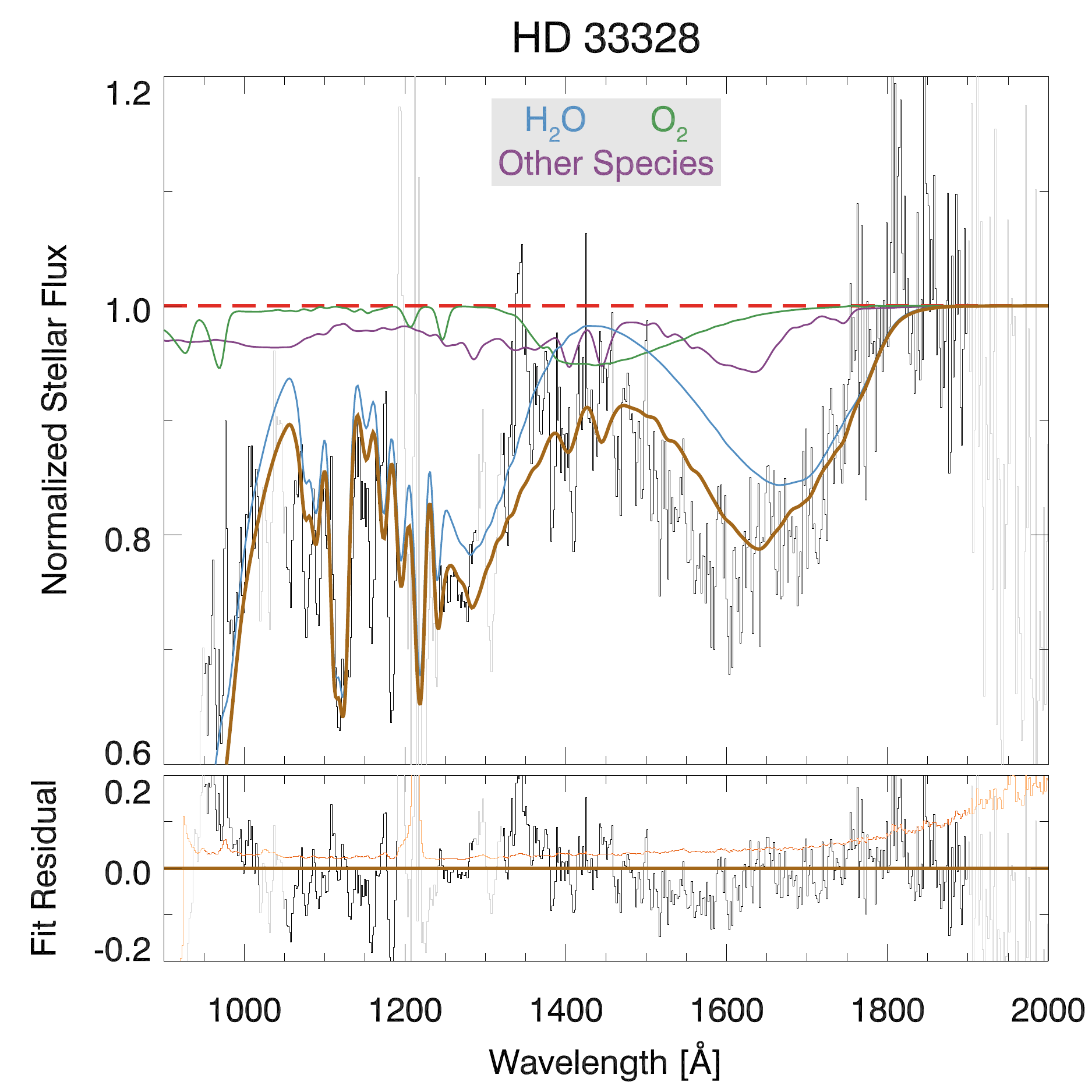}
\vspace{-1em}
\caption{Fits to the appulse absorption of HD~33328 ($\mathrm{FQ}=2$). 
\label{fig:fit_hd33328}}
\end{figure}

\begin{figure}
\centering\includegraphics[width=0.8\columnwidth]{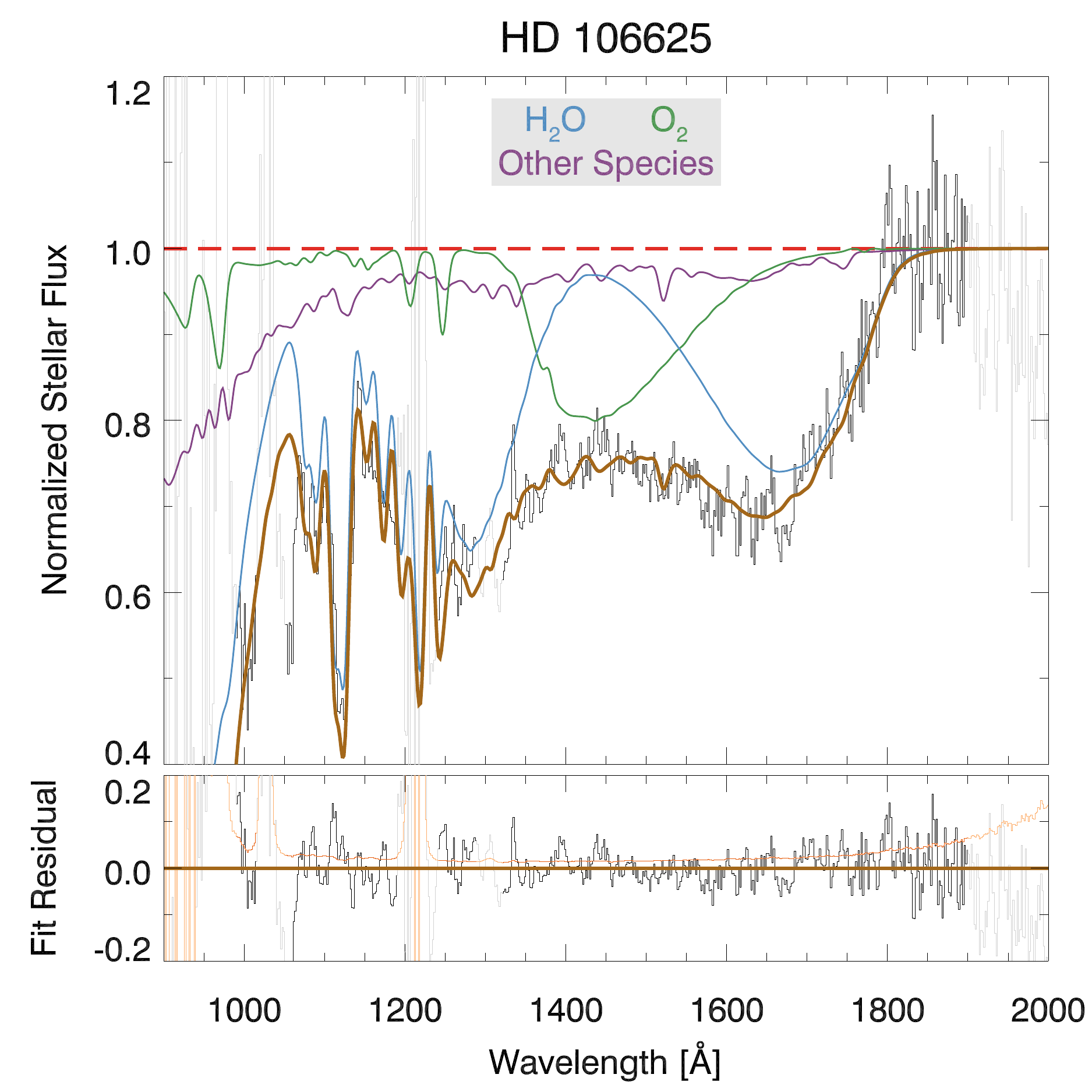}
\vspace{-1em}
\caption{Fits to the appulse absorption of HD~106625 ($\mathrm{FQ}=1$). 
\label{fig:fit_hd106625}}
\end{figure}

\begin{figure}
\centering\includegraphics[width=0.8\columnwidth]{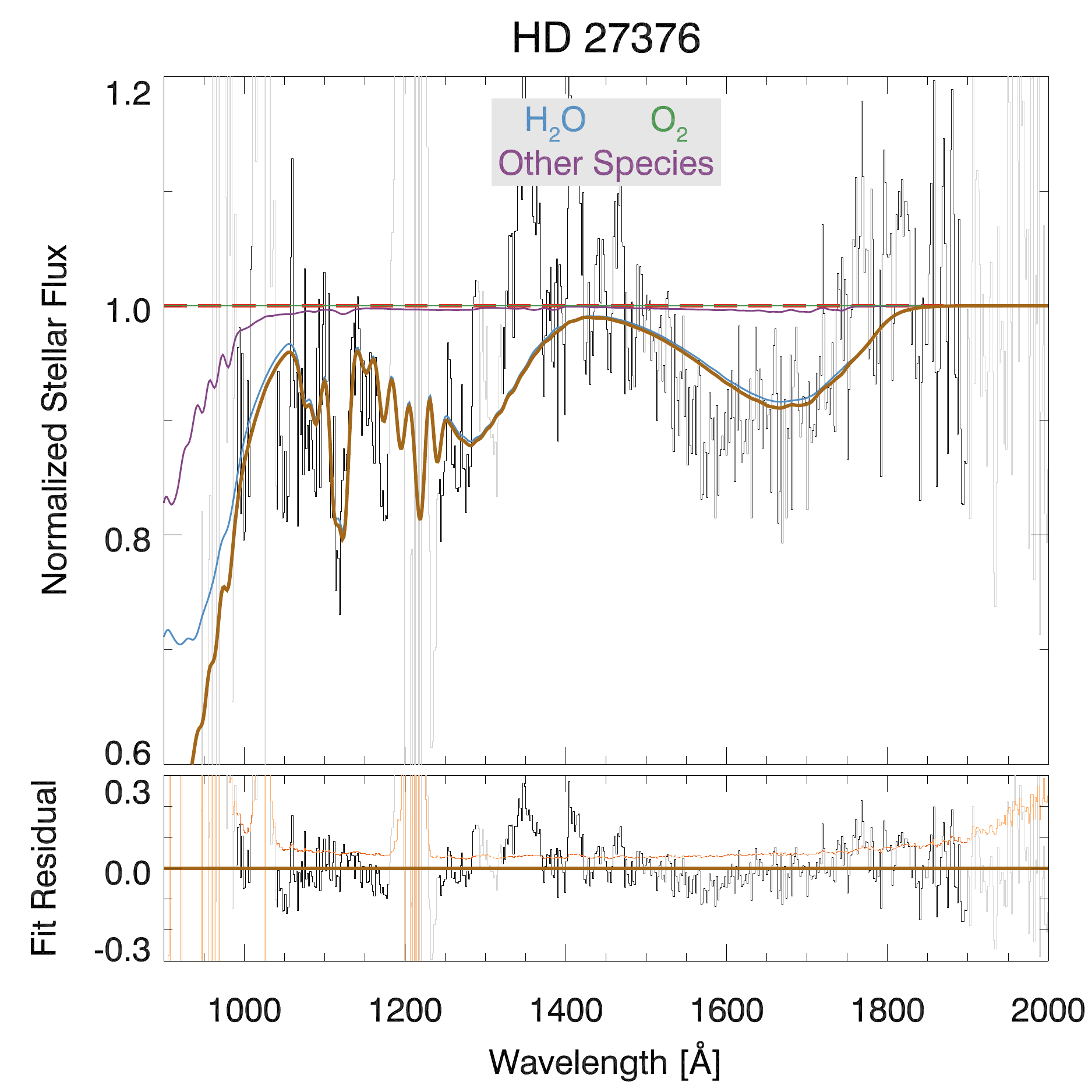}
\vspace{-1em}
\caption{Fits to the appulse absorption of HD~27376 ($\mathrm{FQ}=3$). 
\label{fig:fit_hd27376}}
\end{figure}

\begin{figure}
\centering\includegraphics[width=0.8\columnwidth]{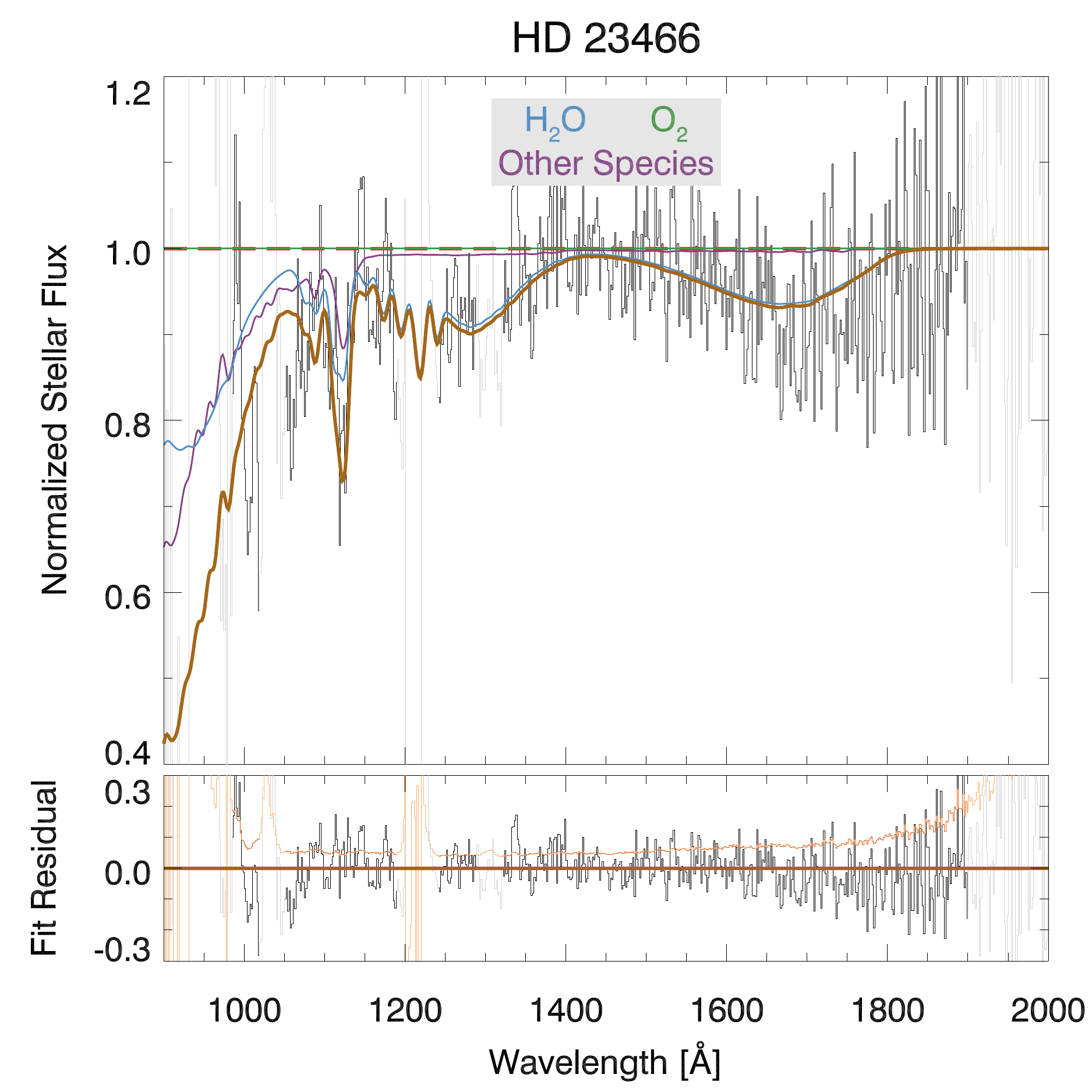}
\vspace{-1em}
\caption{Fits to the appulse absorption of HD~23466 ($\mathrm{FQ}=4$). 
\label{fig:fit_hd23466}}
\end{figure}

\begin{figure}
\centering\includegraphics[width=0.8\columnwidth]{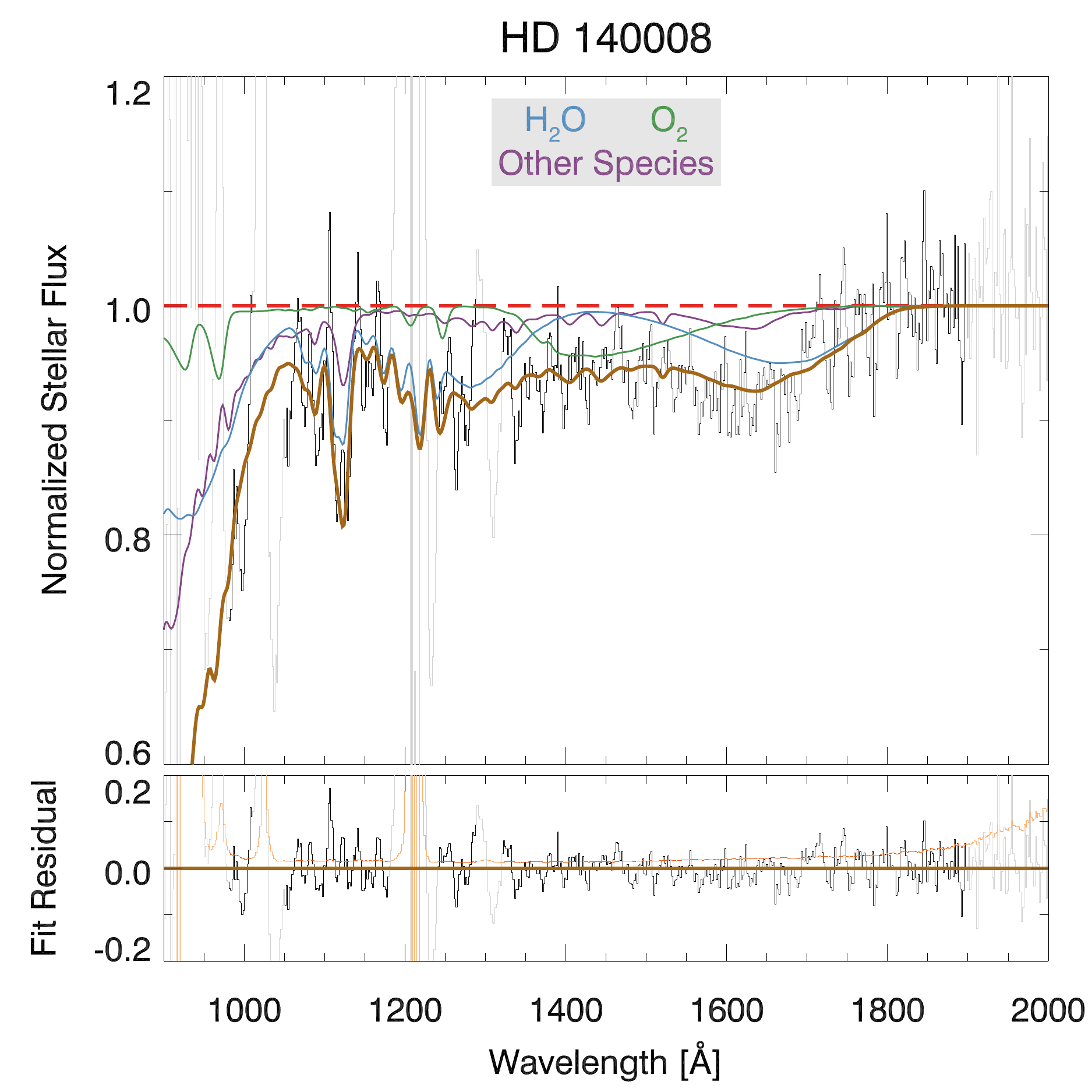}
\vspace{-1em}
\caption{Fits to the appulse absorption of HD~140008 ($\mathrm{FQ}=2$). 
\label{fig:fit_hd140008}}
\end{figure}

\begin{figure}
\centering\includegraphics[width=0.8\columnwidth]{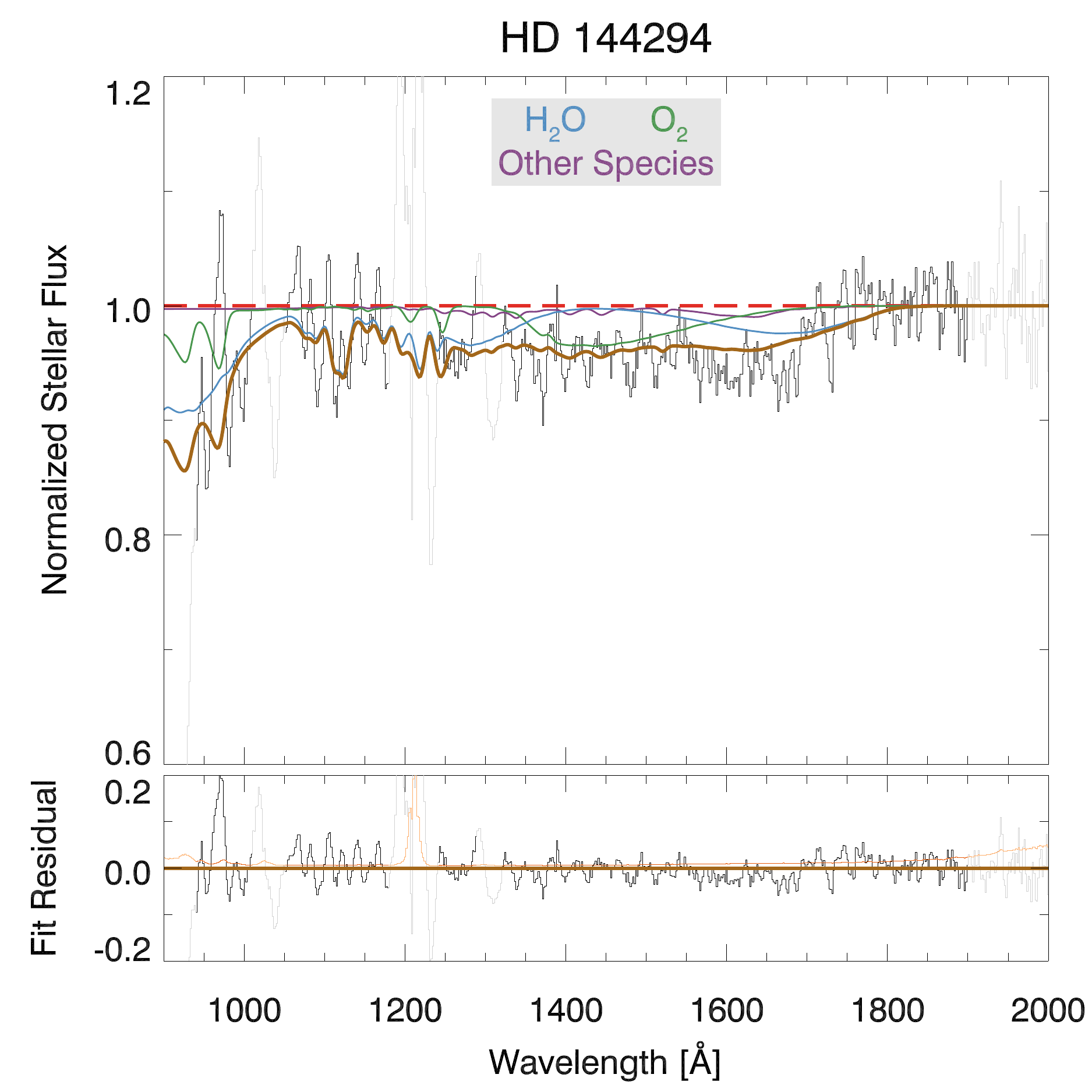}
\vspace{-1em}
\caption{Fits to the appulse absorption of HD~144294 ($\mathrm{FQ}=3$). 
\label{fig:fit_hd144294}}
\end{figure}

\begin{figure}
\centering\includegraphics[width=0.8\columnwidth]{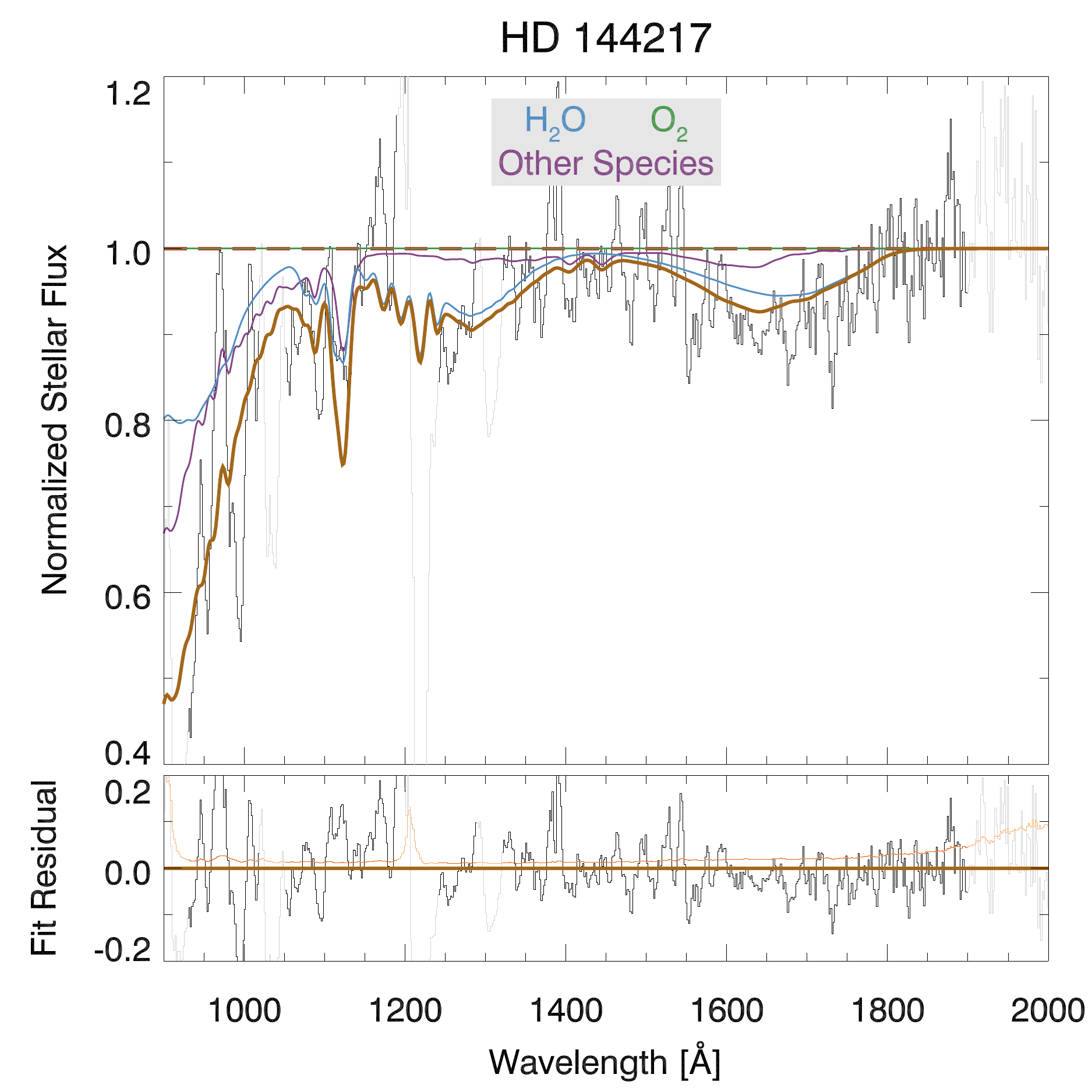}
\vspace{-1em}
\caption{Fits to the appulse absorption of HD~144217 ($\mathrm{FQ}=3$). 
\label{fig:fit_hd144217}}
\end{figure}

\begin{figure}
\centering\includegraphics[width=0.8\columnwidth]{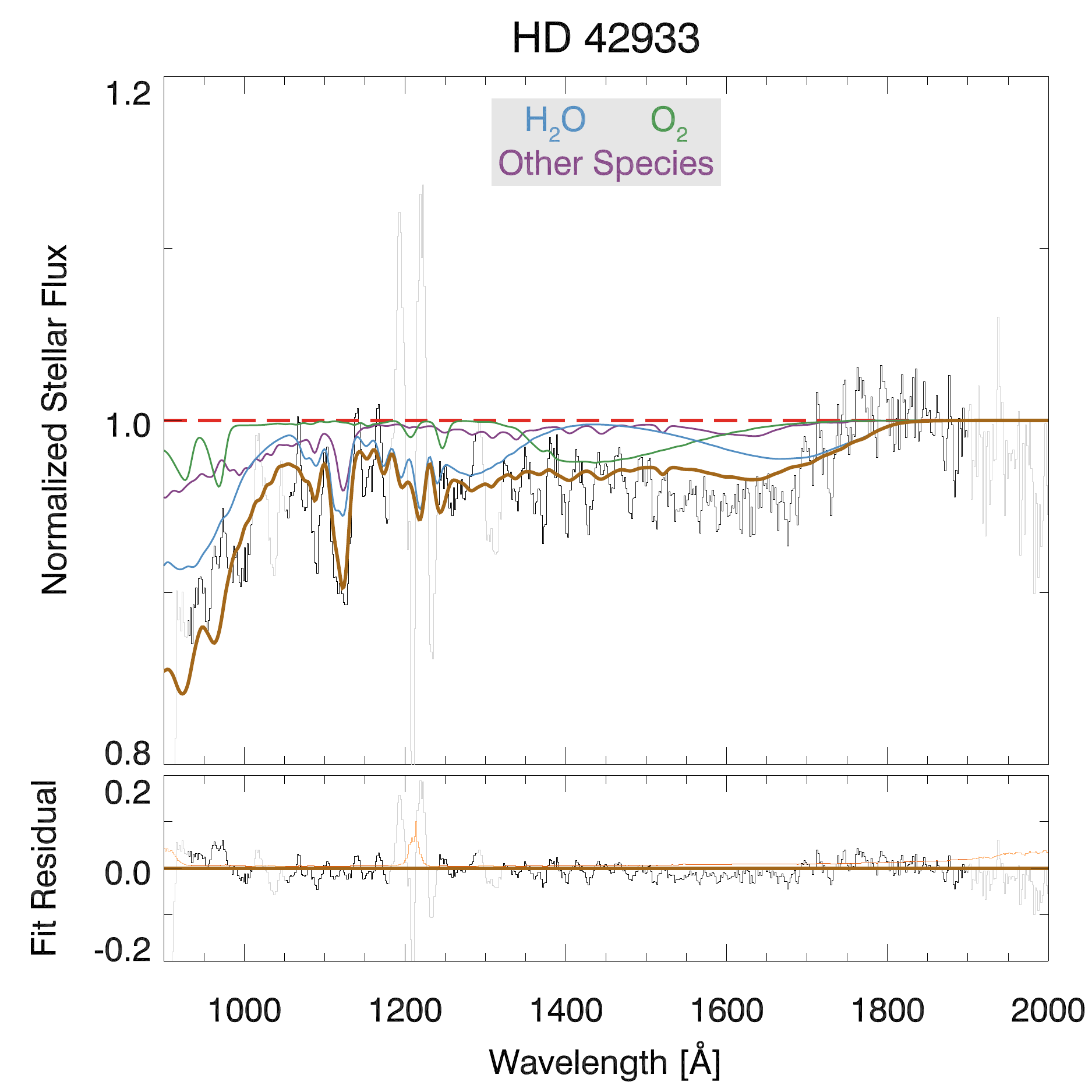}
\vspace{-1em}
\caption{Fits to the appulse absorption of HD~42933 ($\mathrm{FQ}=3$). 
\label{fig:fit_hd42933}}
\end{figure}

\begin{figure}
\centering\includegraphics[width=0.8\columnwidth]{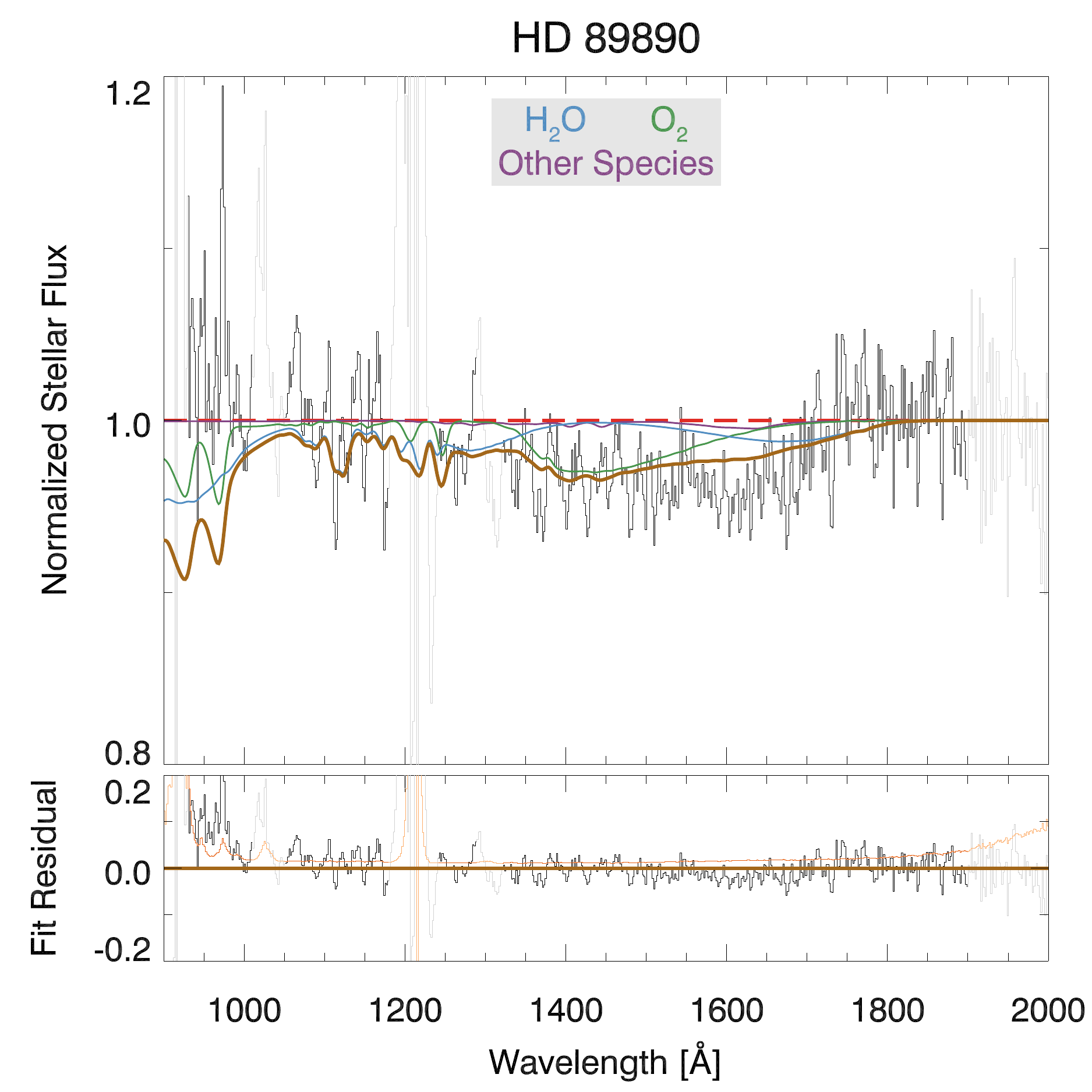}
\vspace{-1em}
\caption{Fits to the appulse absorption of HD~89890 ($\mathrm{FQ}=4$). 
\label{fig:fit_hd89890}}
\end{figure}

\begin{figure}
\centering\includegraphics[width=0.8\columnwidth]{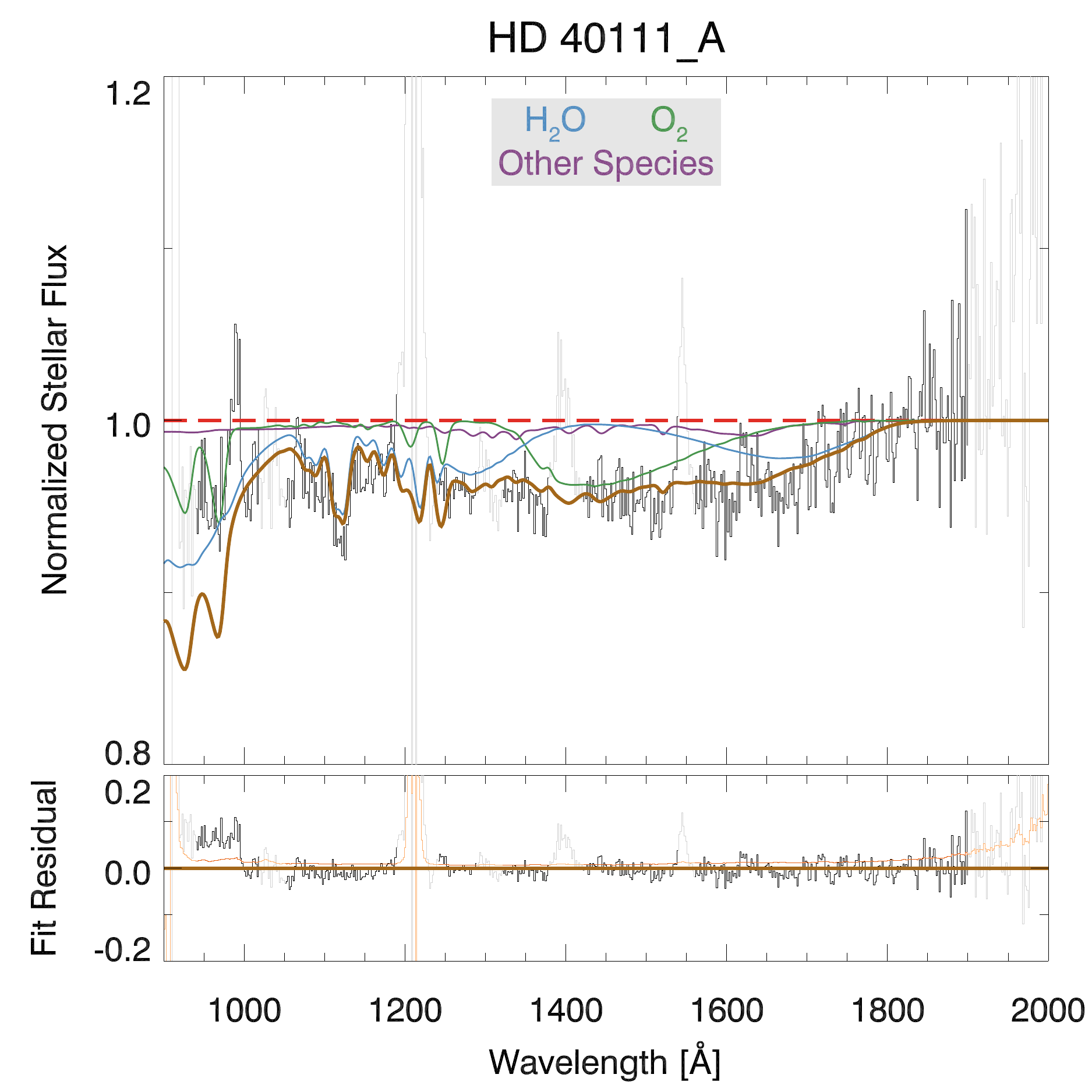}
\vspace{-1em}
\caption{Fits to the first appulse absorption of HD~40111 ($\mathrm{FQ}=3$). 
\label{fig:fit_hd40111_a}}
\end{figure}

\begin{figure}
\centering\includegraphics[width=0.8\columnwidth]{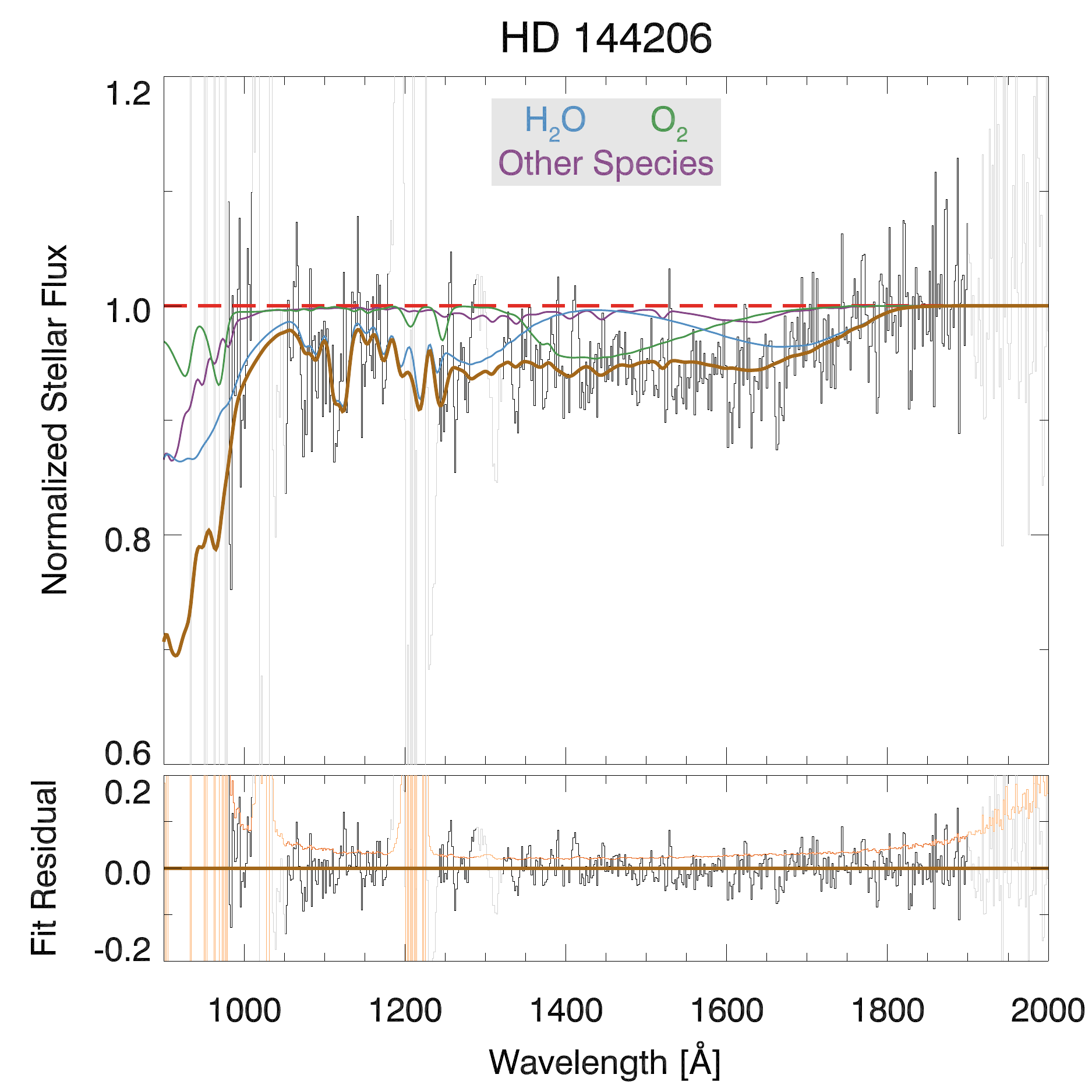}
\vspace{-1em}
\caption{Fits to the appulse absorption of HD~144206 ($\mathrm{FQ}=2$). 
\label{fig:fit_hd144206}}
\end{figure}

\begin{figure}
\centering\includegraphics[width=0.8\columnwidth]{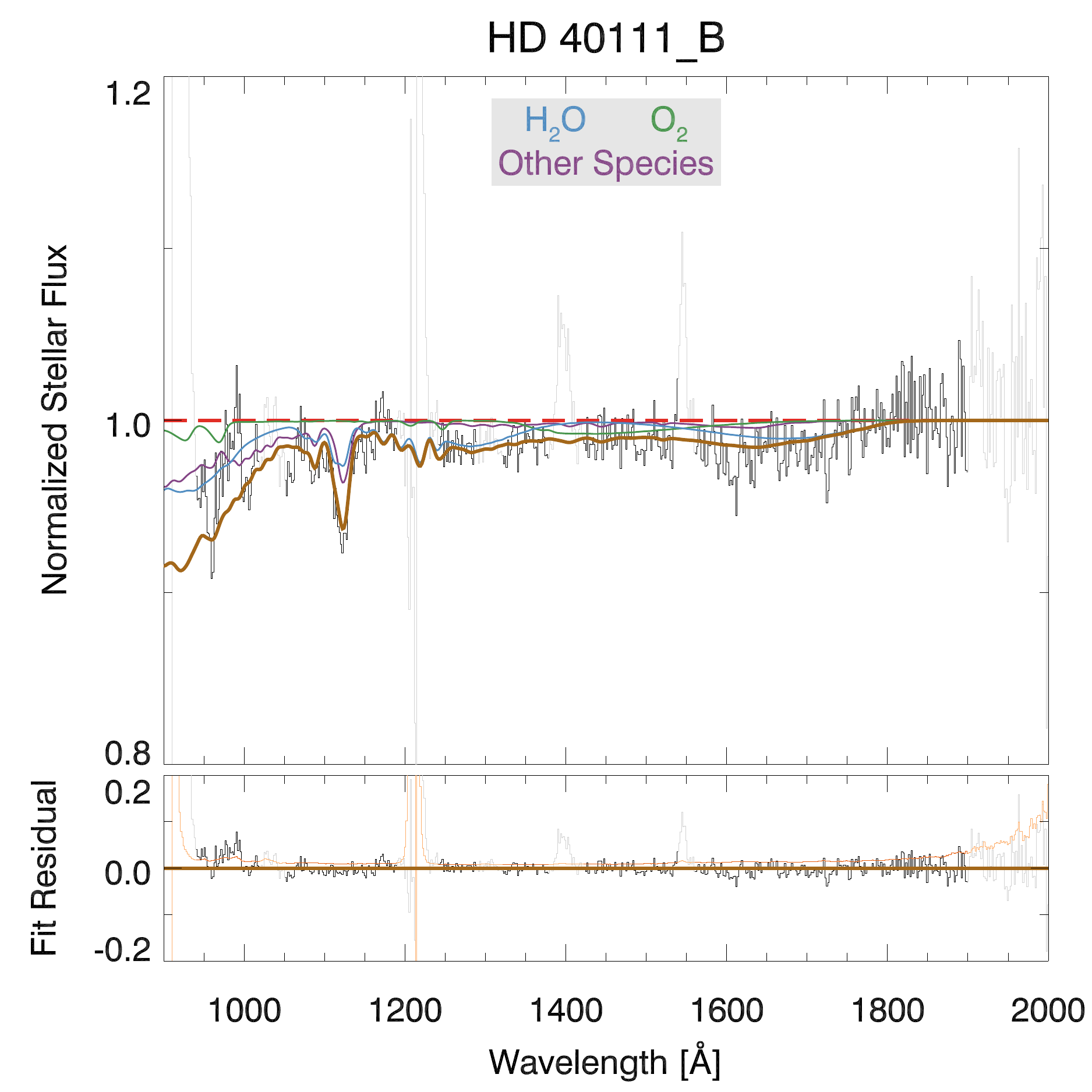}
\vspace{-1em}
\caption{Fits to the second appulse absorption of HD~40111 ($\mathrm{FQ}=3$). 
\label{fig:fit_hd40111_b}}
\end{figure}

Figures~\ref{fig:fit_hd26912}-\ref{fig:fit_hd40111_b} present best-fit absorption profiles for all targeted and archival stellar appulses, arranged chronologically. The top panel of each figure displays the normalized stellar flux (black) and $1\sigma$ uncertainty (gray), with best-fit ensemble absorption (solid brown line) overlaid. Absorption from individual species is shown with solid (\ce{H2O}, \ce{O2}, \ce{CO}, \ce{CO2}, \ce{CH4}) or dashed (\ce{C2H2}, \ce{C2H6}, \ce{C2H4}, \ce{C4H2}, \ce{H2CO}) lines. The bottom panel of each figure displays the residual of the ensemble fit.

\newpage
\section{Adopted Absorption Profiles}
\label{app:conf}

\begin{figure}
\centering\includegraphics[width=0.8\columnwidth]{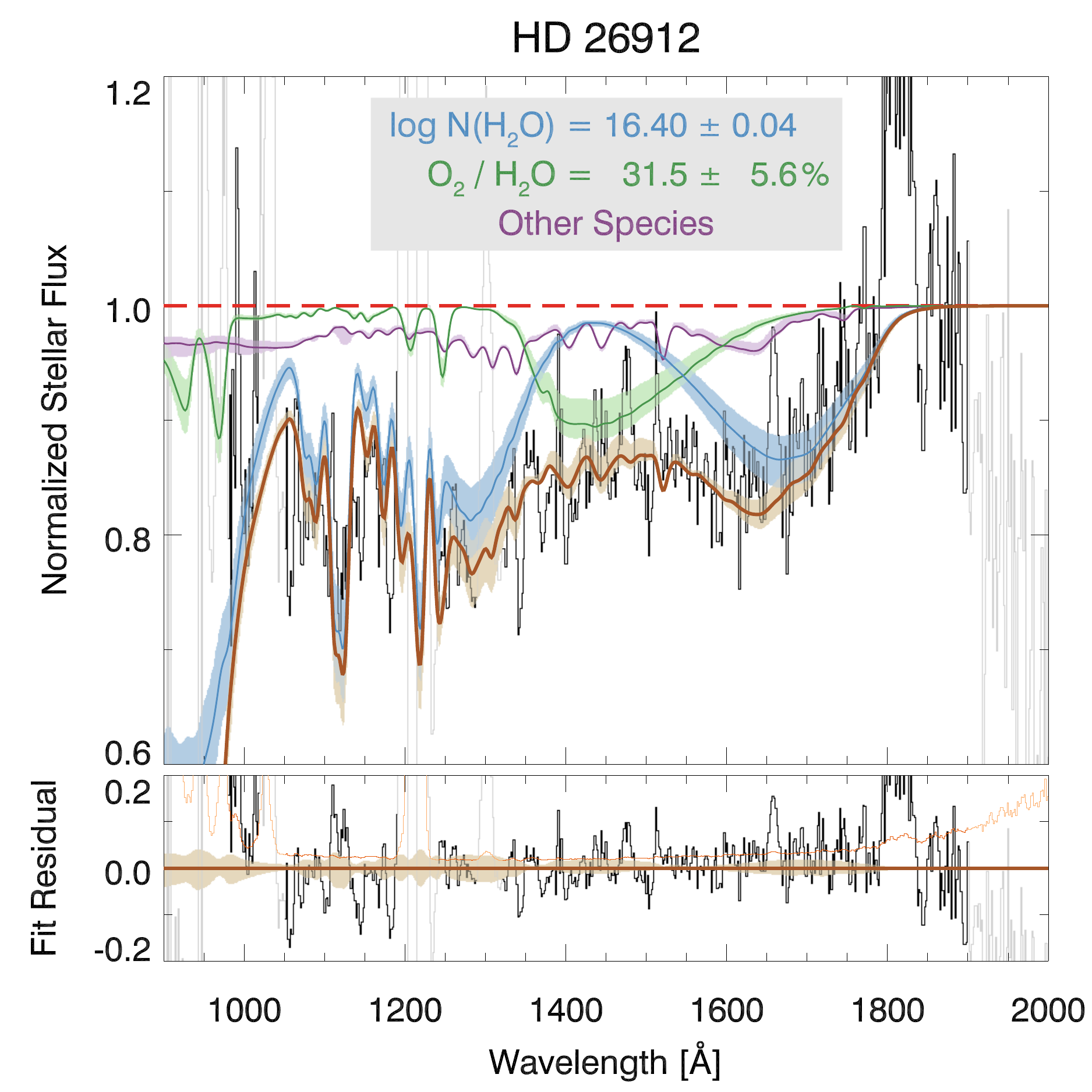}
\vspace{-1em}
\caption{Adopted column densities for the appulse of HD~26912 ($\mathrm{FQ}=2$), with 95\% ($2\sigma$) confidence bands.
\label{fig:conf_hd26912}}
\end{figure}

\begin{figure}
\centering\includegraphics[width=0.8\columnwidth]{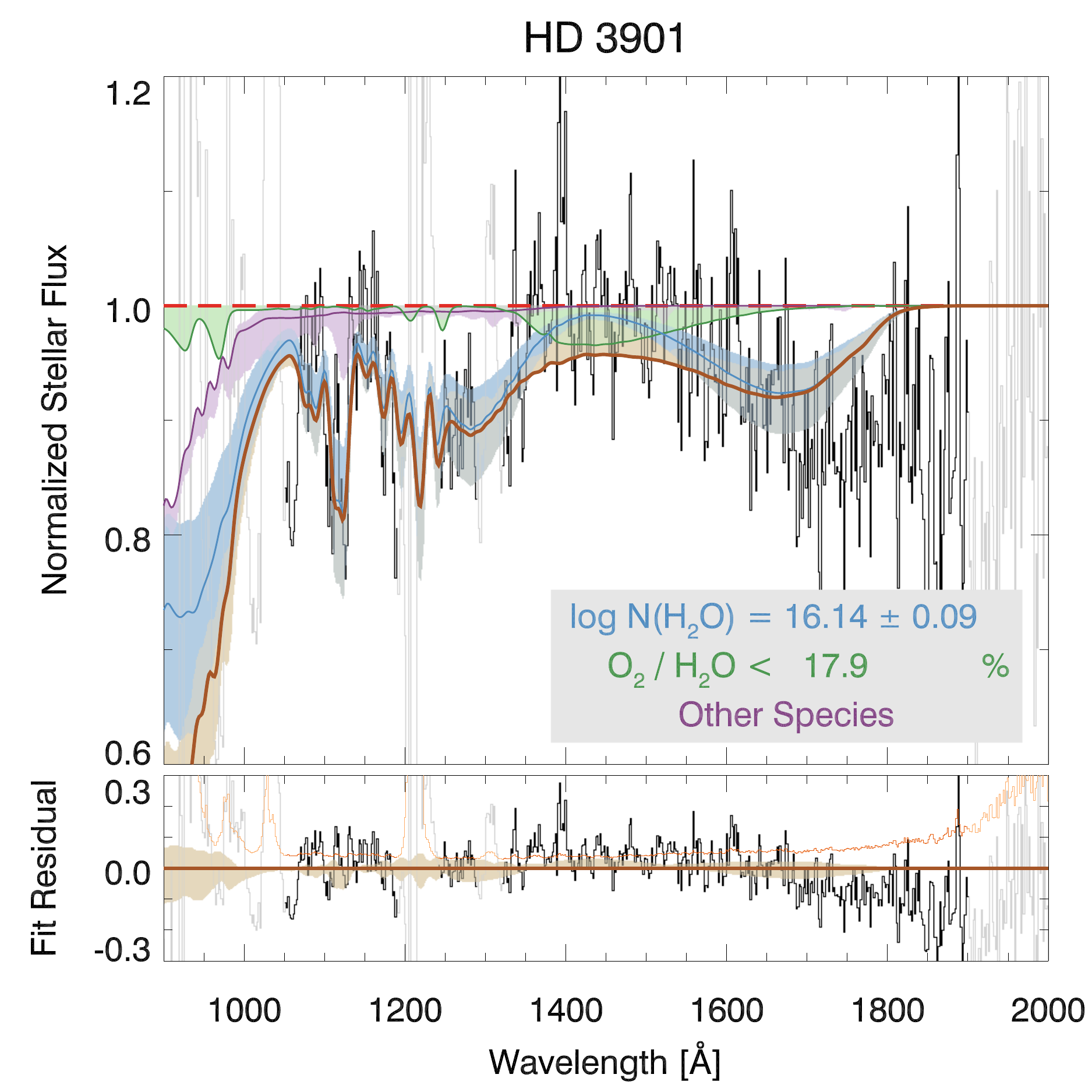}
\vspace{-1em}
\caption{Adopted column densities for the appulse of HD~3901 ($\mathrm{FQ}=4$), with 95\% ($2\sigma$) confidence bands.
\label{fig:conf_hd3901}}
\end{figure}

\begin{figure}
\centering\includegraphics[width=0.8\columnwidth]{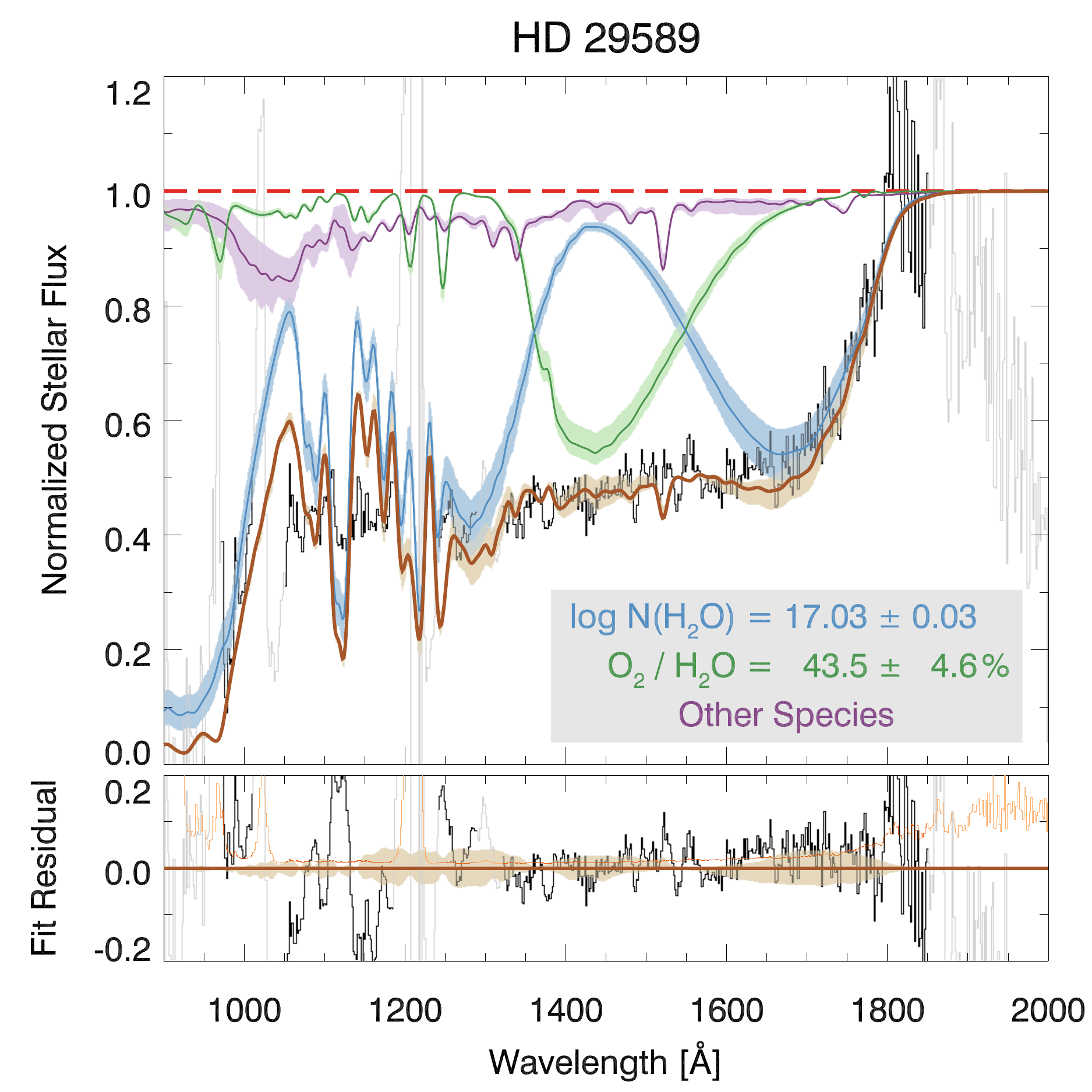}
\vspace{-1em}
\caption{Adopted column densities for the appulse of HD~29589 ($\mathrm{FQ}=2$), with 95\% ($2\sigma$) confidence bands.
\label{fig:conf_hd29589}}
\end{figure}

\begin{figure}
\centering\includegraphics[width=0.8\columnwidth]{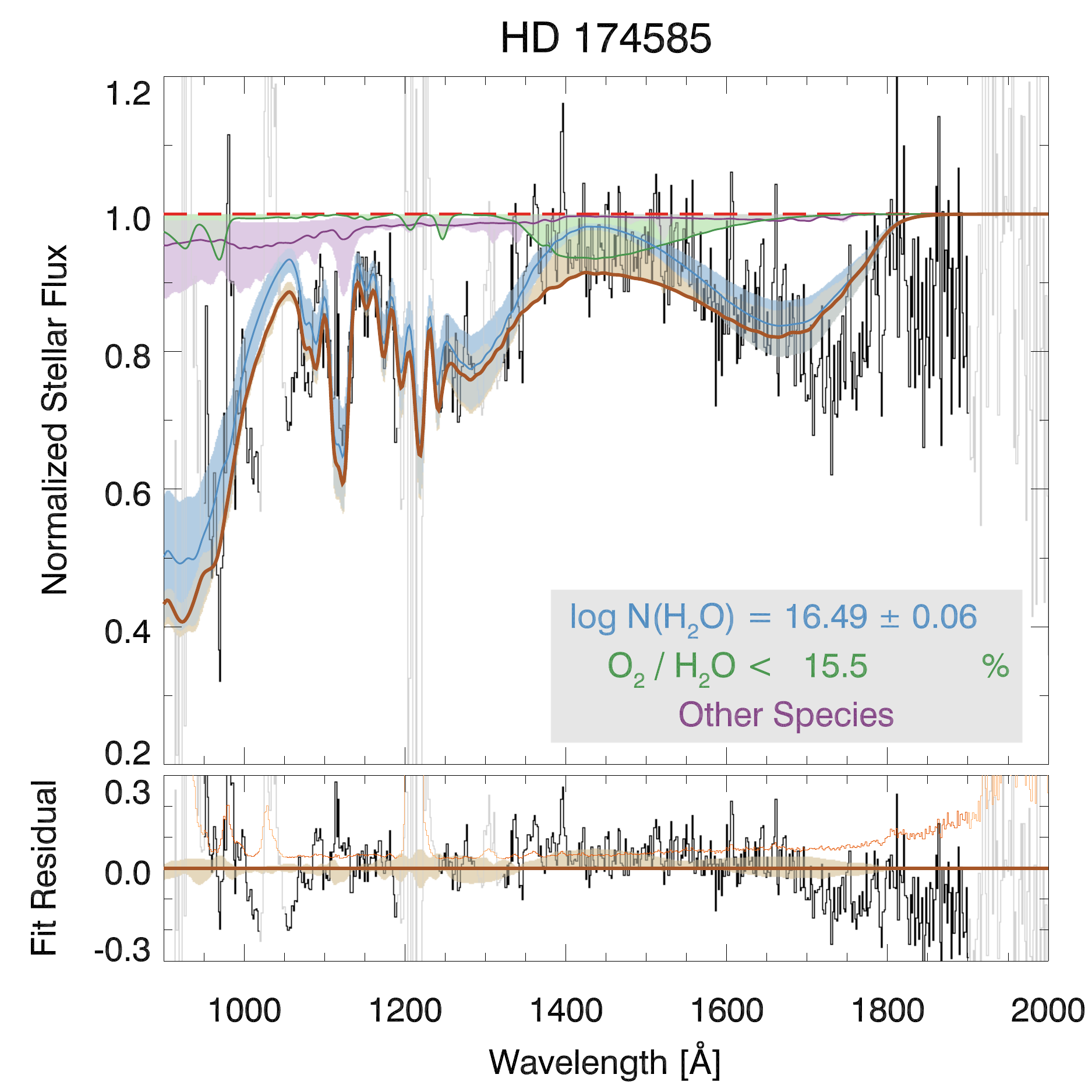}
\vspace{-1em}
\caption{Adopted column densities for the appulse of HD~174585 ($\mathrm{FQ}=3$), with 95\% ($2\sigma$) confidence bands.
\label{fig:conf_hd174585}}
\end{figure}

\begin{figure}
\centering\includegraphics[width=0.8\columnwidth]{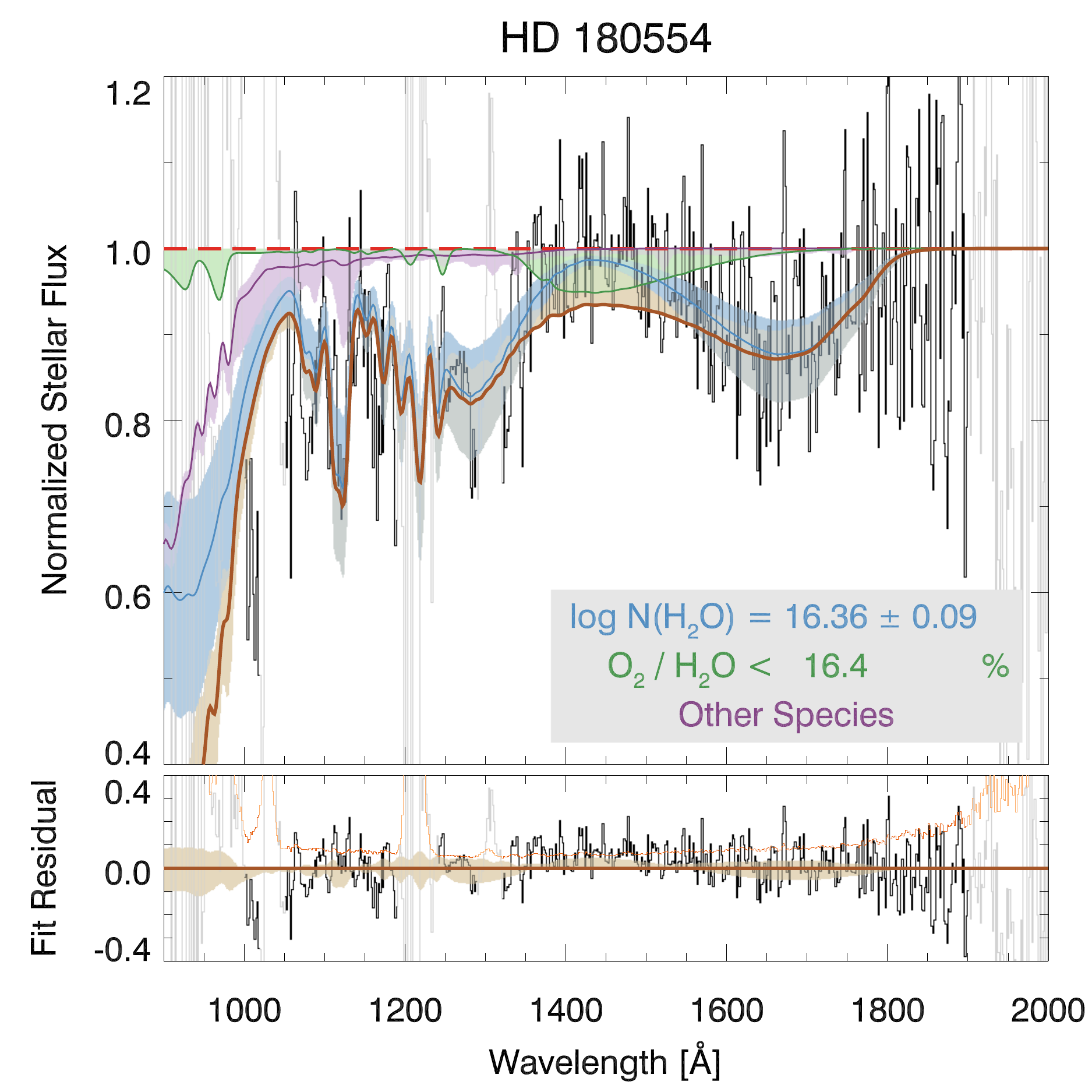}
\vspace{-1em}
\caption{Adopted column densities for the appulse of HD~180554 ($\mathrm{FQ}=4$), with 95\% ($2\sigma$) confidence bands.
\label{fig:conf_hd180554}}
\end{figure}

\begin{figure}
\centering\includegraphics[width=0.8\columnwidth]{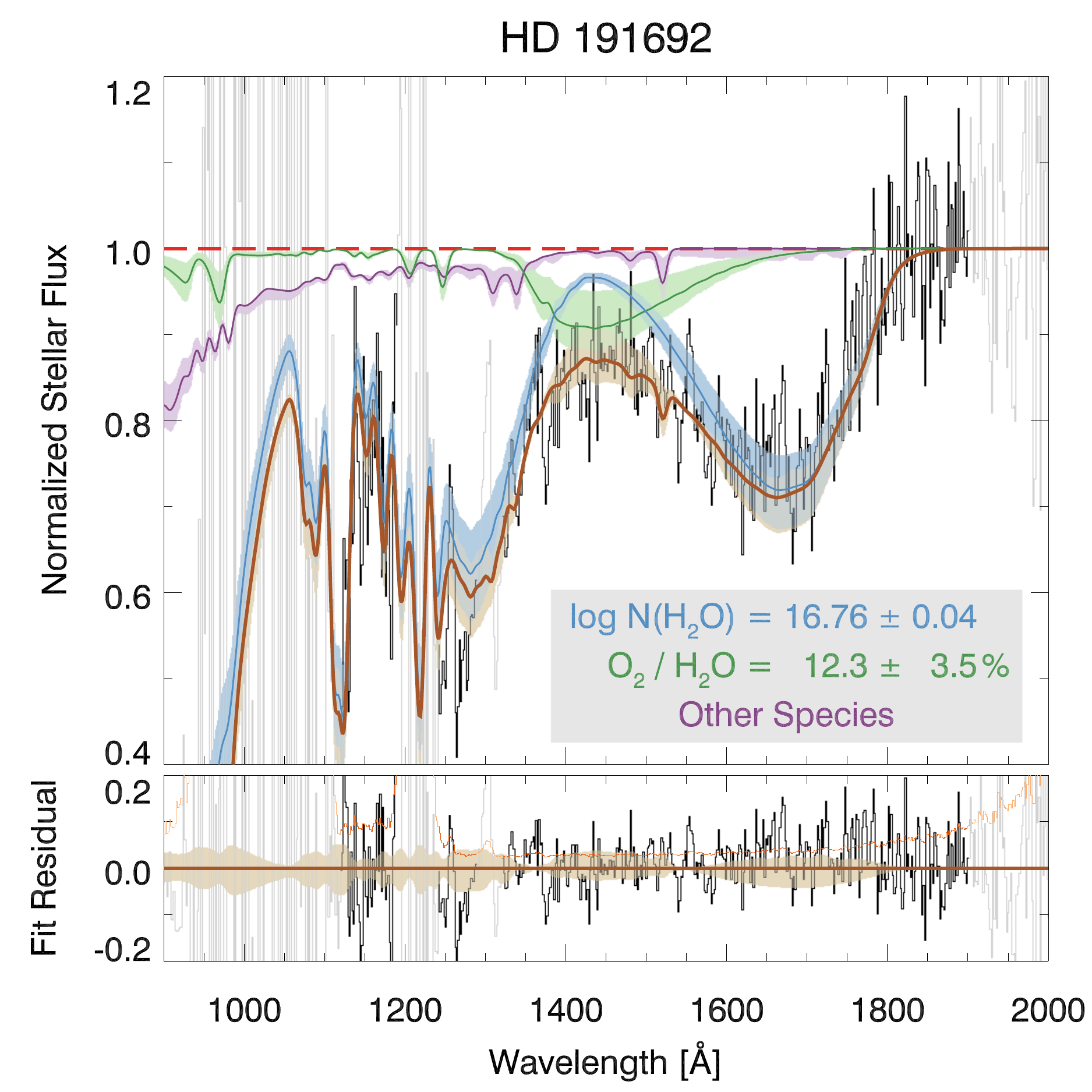}
\vspace{-1em}
\caption{Adopted column densities for the appulse of HD~191692 ($\mathrm{FQ}=2$), with 95\% ($2\sigma$) confidence bands.
\label{fig:conf_hd191692}}
\end{figure}

\begin{figure}
\centering\includegraphics[width=0.8\columnwidth]{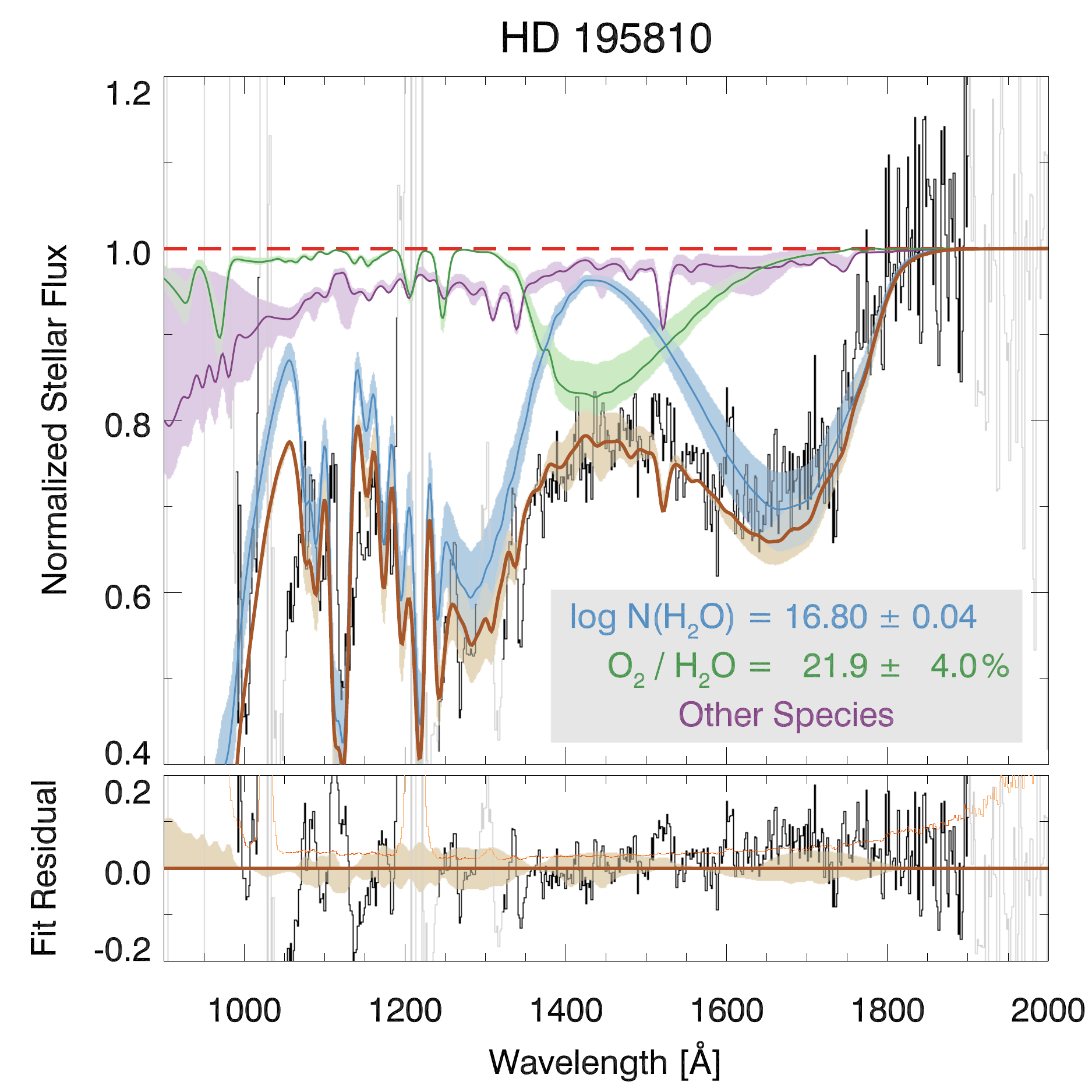}
\vspace{-1em}
\caption{Adopted column densities for the appulse of HD~195810 ($\mathrm{FQ}=2$), with 95\% ($2\sigma$) confidence bands.
\label{fig:conf_hd195810}}
\end{figure}

\begin{figure}
\centering\includegraphics[width=0.8\columnwidth]{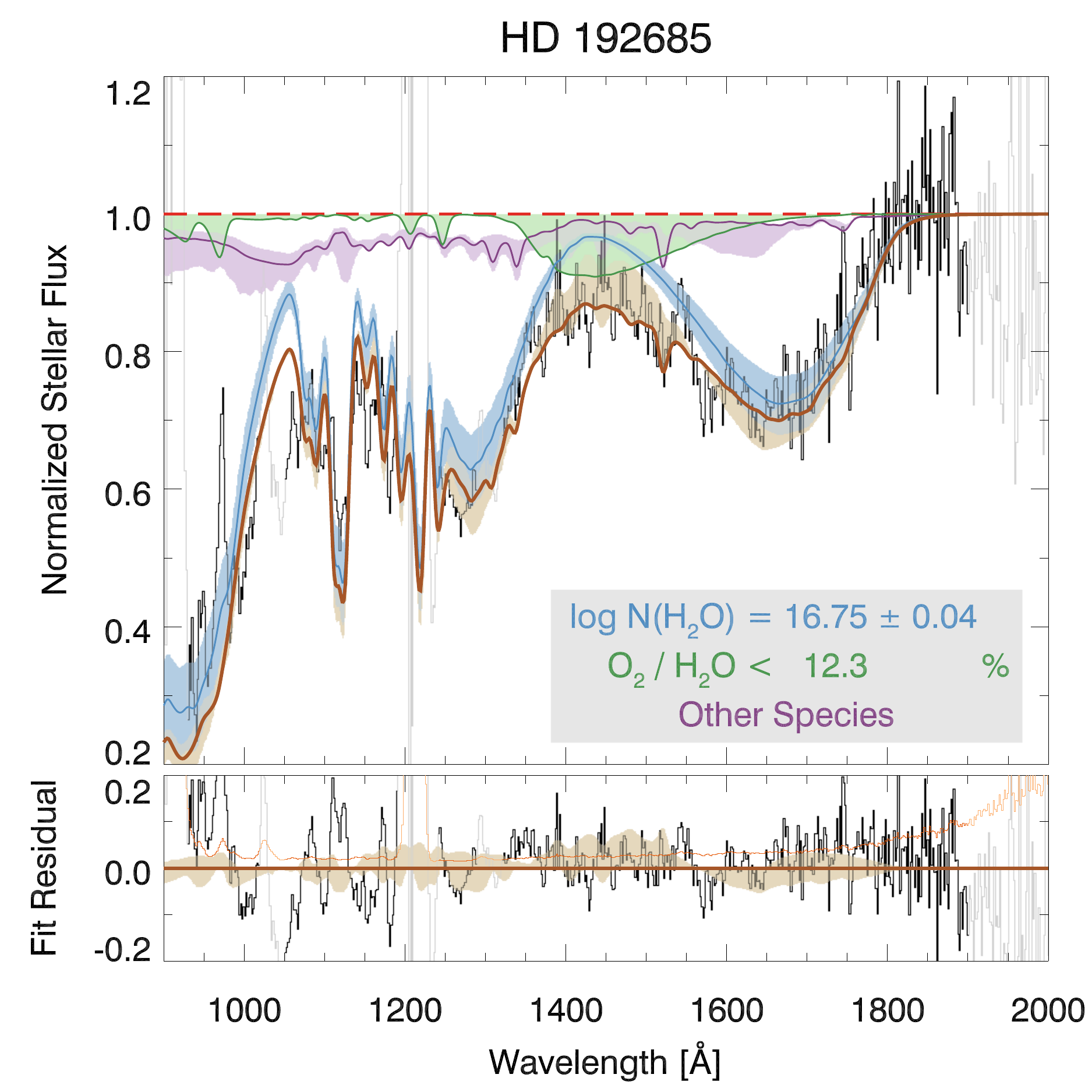}
\vspace{-1em}
\caption{Adopted column densities for the appulse of HD~192685 ($\mathrm{FQ}=1$), with 95\% ($2\sigma$) confidence bands.
\label{fig:conf_hd192685}}
\end{figure}

\begin{figure}
\centering\includegraphics[width=0.8\columnwidth]{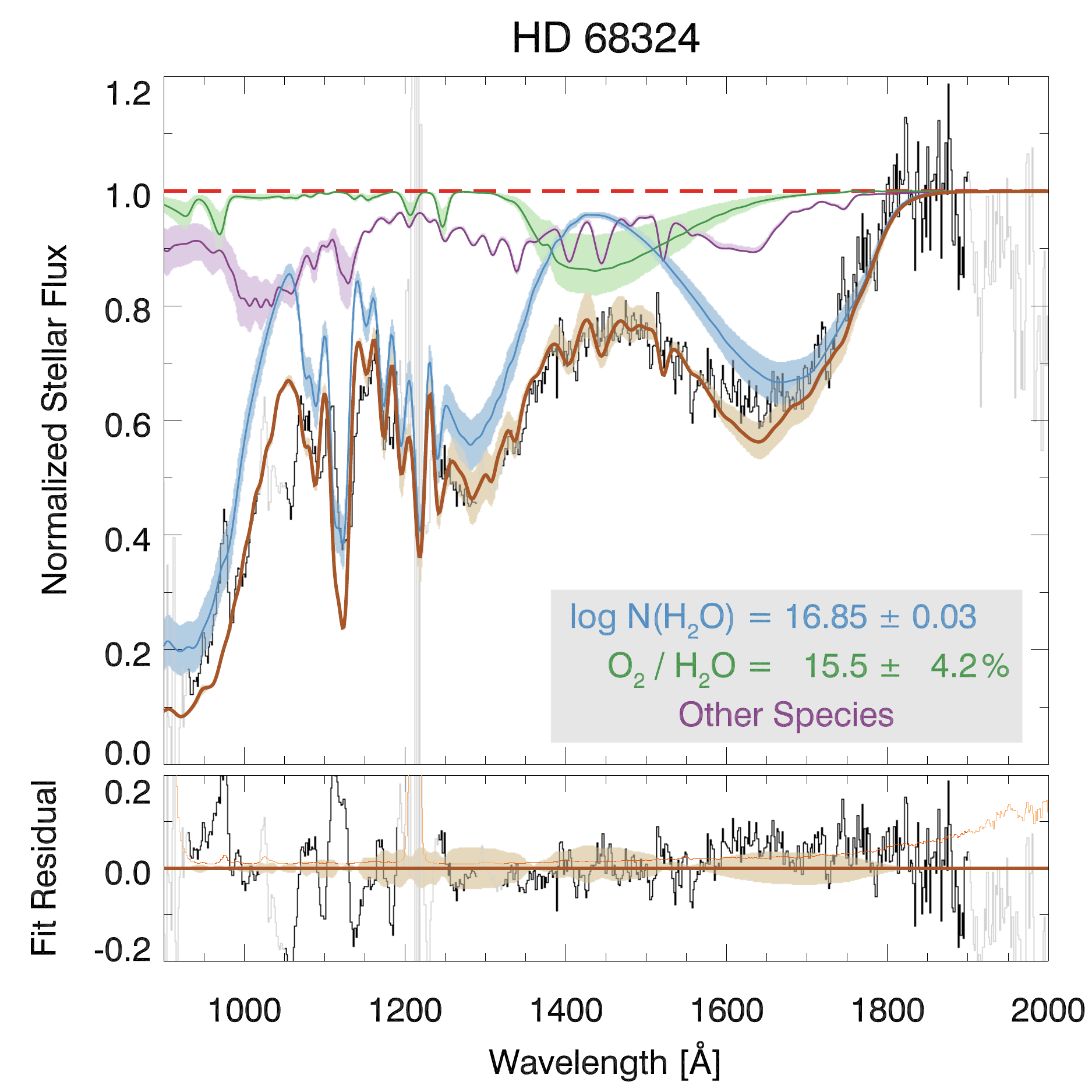}
\vspace{-1em}
\caption{Adopted column densities for the appulse of HD~68324 ($\mathrm{FQ}=2$), with 95\% ($2\sigma$) confidence bands.
\label{fig:conf_hd68324}}
\end{figure}

\begin{figure}
\centering\includegraphics[width=0.8\columnwidth]{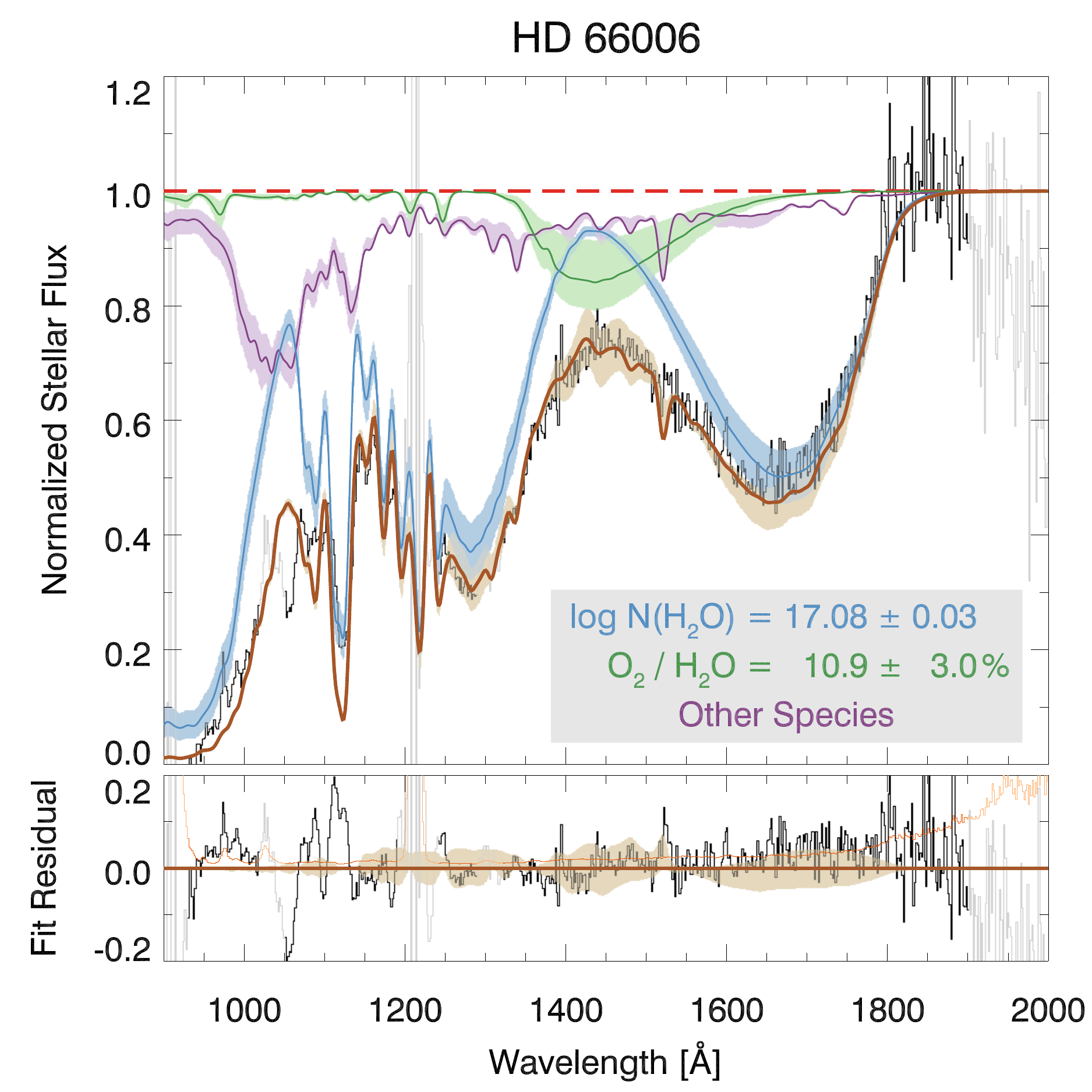}
\vspace{-1em}
\caption{Adopted column densities for the appulse of HD~66006 ($\mathrm{FQ}=1$), with 95\% ($2\sigma$) confidence bands.
\label{fig:conf_hd66006}}
\end{figure}

\begin{figure}
\centering\includegraphics[width=0.8\columnwidth]{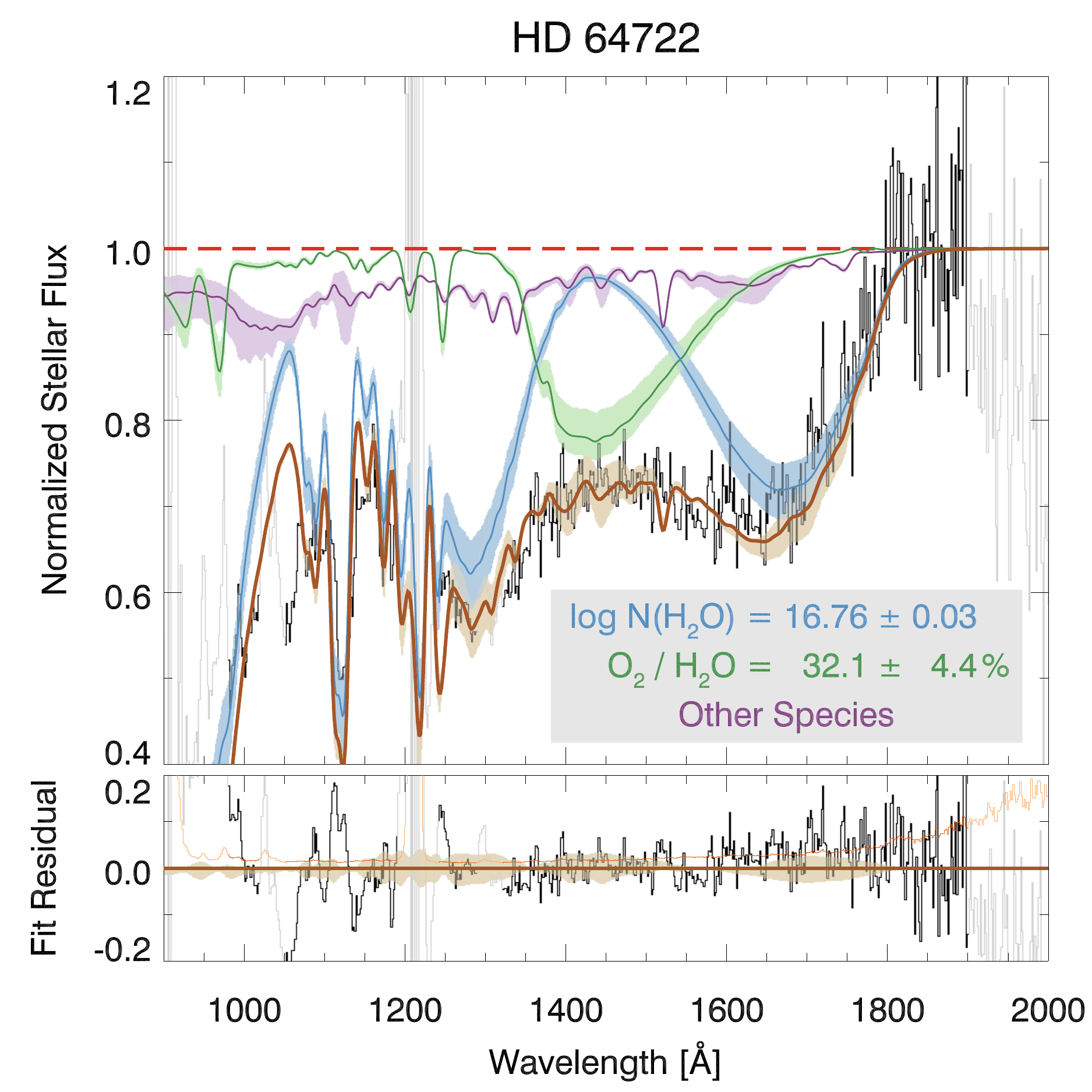}
\vspace{-1em}
\caption{Adopted column densities for the appulse of HD~64722 ($\mathrm{FQ}=2$), with 95\% ($2\sigma$) confidence bands.
\label{fig:conf_hd64722}}
\end{figure}

\begin{figure}
\centering\includegraphics[width=0.8\columnwidth]{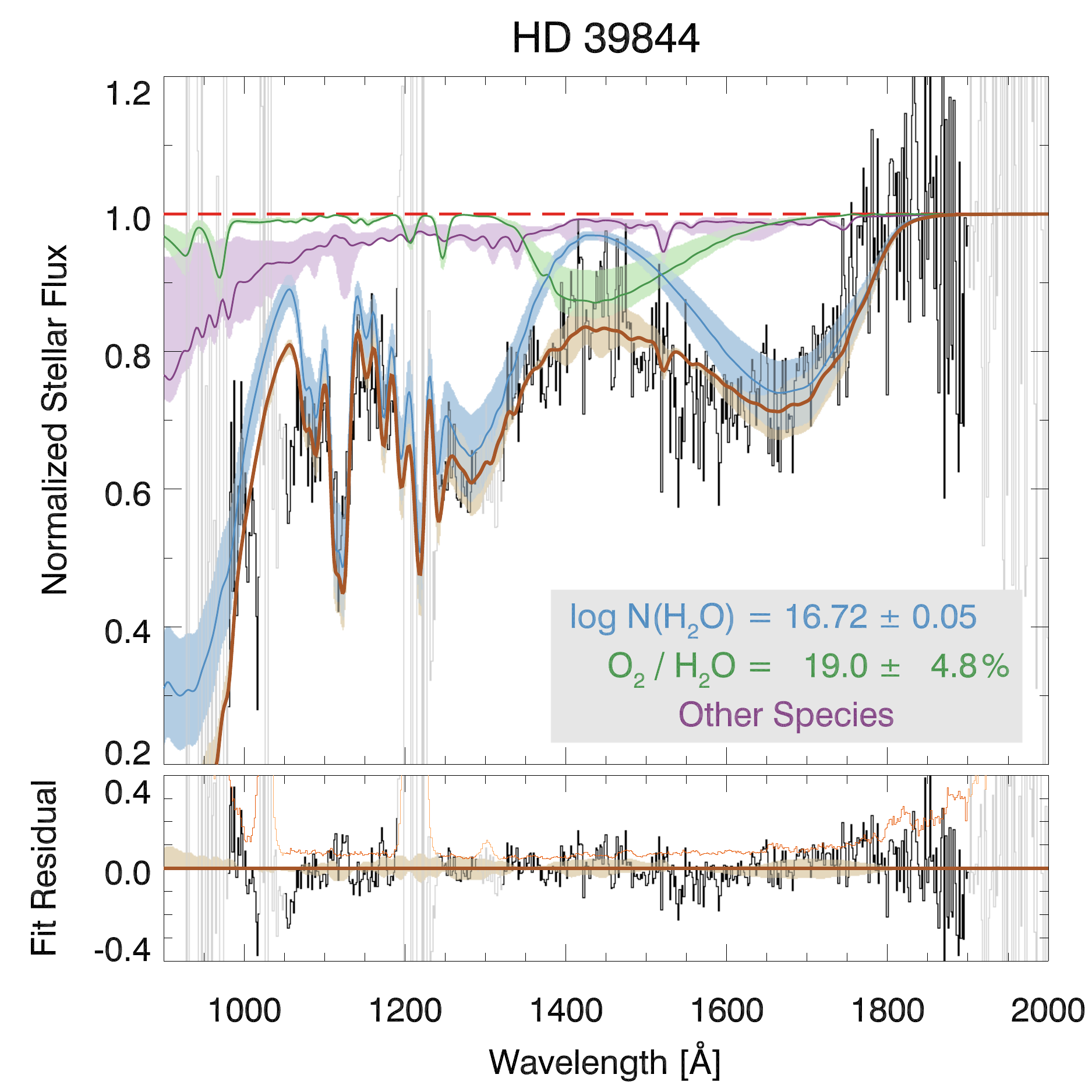}
\vspace{-1em}
\caption{Adopted column densities for the appulse of HD~39844 ($\mathrm{FQ}=2$), with 95\% ($2\sigma$) confidence bands.
\label{fig:conf_hd39844}}
\end{figure}

\begin{figure}
\centering\includegraphics[width=0.8\columnwidth]{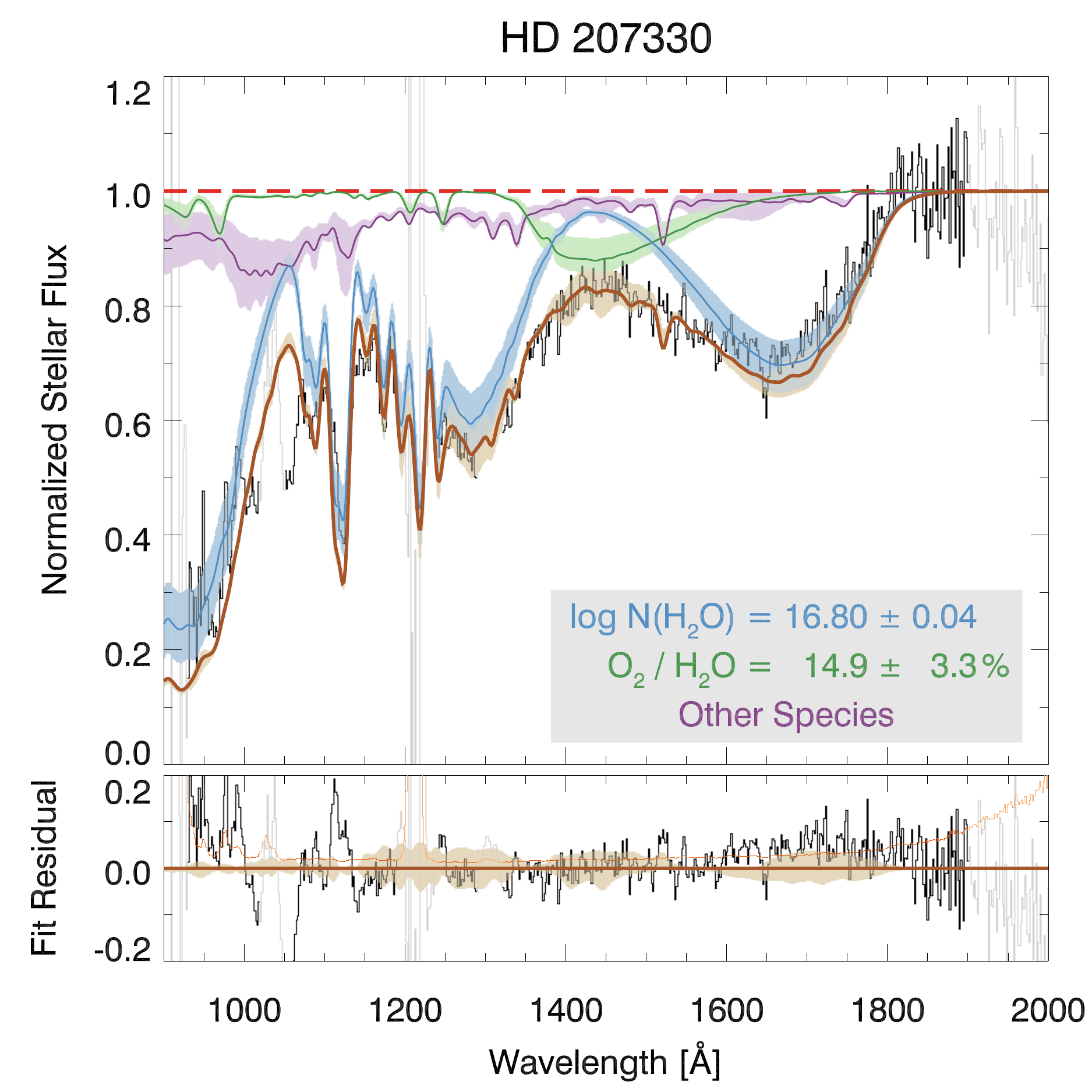}
\vspace{-1em}
\caption{Adopted column densities for the appulse of HD~207330 ($\mathrm{FQ}=2$), with 95\% ($2\sigma$) confidence bands.
\label{fig:conf_hd207330}}
\end{figure}

\begin{figure}
\centering\includegraphics[width=0.8\columnwidth]{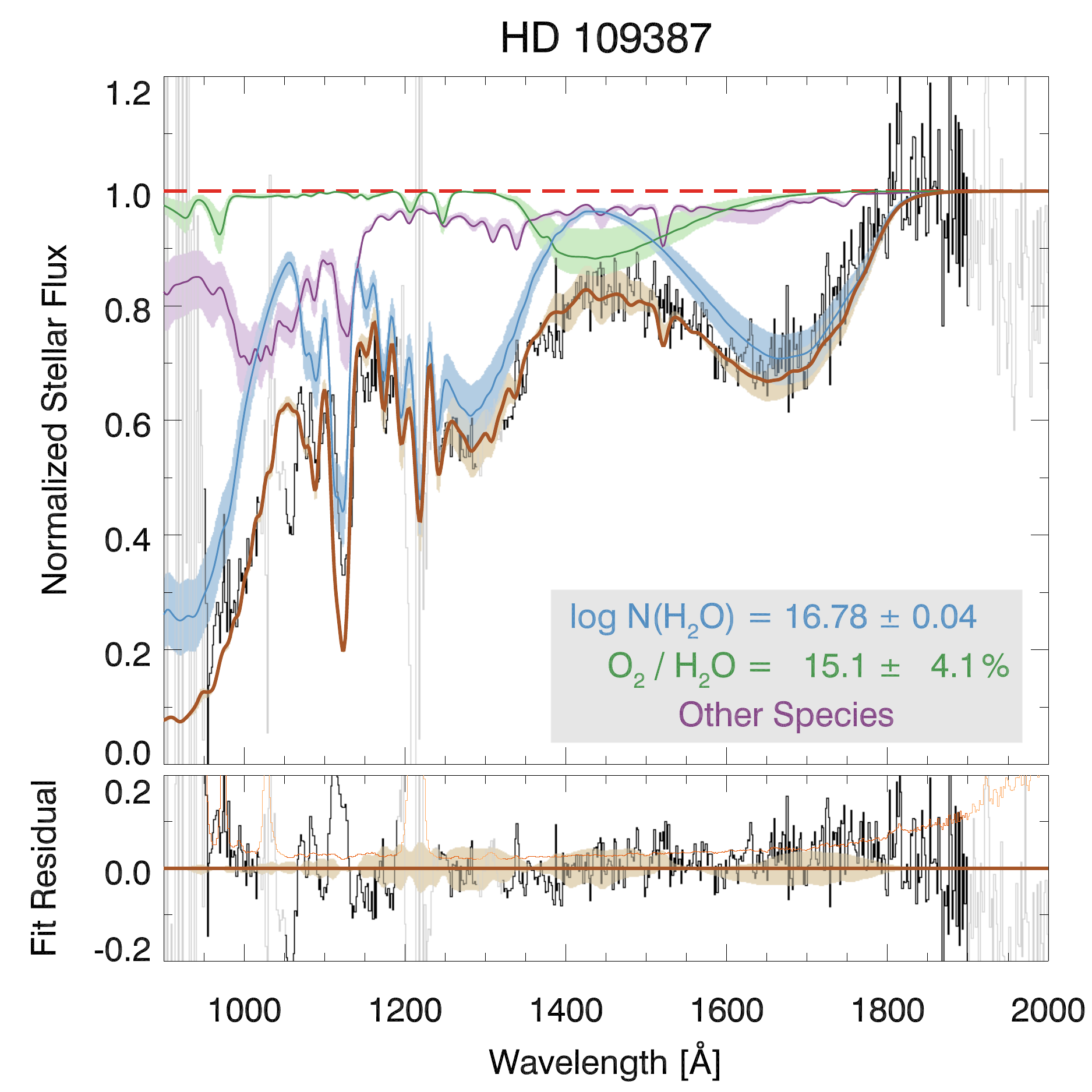}
\vspace{-1em}
\caption{Adopted column densities for the appulse of HD~109387 ($\mathrm{FQ}=1$), with 95\% ($2\sigma$) confidence bands.
\label{fig:conf_hd109387}}
\end{figure}

\begin{figure}
\centering\includegraphics[width=0.8\columnwidth]{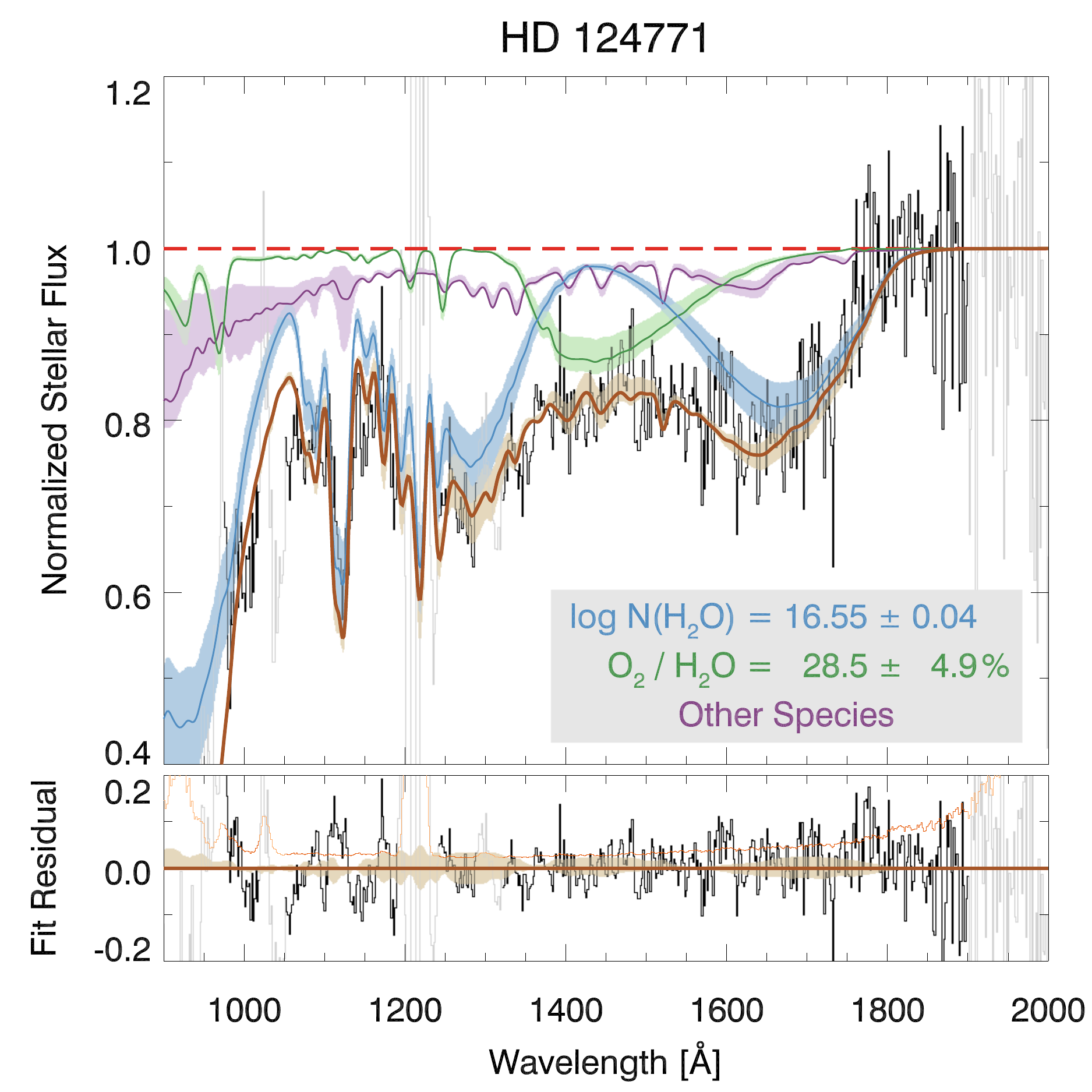}
\vspace{-1em}
\caption{Adopted column densities for the appulse of HD~124771 ($\mathrm{FQ}=1$), with 95\% ($2\sigma$) confidence bands.
\label{fig:conf_hd124771}}
\end{figure}

\begin{figure}
\centering\includegraphics[width=0.8\columnwidth]{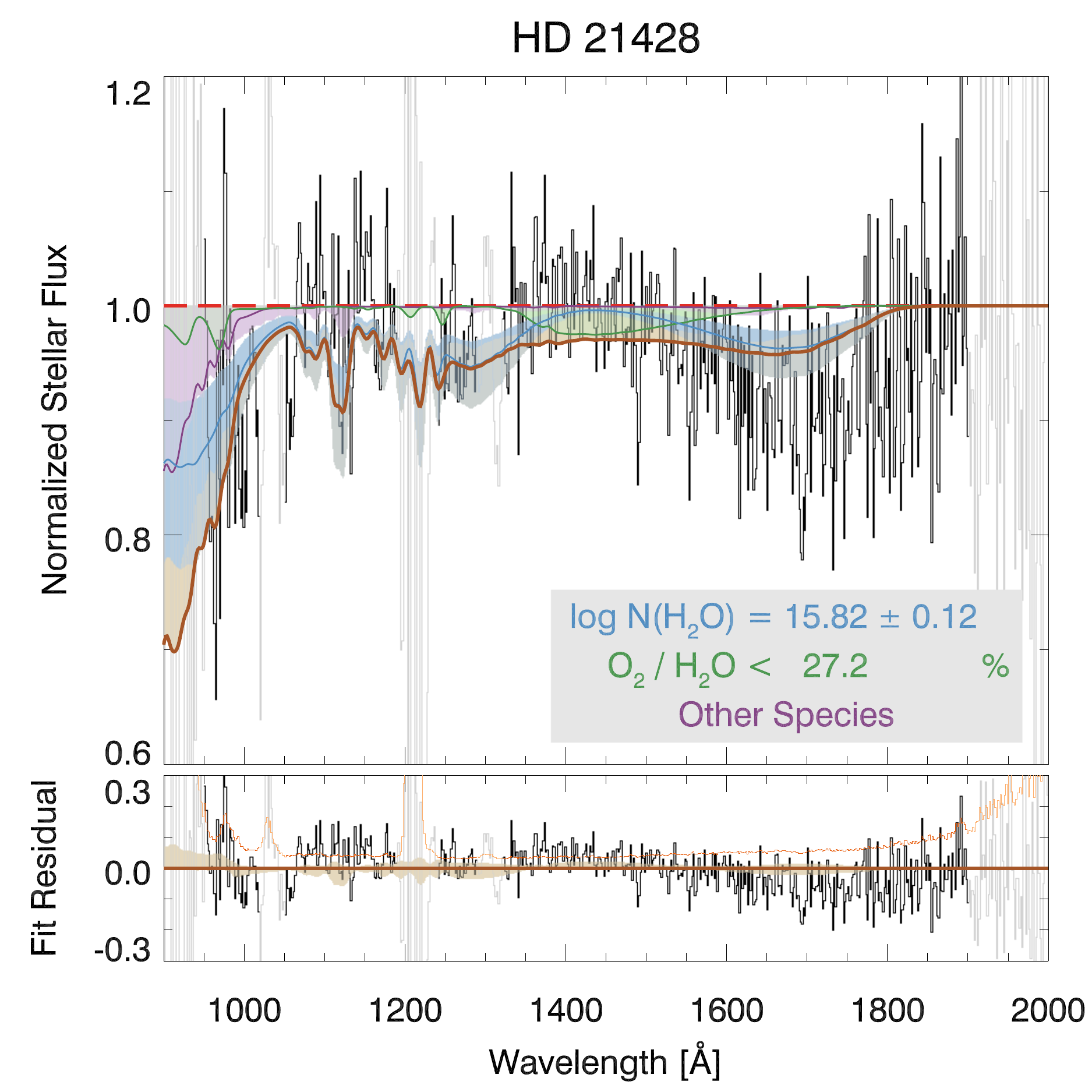}
\vspace{-1em}
\caption{Adopted column densities for the appulse of HD~21428 ($\mathrm{FQ}=4$), with 95\% ($2\sigma$) confidence bands.
\label{fig:conf_hd21428}}
\end{figure}

\begin{figure}
\centering\includegraphics[width=0.8\columnwidth]{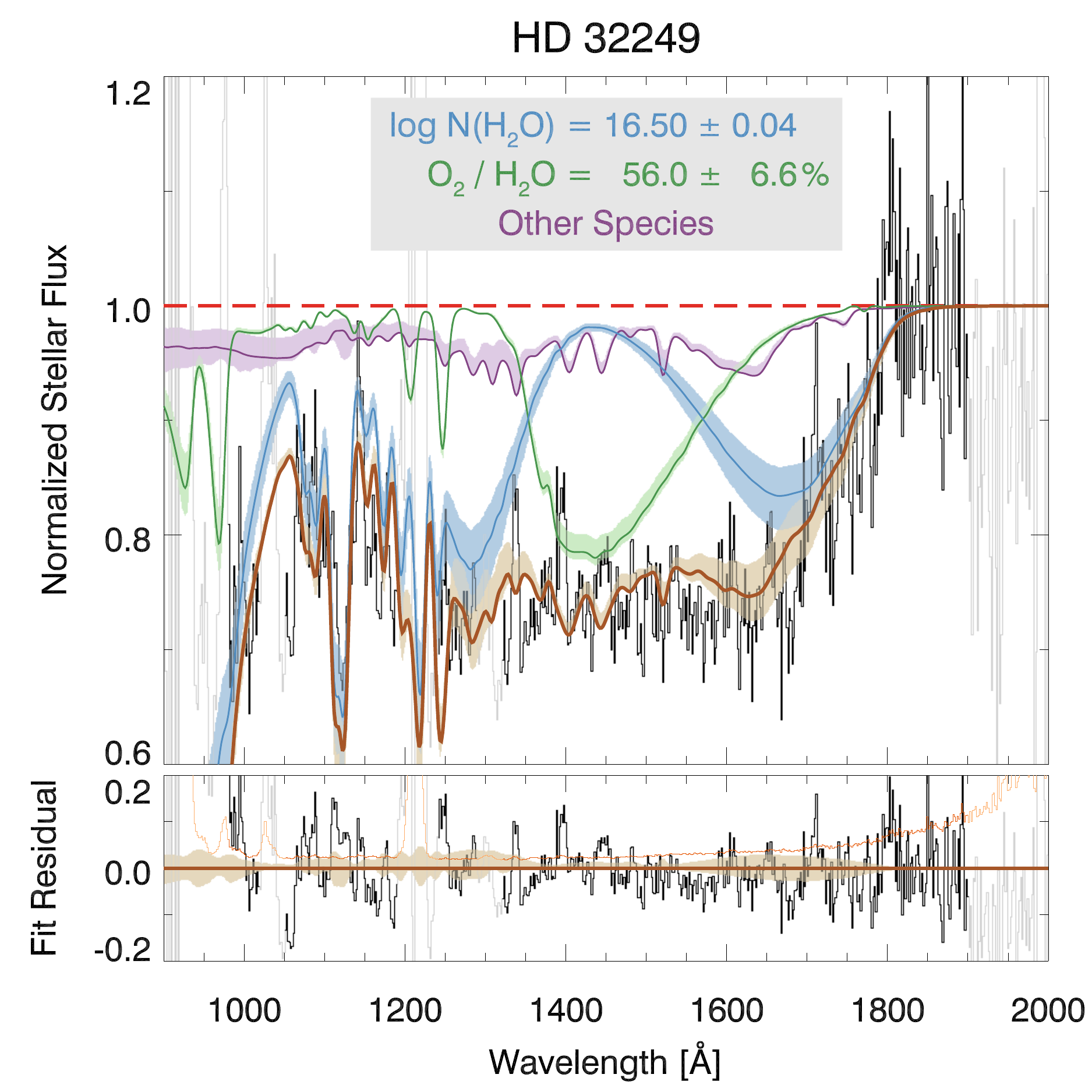}
\vspace{-1em}
\caption{Adopted column densities for the appulse of HD~32249 ($\mathrm{FQ}=2$), with 95\% ($2\sigma$) confidence bands.
\label{fig:conf_hd32249}}
\end{figure}

\begin{figure}
\centering\includegraphics[width=0.8\columnwidth]{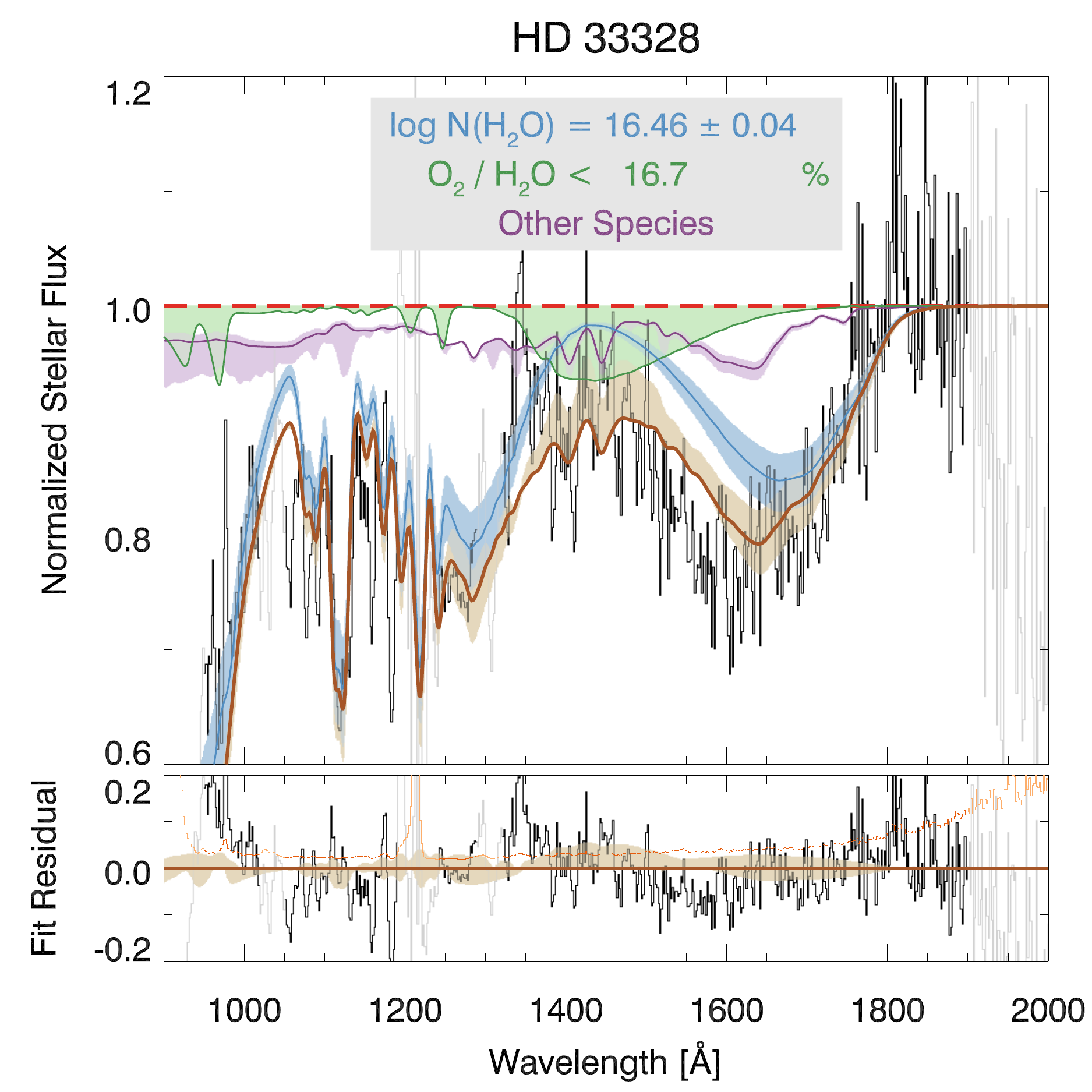}
\vspace{-1em}
\caption{Adopted column densities for the appulse of HD~33328 ($\mathrm{FQ}=2$), with 95\% ($2\sigma$) confidence bands.
\label{fig:conf_hd33328}}
\end{figure}

\begin{figure}
\centering\includegraphics[width=0.8\columnwidth]{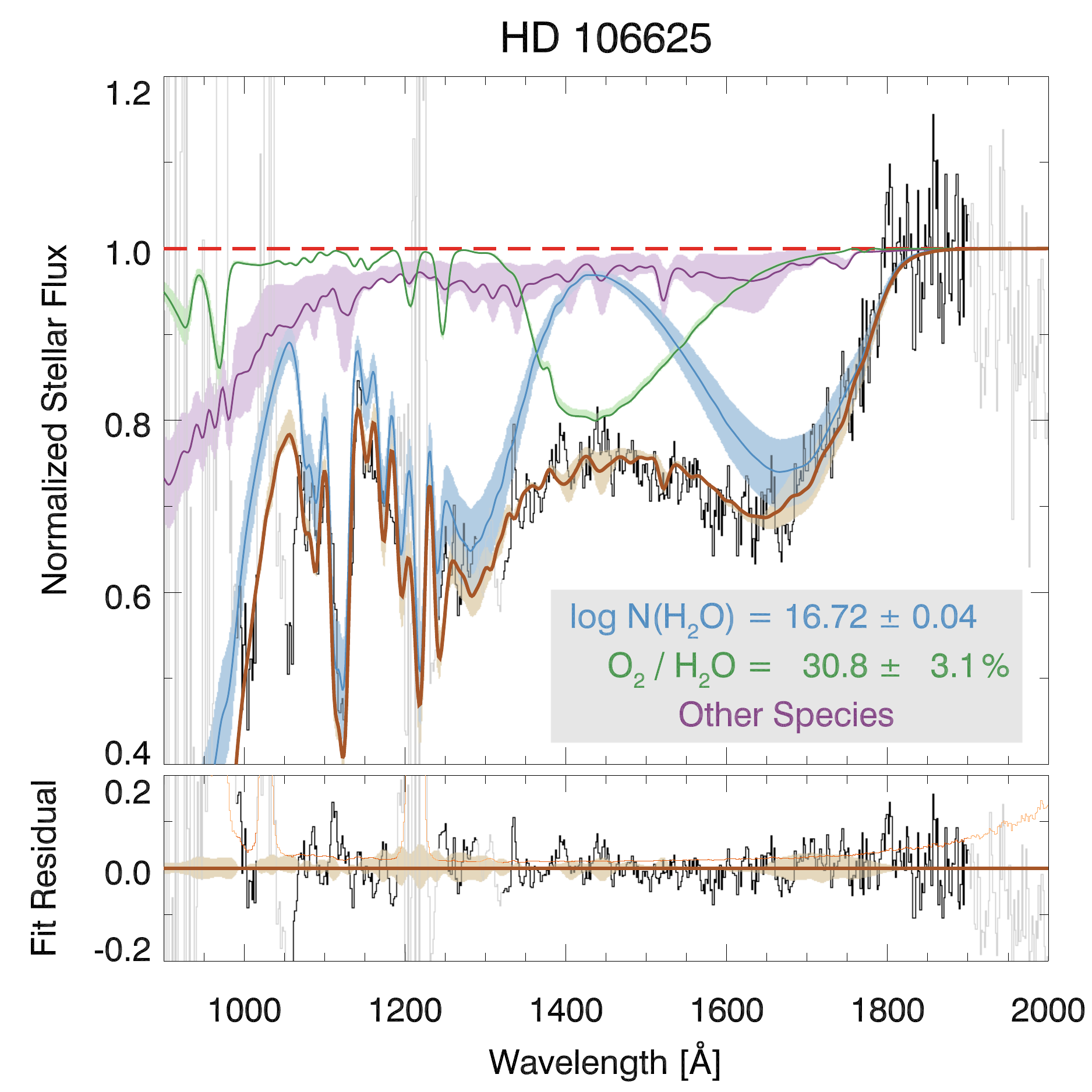}
\vspace{-1em}
\caption{Adopted column densities for the appulse of HD~106625 ($\mathrm{FQ}=1$), with 95\% ($2\sigma$) confidence bands.
\label{fig:conf_hd106625}}
\end{figure}

\begin{figure}
\centering\includegraphics[width=0.8\columnwidth]{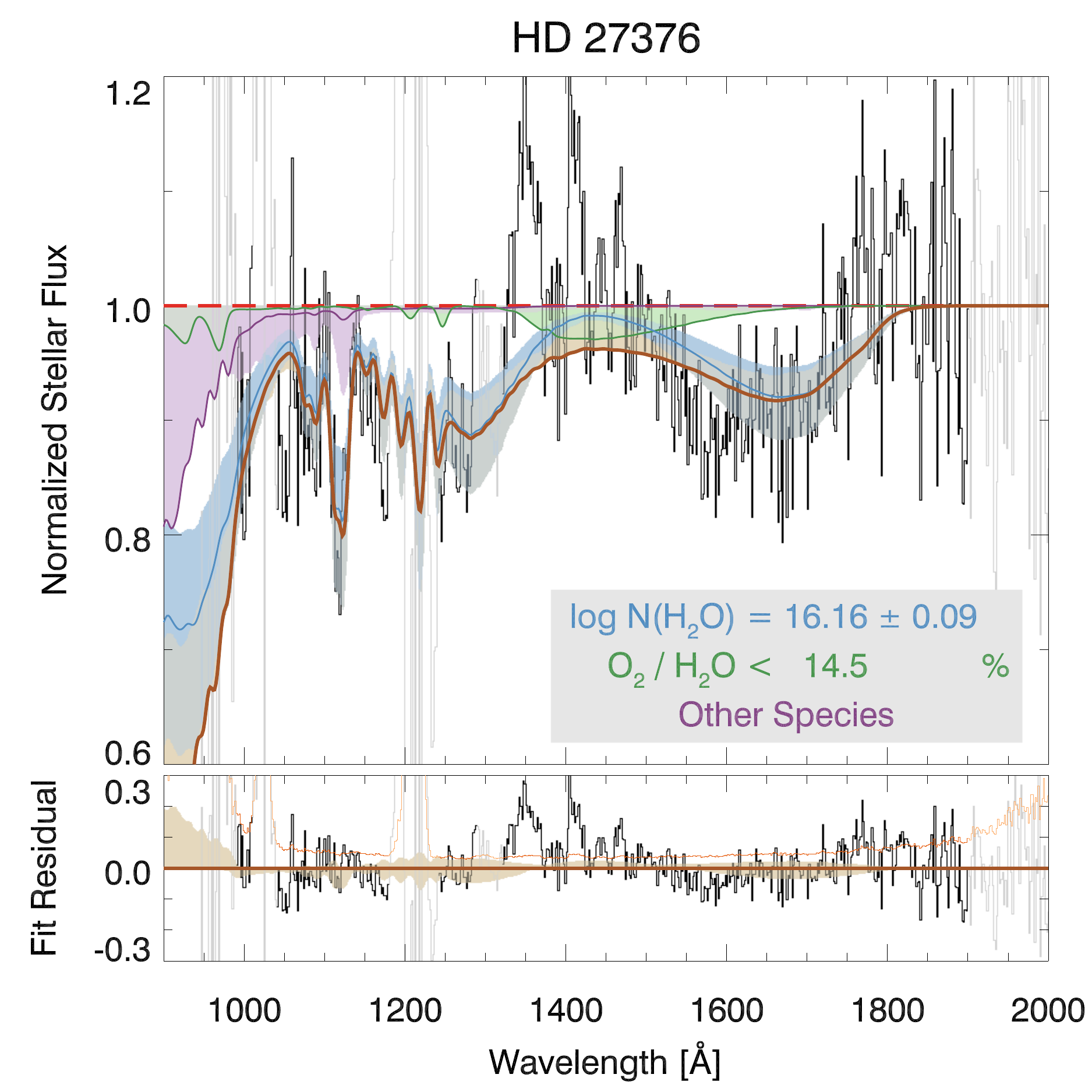}
\vspace{-1em}
\caption{Adopted column densities for the appulse of HD~27376 ($\mathrm{FQ}=3$), with 95\% ($2\sigma$) confidence bands.
\label{fig:conf_hd27376}}
\end{figure}

\begin{figure}
\centering\includegraphics[width=0.8\columnwidth]{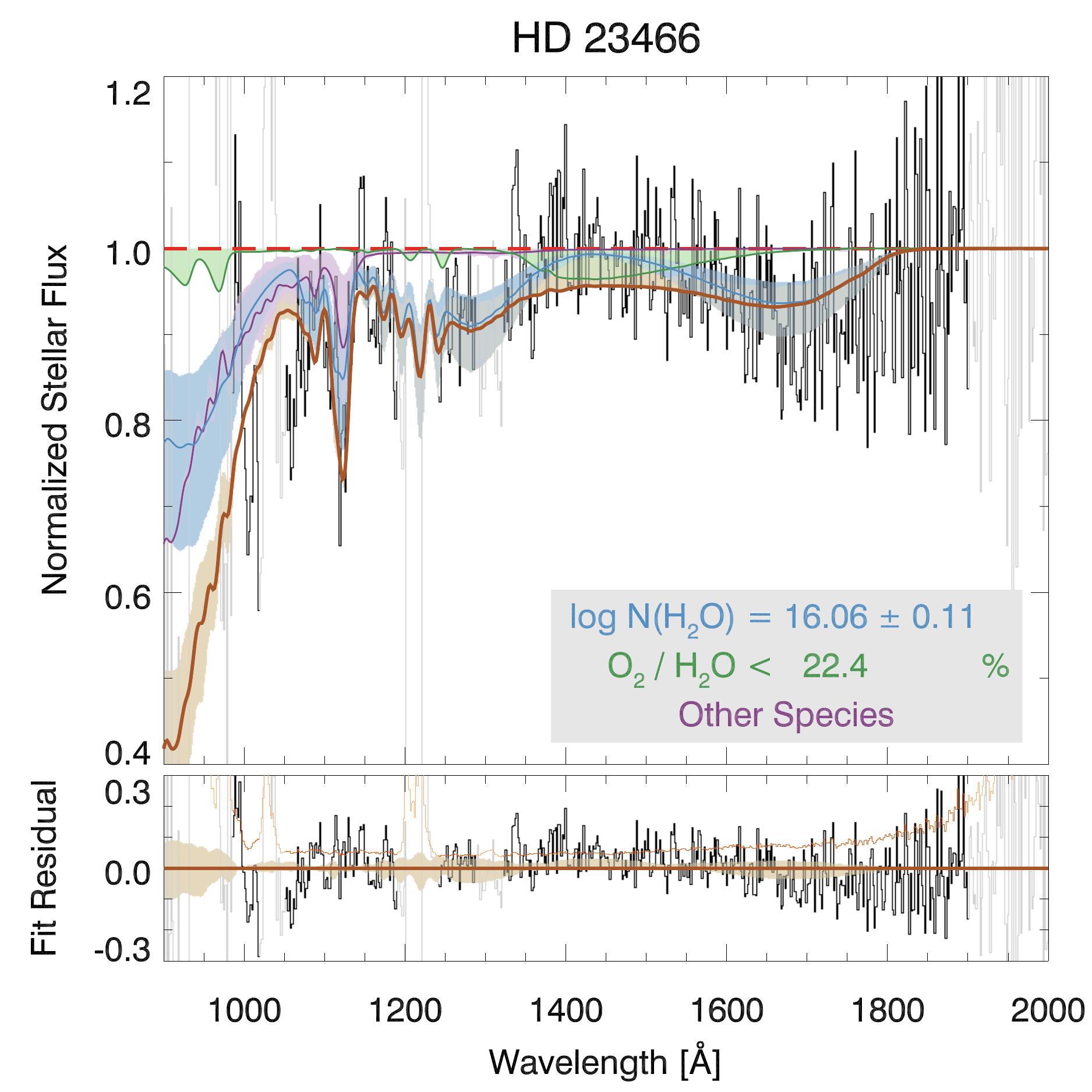}
\vspace{-1em}
\caption{Adopted column densities for the appulse of HD~23466 ($\mathrm{FQ}=4$), with 95\% ($2\sigma$) confidence bands.
\label{fig:conf_hd23466}}
\end{figure}

\begin{figure}
\centering\includegraphics[width=0.8\columnwidth]{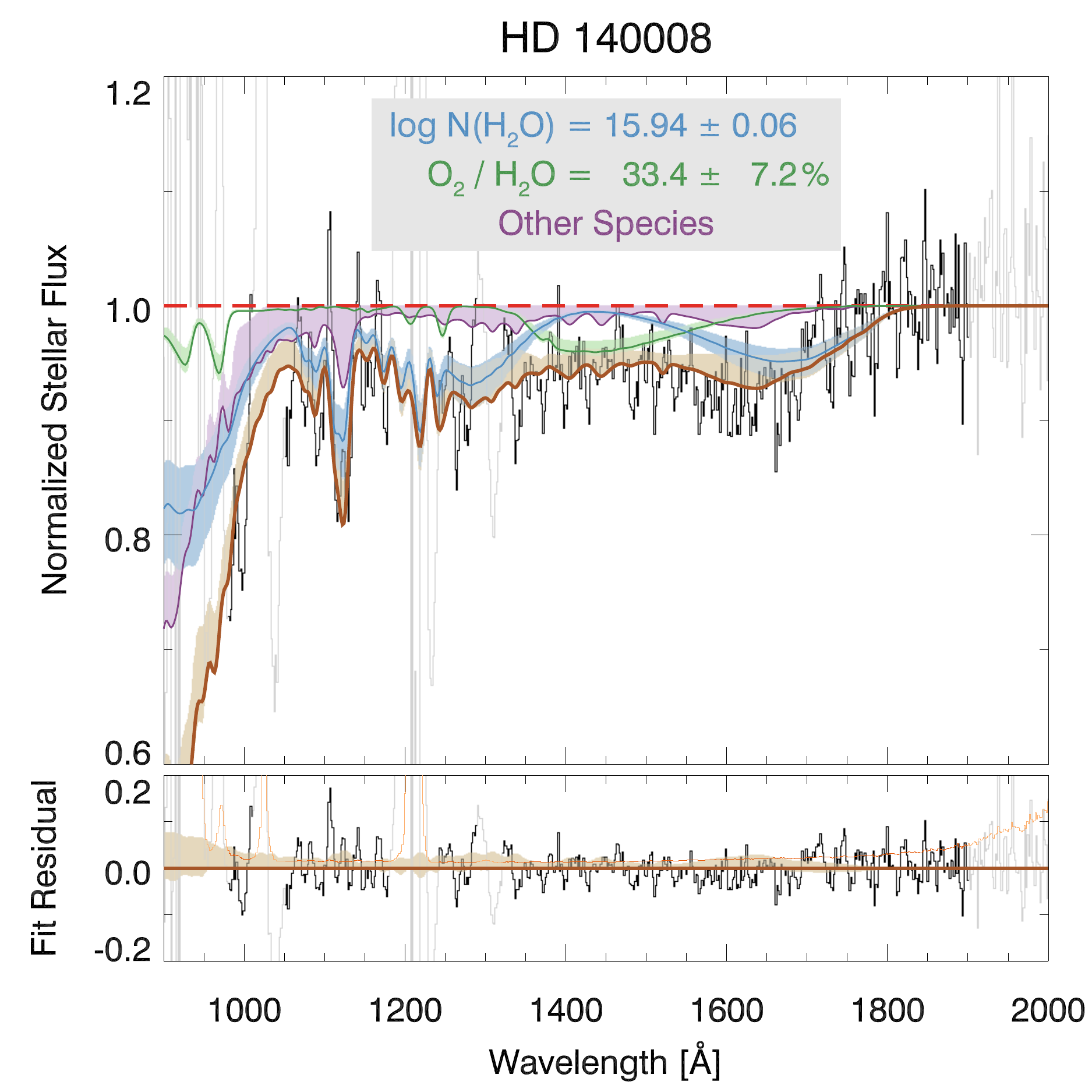}
\vspace{-1em}
\caption{Adopted column densities for the appulse of HD~140008 ($\mathrm{FQ}=2$), with 95\% ($2\sigma$) confidence bands.
\label{fig:conf_hd140008}}
\end{figure}

\begin{figure}
\centering\includegraphics[width=0.8\columnwidth]{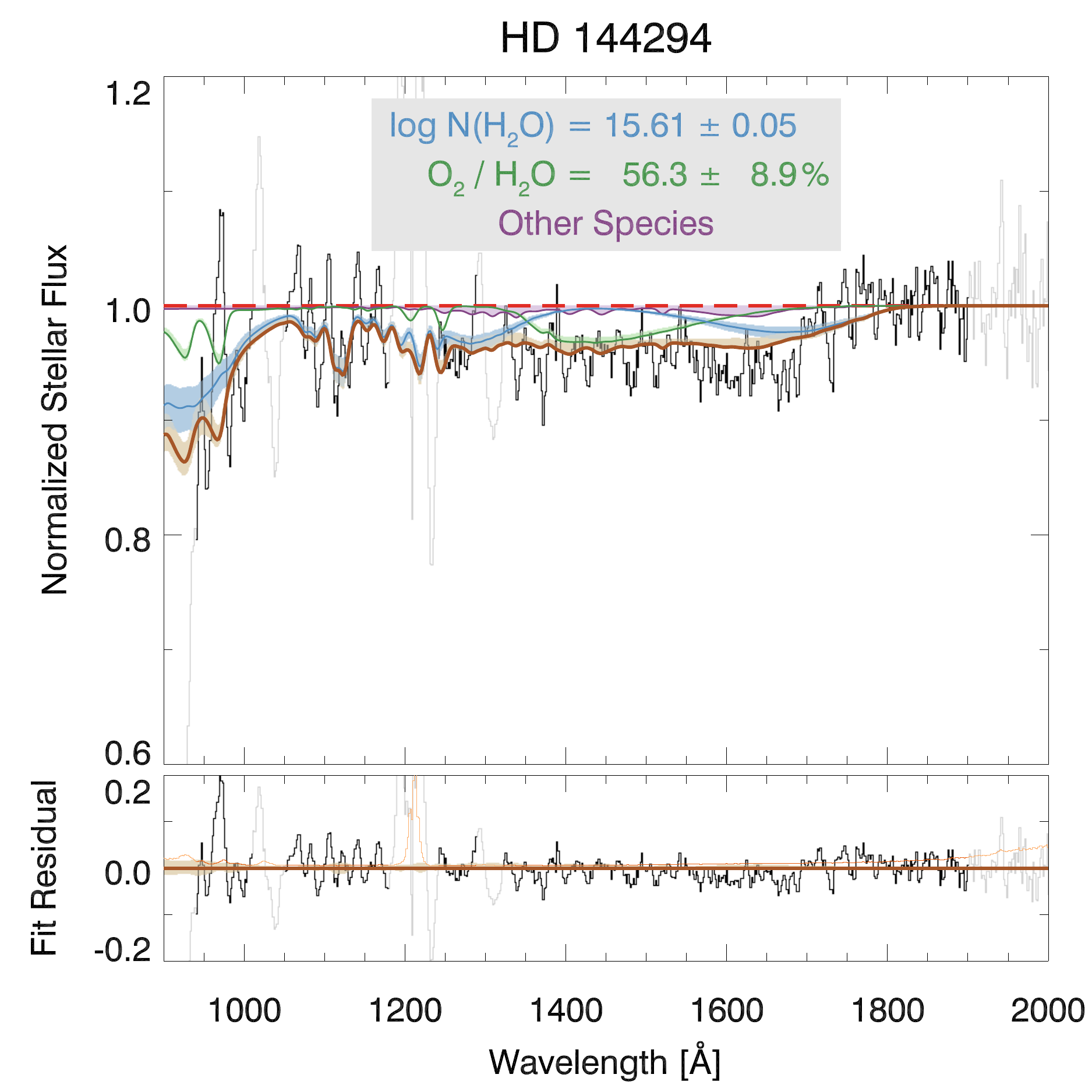}
\vspace{-1em}
\caption{Adopted column densities for the appulse of HD~144294 ($\mathrm{FQ}=3$), with 95\% ($2\sigma$) confidence bands.
\label{fig:conf_hd144294}}
\end{figure}

\begin{figure}
\centering\includegraphics[width=0.8\columnwidth]{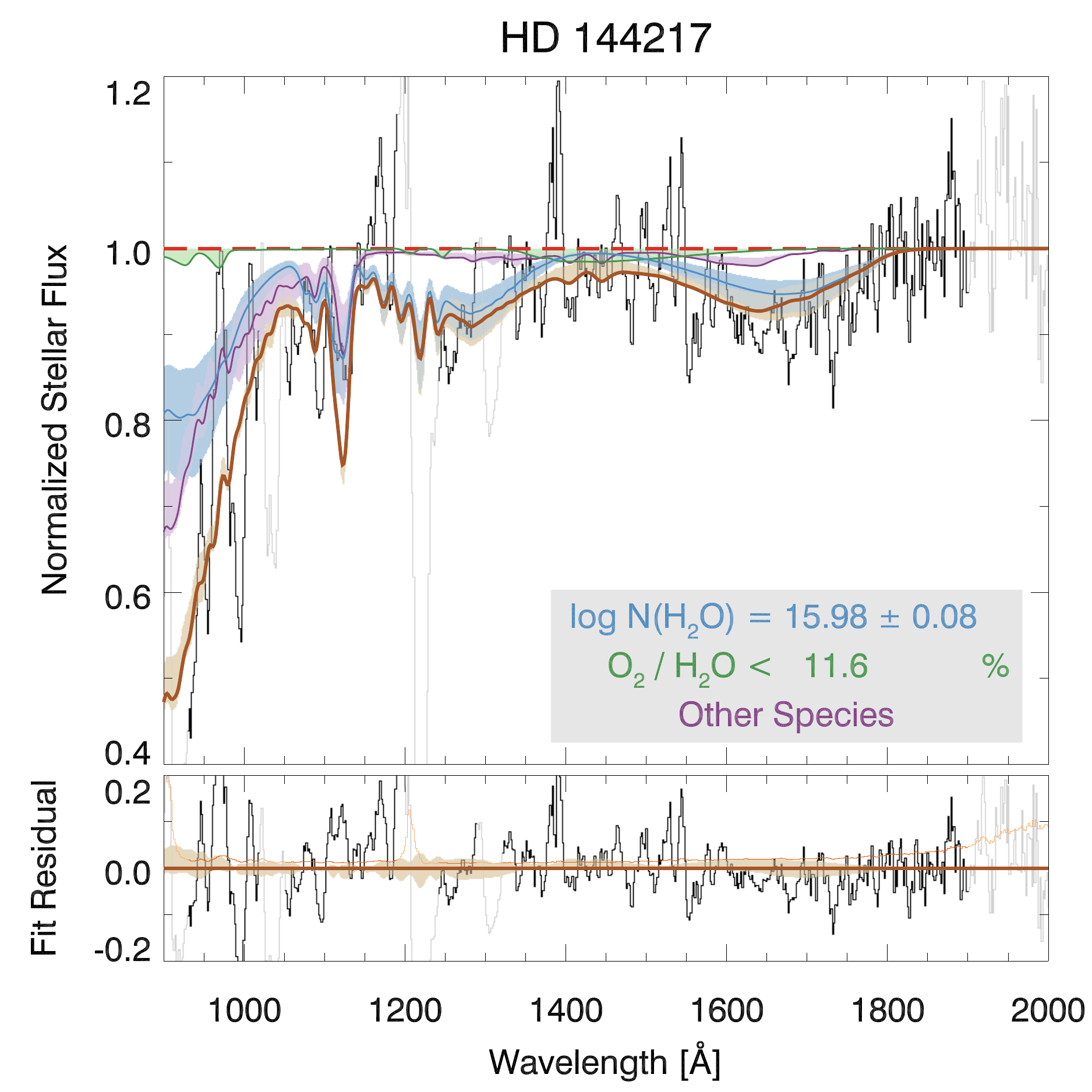}
\vspace{-1em}
\caption{Adopted column densities for the appulse of HD~144217 ($\mathrm{FQ}=3$), with 95\% ($2\sigma$) confidence bands.
\label{fig:conf_hd144217}}
\end{figure}

\begin{figure}
\centering\includegraphics[width=0.8\columnwidth]{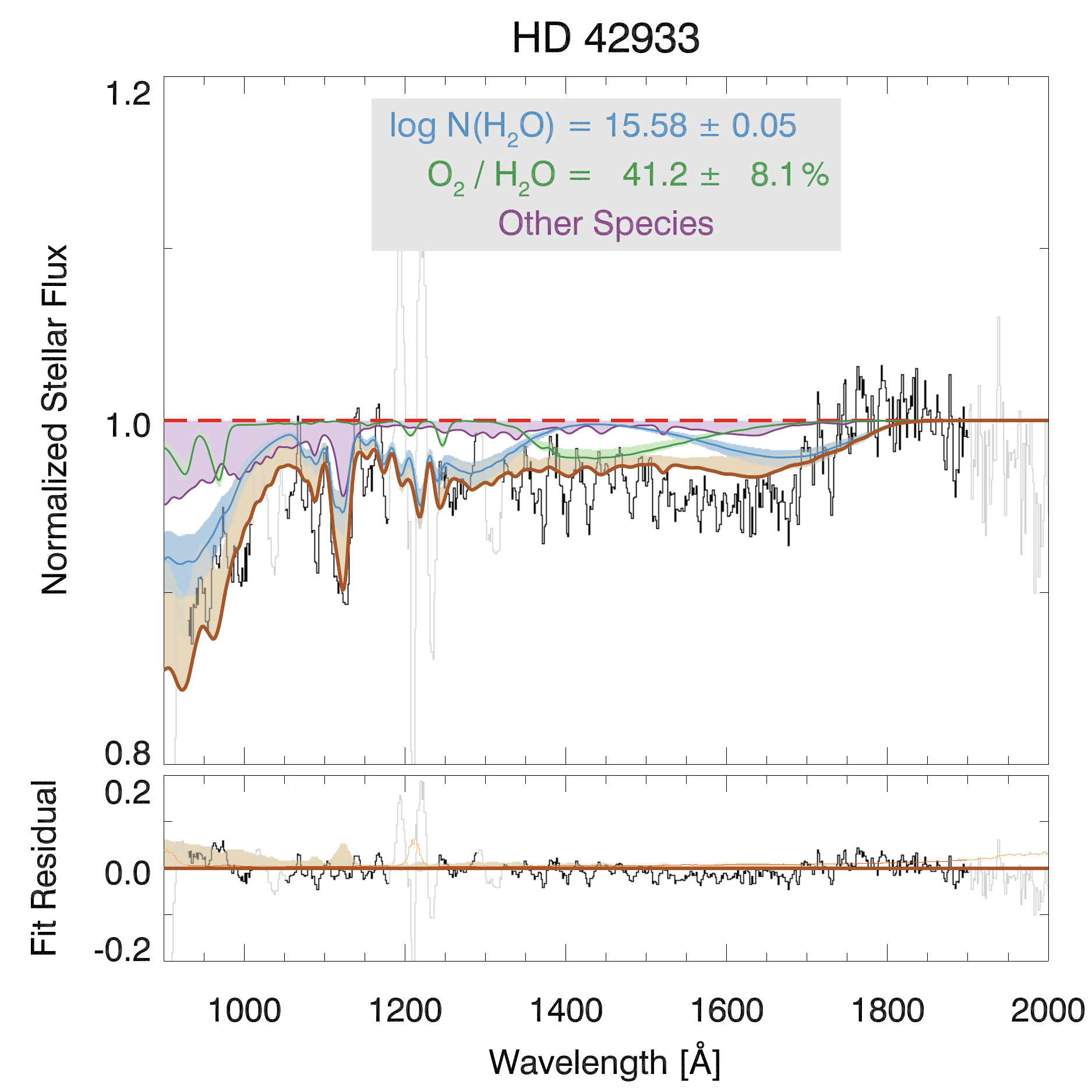}
\vspace{-1em}
\caption{Adopted column densities for the appulse of HD~42933 ($\mathrm{FQ}=3$), with 95\% ($2\sigma$) confidence bands.
\label{fig:conf_hd42933}}
\end{figure}

\begin{figure}
\centering\includegraphics[width=0.8\columnwidth]{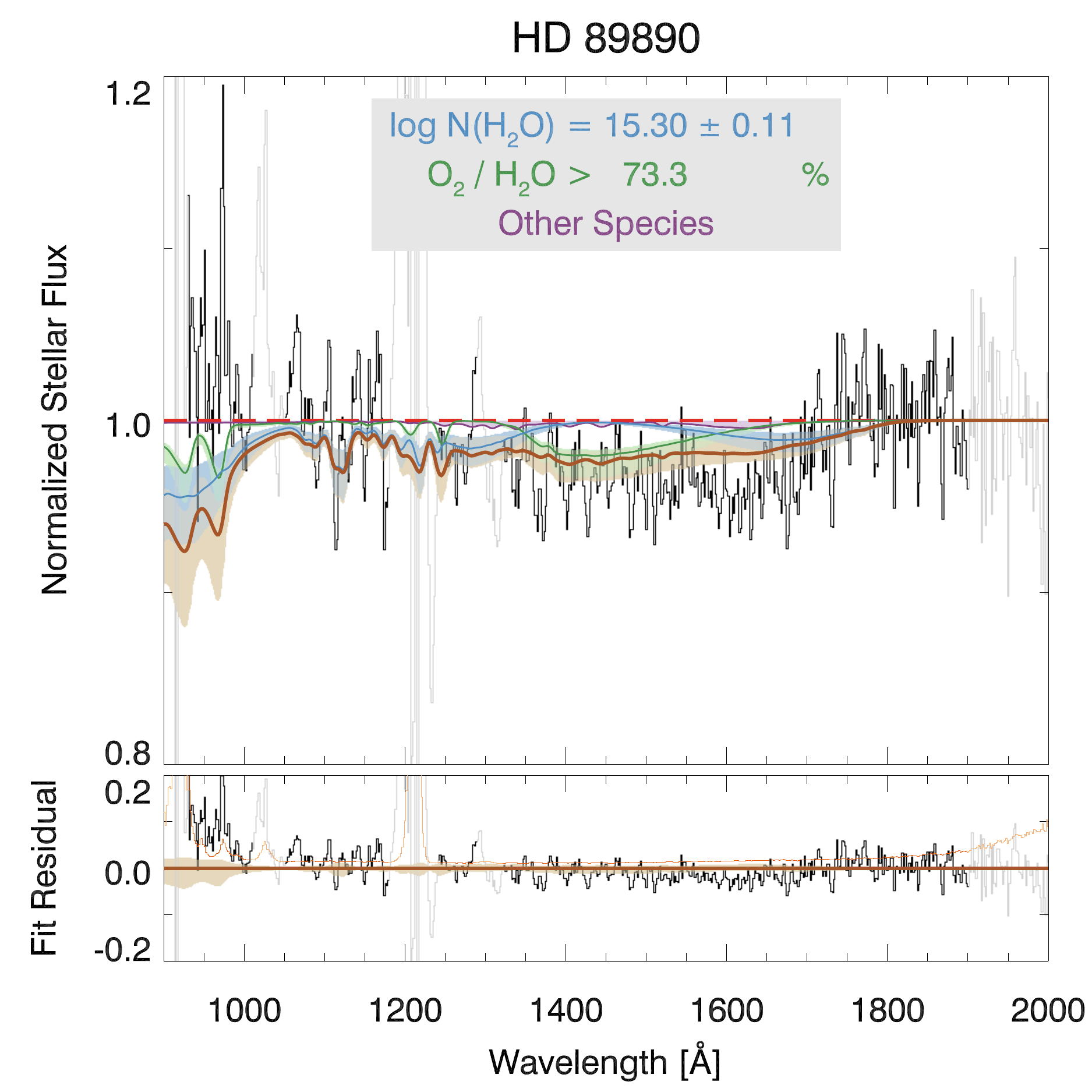}
\vspace{-1em}
\caption{Adopted column densities for the appulse of HD~89890 ($\mathrm{FQ}=4$), with 95\% ($2\sigma$) confidence bands.
\label{fig:conf_hd89890}}
\end{figure}

\begin{figure}
\centering\includegraphics[width=0.8\columnwidth]{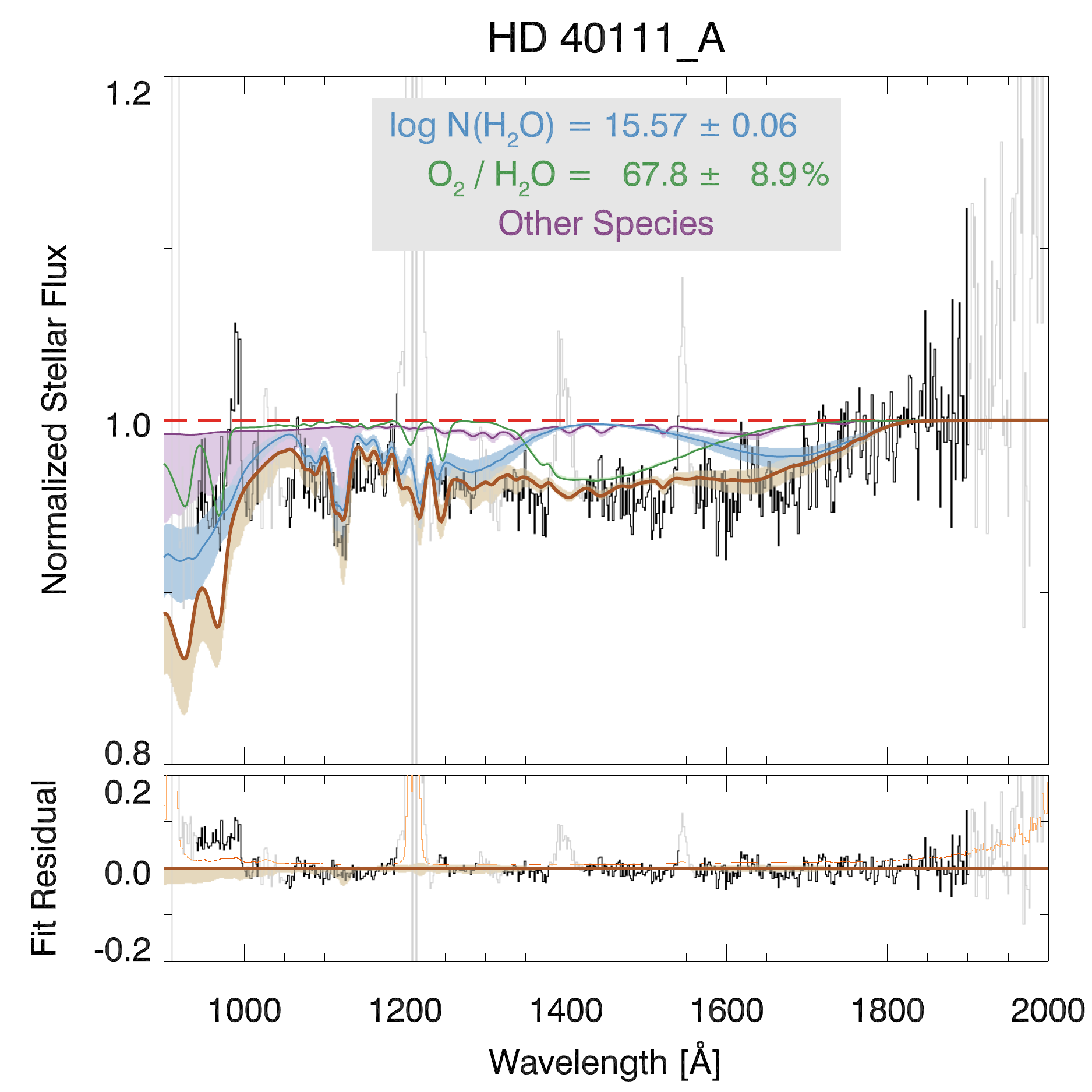}
\vspace{-1em}
\caption{Adopted column densities for the first appulse of HD~40111 ($\mathrm{FQ}=3$), with 95\% ($2\sigma$) confidence bands.
\label{fig:conf_hd40111_a}}
\end{figure}

\begin{figure}
\centering\includegraphics[width=0.8\columnwidth]{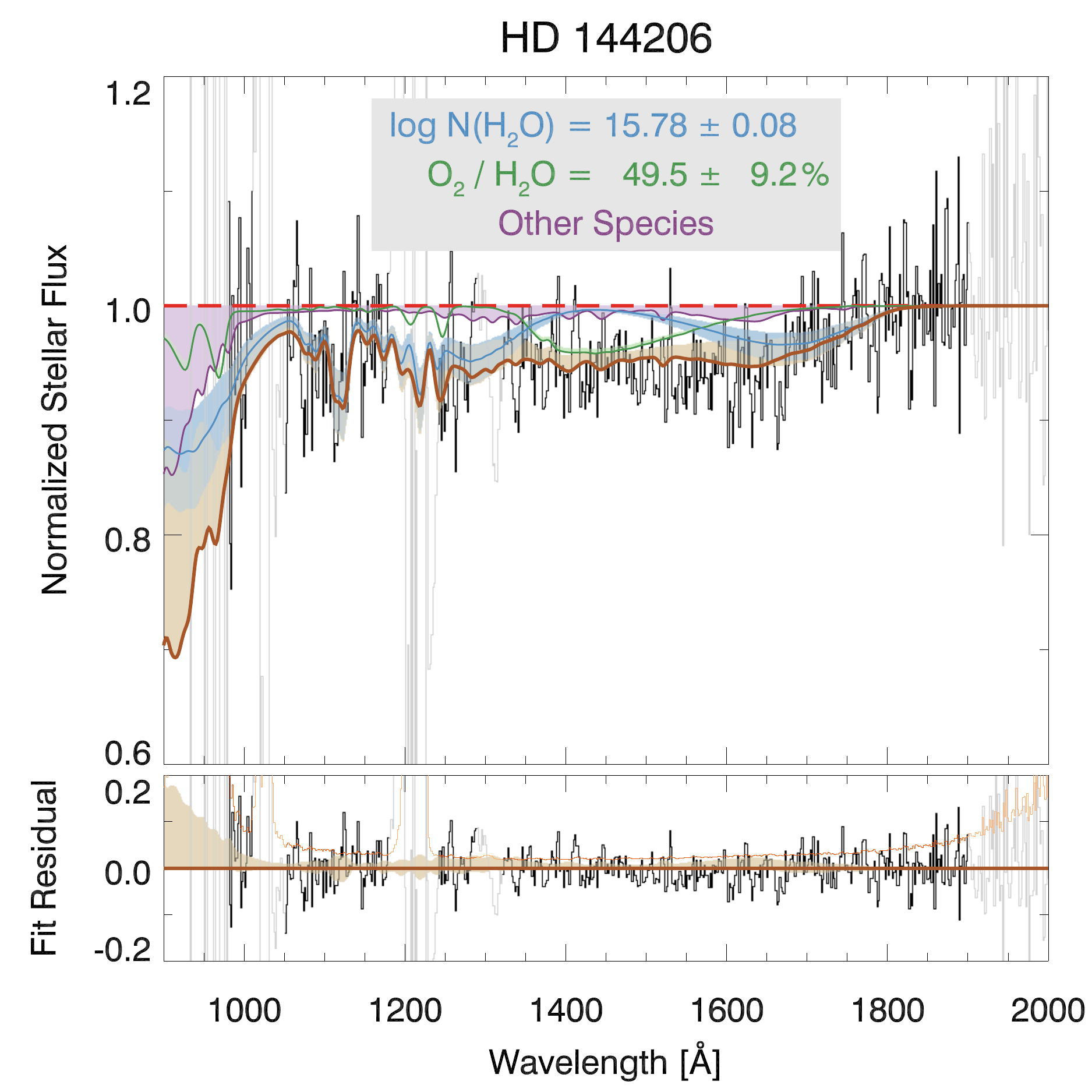}
\vspace{-1em}
\caption{Adopted column densities for the appulse of HD~144206 ($\mathrm{FQ}=2$), with 95\% ($2\sigma$) confidence bands.
\label{fig:conf_hd144206}}
\end{figure}

\begin{figure}
\centering\includegraphics[width=0.8\columnwidth]{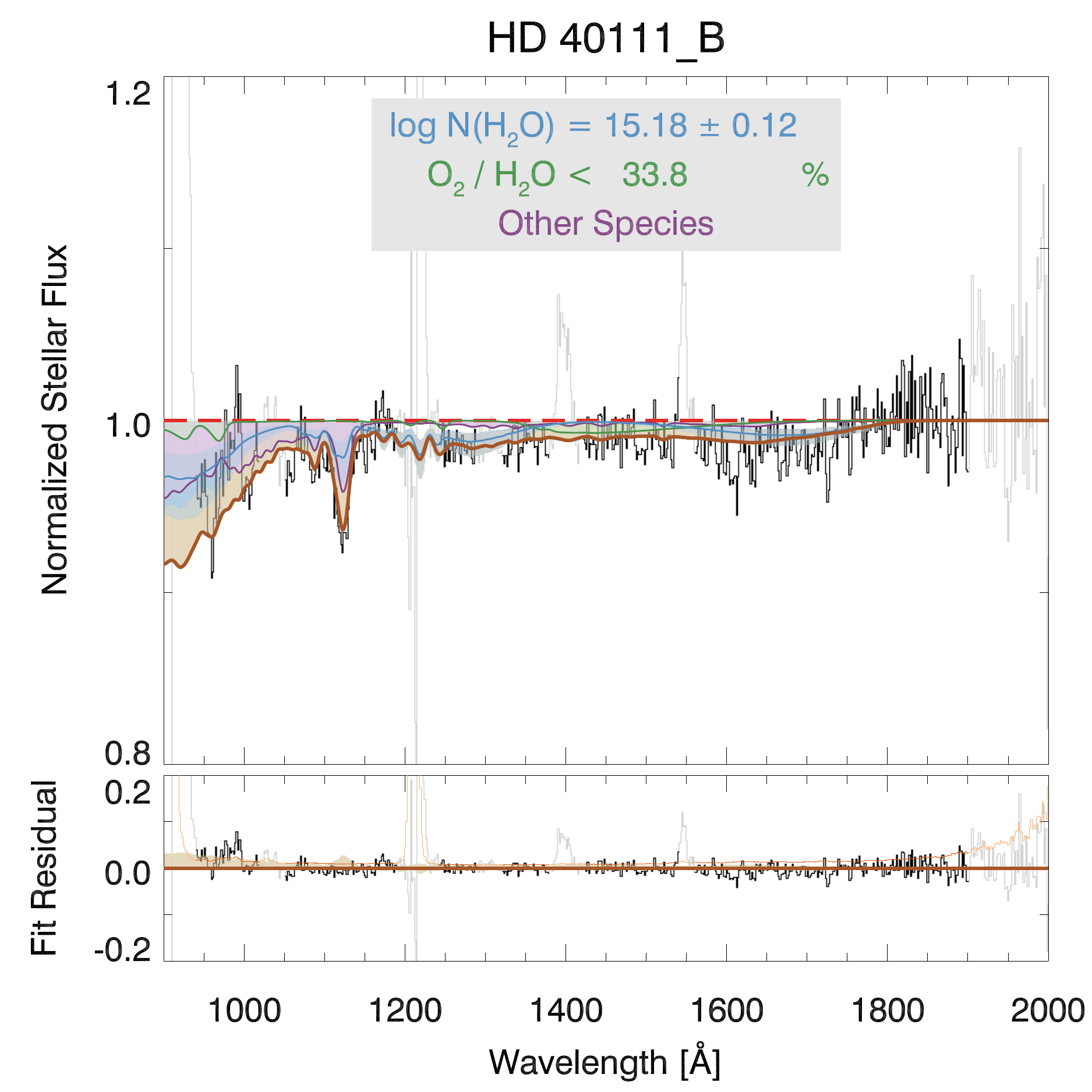}
\vspace{-1em}
\caption{Adopted column densities for the second appulse of HD~40111 ($\mathrm{FQ}=3$), with 95\% ($2\sigma$) confidence bands.
\label{fig:conf_hd40111_b}}
\end{figure}

Figures~\ref{fig:conf_hd26912}-\ref{fig:conf_hd40111_b} present the adopted column densities for all targeted and archival stellar appulses, with 95\% ($2\sigma$) confidence bands. The top panel of each figure displays the normalized stellar flux and associated 95\% confidence band (gray), with ensemble fit (brown) and individual-species absorption overlaid using the adopted column densities of \ce{H2O} and \ce{O2} from \autoref{tab:fits}. The bottom panel of each figure displays the residual of the ensemble fit.

\newpage

% Don't change these lines
\bsp	% typesetting comment
\label{lastpage}
\end{document}